\documentclass[12pt]{article}

\usepackage{amsmath}
\usepackage{amssymb} 
\numberwithin{equation}{section}
\usepackage{a4wide}
\usepackage{url}
\usepackage{hyperref}
\usepackage{graphicx}
\usepackage{xspace}

\newcommand{\vProb}{f}
\newcommand{\cH}{{\cal H}}
\newcommand{\cF}{{\cal F}}
\newcommand{\subProc}{\delta}
\newcommand{\momConf}{{\cal B}}

\def\CF{C_F}
\def\TR{T_R}
\def\CA{C_A}

\newcommand{\ee}{e^+e^-}
\newcommand{\eg}{e.g.\ }
\newcommand{\ie}{i.e.\ }
\newcommand{\order}[1]{{\cal O}\left(#1\right)}
\newcommand{\comment}[1]{\textbf{[#1]}}
\newcommand{\as}{\alpha_s}
\newcommand{\pb}{\;\mathrm{pb}}
\newcommand{\nb}{\;\mathrm{nb}}
\newcommand{\GeV}{\;\mathrm{GeV}}
\newcommand{\TeV}{\;\mathrm{TeV}}

\newcommand{\spherig}{S^\mathit{pheri}_{\perp,g}}
\newcommand{\spherog}{S^\mathit{phero}_{\perp,g}}

\newcommand{\Qperp}{Q_\perp}

\newcommand{\cC}{{\cal{C}}}
\newcommand{\cE}{{\cal{E}}}
\newcommand{\cR}{{\cal{R}}}
\newcommand{\CU}{\cC_U}
\newcommand{\CD}{\cC_D}
\newcommand{\gap}{\mathrm{gap}}
\newcommand{\evshp}{\text{ev-shp}}
\newcommand{\tw}{\textwidth}
\def\barcC{{\cal{\bar C}}}
\def\cO#1{{\cal{O}}\left(#1\right)}
\newcommand{\VEV}[1]{\langle #1 \rangle}
\newcommand{\NLL}{\mathrm{NLL}}

\newcommand{\hoppet}{\textsc{hoppet}\xspace}
\newcommand{\caesar}{\textsc{caesar}\xspace}
\newcommand{\nlojet}{\textsc{nlojet}\texttt{++}\xspace}

\date{}

\title{Phenomenology of event shapes at hadron colliders}
\author{Andrea Banfi,$^1$ Gavin P. Salam$^2$ and Giulia
  Zanderighi$^3$
  \\[0.5em]\normalsize
  $^1$ Institute for Theoretical Physics, ETH Zurich, 8093 Zurich,
  Switzerland.
  \\[0.1em]\normalsize
  $^2$ LPTHE; UPMC Univ.\ Paris 6; CNRS UMR 7589; Paris, France.
  \\[0.1em]\normalsize
  $^3$ Rudolf Peierls Centre for Theoretical Physics,
  1 Keble Road, University of Oxford, UK.
}
\begin{document}

\maketitle
\vspace{-22em}
\begin{flushright}
  January 2010
\end{flushright}
\vspace{20em}

\begin{abstract}
  We present results for matched distributions of a range of dijet event
  shapes at hadron colliders, combining next-to-leading logarithmic
  (NLL) accuracy in the resummation exponent, next-to-next-to leading
  logarithmic (NNLL) accuracy in its expansion and next-to-leading
  order (NLO) accuracy in a pure $\as$ expansion.
  This is the first time that such a matching has been carried out for
  hadronic final-state observables at hadron colliders. 
  We compare our results to Monte Carlo predictions, with and without
  matching to multi-parton tree-level fixed-order calculations.
  These studies suggest that hadron-collider event shapes have
  significant scope for constraining both perturbative and
  non-perturbative aspects of hadron-collider QCD.
  The differences between various calculational methods also highlight
  the limits of relying on simultaneous variations of renormalisation
  and factorisation scale in making reliable estimates of
  uncertainties in QCD predictions.
  We also discuss the sensitivity of event shapes to the topology of
  multi-jet events, which are expected to appear in many New Physics
  scenarios.
\end{abstract}
\newpage

%======================================================================
\tableofcontents

%======================================================================
\section{Introduction}
\label{sec:introduction}

Event shapes measure the geometrical properties of the energy flow in
QCD events and, notably, its deviation from that expected based on
pure lowest order partonic predictions.
Event shapes, as well as being among the first observables proposed to
test QCD~\cite{SU3}, have been inextricably tied with the progress of
QCD. 
They have played a crucial role in the extraction of the strong
coupling from properties of the final-state~\cite{Bethke:2002rv}.
They have been essential in tuning the parton showers and
non-perturbative components of Monte Carlo event
generators~\cite{Herwig,Sjostrand:2006za,Ariadne,TuningDelphi} and
have also provided a laboratory for developing and testing analytical
insight into the hadronisation process (e.g. refs.~\cite{DMW} and the reviews
\cite{Beneke,DasSalReview}).
From a technical point of view, the development of resummations and
fixed-order calculations has benefited from comparisons of predictions
for event-shape distributions obtained with both kinds of
methods~\cite{Broad,NG,y3,DISENTError,NNLOZurich,NNNLLthrust,NNLOWeinzierl}.
Additionally, they are one of the several tools that are used for
classifying hadronic final states in new physics searches.

The majority of investigations of event shapes has been performed for $\ee$ colliders,
with significant work also in DIS. A review of some of that work is
given in~\cite{DSreview}. In contrast, few dedicated studies have been
performed on them at hadron colliders, with a handful of measurements
at the Tevatron~\cite{CDF-Broadening,D0Thrust,NewCDFMesropian}, a pure
fixed order study~\cite{NLOJET}, pure resummations
in~\cite{KoutZ0,Banfi:2004nk,Stewart:2009yx}, and a recent experimental simulation study by
CMS~\cite{Dissertori:2008es}, as well as some investigation of the use
of event shapes applied to jet contents for the identification of
hadronic decays of boosted massive particles
\cite{Thaler:2008ju,Almeida:2008tp,Almeida:2008yp} (other approaches
are reviewed e.g.\ in ref.~\cite{Salam:2009jx}).
The purpose of this article is to help bring our understanding of
hadron-collider event-shape phenomenology closer to the level of
sophistication that is standard in the $\ee$ and DIS cases,
concentrating specifically on event shapes in hard QCD (dijet) events.

As is well known from the $\ee$ and DIS cases, accurate studies of
event shapes involve the simultaneous use of two kinds of
calculation.
Fixed-order calculations provide expansions of event-shape
distributions in powers of the strong coupling, $\as$. 
They are available up
to next-leading-order (NLO) for hadron-collider event shapes, through
the \nlojet~\cite{NLOJET} program.
When the event shape has a value $v \ll 1$, for each power of $\as$ in
the distribution there can be up two powers of a large logarithm,
$L = \ln 1/v$, associated with soft and collinear enhancements. This
compromises the convergence of the perturbative series.
The enhanced terms can however be resummed to all orders, providing
the dominant contribution to the distribution for $v\ll 1$. 
Such resummed predictions tend to be carried out for the distribution
integrated up to some value $v$, which generally has an exponentiated
structure $\exp( L g_1(\as L) + g_2(\as L) + \as g_3(\as L) +
\ldots)$.\,\footnote{Not all event-shape
distributions exponentiate, see e.g.~\cite{Brown:1990nm,caesar}.}
The $L g_1(\as L)$ term gives leading logarithmic (LL) accuracy in the
exponent, $g_2(\as L)$ is next-to-leading-logarithmic (NLL), etc.
For suitable (recursively infrared and collinear safe, global)
observables, the \caesar program \cite{caesar} calculates both $L
g_1(\as L)$ and $g_2(\as L)$.
To obtain a reliable prediction for an event shape distribution, one
must combine both types of calculations, via a ``matching''
procedure~\cite{CTTW}.
An appropriately performed matching of NLO fixed order and NLL
exponentiated resummation allows one to ensure that in the expansion
of the resummation one correctly accounts for all terms $\as^n L^p$,
with $2n-2 \le p \le 2n$, which is NNLL in the expansion.
While NLO+NLL with NNLL in the expansion is the state-of-the-art for
generic $\ee$ and DIS event shapes,\footnote{
  For specific observables, higher logarithmic and/or
  fixed-order accuracies have been reached.  This is e.g.\ the case
  with NNLL accuracy in the exponent for the $e^+ e^-$ energy-energy
  correlation~\cite{deFlorian:2004mp}, NNNLL for the thrust
  distribution in $e^+e^-$~\cite{NNNLLthrust} and NNLL for the Higgs
  or vector boson transverse momentum spectrum at hadron
  colliders~\cite{deFlorian:2000pr,Bozzi:2003jy}. Additionally NNLO
  accuracy has been achieved for a range of $\ee$ event
  shapes~\cite{NNLOZurich}, with NNLO+NLL matching
  in~\cite{Gehrmann:2008kh} and NNLO+NNNLL in~\cite{NNNLLthrust}.}
matching of this kind had not so far been achieved for hadron-collider
event shapes.

Our study here will give NLL+NLO predictions for event shape
distributions both at Tevatron ($p\bar p$, $\sqrt{s} = 1.96$ TeV) and
at the LHC ($pp$, $\sqrt{s} = 14$ TeV).
We start in Sec.~\ref{sec:event-shape-defin} by
recalling the definitions of three classes of global event shapes for
hadron colliders,\footnote{
  For non-global observables~\cite{NG}, those sensitive to emissions
  only in a restricted phase space region, the angular-ordered
  branching underlying \caesar's resummations does not account for all
  NLL effects. Additional soft, large-angle contributions have to be
  resummed, which is currently possible only in the large-$N_c$
  limit~\cite{NG,Banfi:2002hw} (though see also
  \cite{Weigert:2003mm}). This is the reason why, in this work, we
  consider only the global case.  } %
addressing also the question of event shapes defined in terms of jets
rather than particles.
In Sec.~\ref{sec:pert-struct} we describe the structure of the
perturbative resummation as well as its matching to the NLO result. We
also discuss possible general event-shape resummation issues
associated with ``super-leading'' logarithms
\cite{Forshaw:2006fk,Forshaw:2008cq}. 
In Sec. \ref{sec:pert-res} we present our results for
matched distributions, paying particular attention to the issue of
theoretical uncertainties. We also compare our results to those
obtained with parton-level shower Monte Carlo event generators, in
some cases also matched to exact multi-parton tree-level matrix elements. In
Sec.~\ref{sec:study-np-effects} we briefly discuss the impact of
non-perturbative effects, the hadronisation and the underlying
event. 
Finally, switching to more
phenomenological questions, in Sec.~\ref{sec:multi-jet} we compare various event
shapes' ability to distinguish characteristically different event
topologies, and examine their robustness in such tasks, both with
respect to parton showering and to the orientation of the final state
event. Our results are summarised in Sec.~\ref{sec:ressum}. Some
technical details are collected in
Appendices~\ref{sec:appA} and~\ref{sec:appB}. 
Many further additional plots can be obtained from the  URL~\cite{website}.

%---------------------------------------------------------------------------
\section{Event-shape definitions at hadron colliders}
\label{sec:event-shape-defin}

In this article, we shall consider observables that measure the extent
to which an event's energy flow departs from a dijet structure.
The lowest-order contribution to a dijet event consists of just two
incoming and two outgoing partons. Throughout the paper we refer to
these QCD configurations as the ``Born limit''.
Many of the event shapes studied here were presented for the first
time in~\cite{Banfi:2004nk}. 
All share the property of continuous globalness~\cite{DISresum,caesar}, which
is a necessary condition for being able to carry out a resummation to
NLL accuracy without a leading-$N_C$ approximation, and which also
contributes to the simplicity of \caesar's generalised resummation
approach (independently of the question of large-$N_C$
approximations).
For an observable to be
continuously global, it has to be sensitive to all emissions in an
event (this is the requirement of globalness), and moreover it should
have definite scaling properties with respect to secondary emission's
transverse momenta (see sec.~\ref{sec:preCAESAR} for a
mathematical formulation).
The continuously global event shapes we propose fall into three main
classes: observables that are directly global, others that are
supplemented with ``exponentially suppressed forward terms'' and
observables with ``recoil terms''.

%---------------------------------------------------------------------------
\subsection{Directly global observables}
\label{sec:dir-glob}
We first consider observables that are defined in terms of \emph{all}
hadrons in the event, therefore the name `directly' global.
The {\bf global transverse thrust} is defined as 
\begin{equation}
  \label{eq:Ttg}
  T_{\perp,g} \equiv \max_{\vec n_T} \frac{\sum_i |{\vec
      q}_{\perp i}\cdot {\vec
      n_T}|}{\sum_i q_{\perp i}}\,,
\end{equation}
where the sum runs over all particles $q_i$ in the final state, $\vec
q_{\perp i}$ represents the two momentum components transverse to the
beam, $q_{\perp i}$ its modulus, and $\vec n_T$ is the transverse
vector that maximises the sum. The observable which is resummed is
then $\tau_{\perp,g}\equiv 1-T_{\perp,g}$, which vanishes in the Born
limit.
The normalization of event shape observables to a hard transverse
scale of the event is important because it reduces uncertainties
associated with the experimental jet-energy scale, which partially
cancel between numerator and denominator~\cite{mweber}. 
For most event shapes (except $\tau_{\perp,g}$) the choice of specific
hard scale to which one normalises is arbitrary, and could for example
also be the sum of the transverse momenta of the two hardest jets.

The transverse thrust axis $\vec n_T$ and the beam form the so-called
event plane. One can then define a {\bf directly global thrust minor},
which is a measure of the out-of-event-plane energy flow
\begin{equation}
  \label{eq:Tmg}
  T_{m,g} \equiv
\frac{\sum_i |{\vec
      q}_{\perp i}\times {\vec
      n_T}|}{\sum_i q_{\perp i}}\,. 
%
%\frac{\sum_i |q_{xi}|}{\sum_i q_{\perp i}}\,,
\end{equation}

In close analogy with the $\ee$ case~\cite{Farhi}, one can formulate a
{\bf transverse spher\emph{o}city}:
\begin{equation}
  \label{eq:trans-sphero-matrix}
  \spherog \equiv  \frac{\pi^2}{4}\, \min_{{\vec n}
    = (n_x, n_y, 0)
    %_{\perp}
    %\,  n_z=0
  }\,
  \left(\frac{\sum_i | {\vec q}_{\perp, i} \times {\vec n}|}{\sum_i q_{\perp i}} \right)^2
\end{equation}
where the minimisation is carried over all possible unit transverse
$2$-vectors $\vec n$.\footnote{Numerically, the minimisation is
  simplified by the observation (based on extensive numerical tests)
  that the $\vec n$ that provides the minimal sum always coincides with the
  transverse direction of one of the $\vec q_i$.}  This variable
ranges from $0$ for pencil-like events, to a maximum of $1$ for
circularly symmetric events.

An alternative observable, which makes use of a linearised version of
the transverse momentum tensor (with direct analogy to the $C$ and $D$
parameters \cite{FoxWolf} used in $\ee$), is the {\bf
$\boldsymbol{F}$-parameter}:
\begin{equation}
  \label{eq:fpar}
  M^\mathit{lin} = \sum_i \frac1{q_{\perp i}}\left( 
    \begin{array}{cc}
      q_{xi}^2       & q_{xi} q_{yi}\\
      q_{xi} q_{yi}  & q_{yi}^2
    \end{array}
  \right)\,,\qquad\qquad
  F_g \equiv \frac{\lambda_2}{\lambda_1}
\end{equation}
where $\lambda_1 \ge \lambda_2$ are the two eigenvalues of
$M^\mathit{lin}$. Related variables have been considered in the plane
transverse to the thrust axis in $\ee$ (resummed for $3$-jet events in
\cite{BSZFpar}) and in the plane transverse to a jet in the context of
boosted top-quark decays
\cite{Thaler:2008ju,Almeida:2008yp,Almeida:2008tp}, where forms involving the
determinant of $M^\mathit{lin}$, \eg $4\lambda_1
  \lambda_2/(\lambda_1 + \lambda_2)^2 = 4F/(1+F)^2$, have been used. There is a
one-to-one mapping between different forms, and we have chosen
eq.~(\ref{eq:fpar}) because it gives clearer separation between
different kinematic regions.

Finally, we consider the exclusive variant of the
$k_t$-algorithm~\cite{KtHH} (closely related to the inclusive variant
\cite{Kt-EllisSoper} as adopted for Run II of the
Tevatron~\cite{RunII-jet-physics} and expected to be used also at the
LHC)
\begin{enumerate}
\item 
  One defines, for all $n$ final-state (pseudo)particles still in
  the event,  
  \begin{equation}
    \label{eq:diB_hh}
    d_{kB} = q_{\perp k}^2\,,
  \end{equation}
  and for each pair of final state particles
  \begin{equation}
    \label{eq:dij_hh}
    d_{kl} = \min\{{q_{\perp k}^2,q_{\perp l}^2}\}
    \frac{(y_k-y_l)^2+(\phi_k-\phi_l)^2}{R^2}\,,
  \end{equation}
  where $y_i = \frac12 \ln \frac{E_i+p_{zi}}{E_i-p_{zi}}$ is the
  rapidity of particle $i$ and $\phi_i$ its azimuthal angle.
  The jet-radius parameter $R$ sets the angular reach of the jet
  algorithm. Throughout this paper, we will take $R=0.7$.

\item One determines the minimum over $k$ and $l$ of the $d_{kl}$ and
  the $d_{kB}$ and calls it $d^{(n)}$. If the smallest value is
  $d_{iB}$ then particle $q_{i}$ is included in the beam and
  eliminated from the final state particles.  If the smallest value is
  $d_{ij}$ then particles $q_{i}$ and $q_{j}$ are recombined into a
  pseudoparticle (jet). A number of recombination procedures exist. We
  adopt the E-scheme, in which the particle four-momenta are simply
  added together,
\begin{equation}
  q_{i j} = q_{i}+q_{j}\,.
\end{equation}
\item The procedure is repeated until only 3 pseudoparticles are left
  in the final state.

The observable we resum is the {\bf directly global three-jet
resolution parameter} 
\begin{equation}
  \label{eq:hhy23}
  y_{23} = \frac{1}{P_{\perp}^2} \max_{n \ge 3}\{d^{(n)}\}\>,
\end{equation}
where $P_{\perp}$ is defined by further clustering the event until
only two jets
remain and taking $P_{\perp}$ as the sum of the two jet transverse
momenta,
\begin{equation}
  \label{eq:Eperpdef}
  P_{\perp} = p_{\perp,1} + p_{\perp,2}\,.
\end{equation}
\end{enumerate}

While directly global event-shapes are defined in terms of all
particles in the event, experimental measurements can be carried out
only up to some given pseudorapidity $\eta_{\rm max}$ ($\eta_{\rm max}
\sim 3.5$ at the Tevatron and $\eta_{\rm max} \sim 5$ at the LHC).
However, as long as the event-shape's value $v$ is not too
small~\cite{KoutZ0}, $v>v_{\min}$, one can safely neglect the
contribution of hadrons beyond the rapidity cut.  The value $v_{\min}$
up to which a NLL resummation is valid is observable specific. In
particular, it depends on the behaviour of each event shape under a
soft and collinear emission, as derived in~\cite{KoutZ0,Banfi:2004nk}.
Further discussion is given in appendix~\ref{sec:appB}.

%---------------------------------------------------------------------------
\subsection{Observables with exponentially suppressed forward terms}
\label{sec:exp-obs}
One way to address the difficulty in performing measurements
near the beam is to define event-shapes using only particles
in a central region and to add a term sensitive to emissions along the
beam direction, so as to render them global, but with an exponential
suppression in the forward or backward directions.
We define the central region ${\cal C}$ by requiring that the
rapidity of particles in ${\cal C}$ satisfies
%$|\eta_i| < \eta_c + \delta\eta$,
$|\eta_i| < \eta_c = y_{\rm j, \max} + \delta\eta$,
where $y_{\rm j, \max}$ specifies the rapidity region in which the two
highest $p_t$ jets should lie, and $\delta\eta$ is a rapidity buffer
around the jets of size $\sim 1$.

Given the central region $\cC$, we introduce the mean
transverse-energy weighted rapidity $\eta_\cC$ of this 
region,
\begin{equation}
  \label{eq:etabar}
  \eta_\cC = \frac{1}{
    Q_{\perp,\cC}} \sum_{i\in\cC} \eta_i\, q_{\perp i}\,,\qquad\quad
    Q_{\perp,\cC} = \sum_{i\in \cC} q_{\perp i}\,,
\end{equation}
and define the exponentially suppressed (boost-invariant) forward term as 
\begin{equation}
  \label{eq:exp-term}
  \cE_\barcC = \frac{1 }{Q_{\perp,\cC}}
  \sum_{i \notin \cC} q_{\perp i} \,e^{-|\eta_i - \eta_\cC|}\,.
\end{equation}
We can then define non-global variants of the event-shapes defined in
sec.~\ref{sec:dir-glob} by restricting the sums to just the central
region. For example, we have a {\bf central transverse thrust},
\begin{equation}
  \label{eq:Tperp-C}
  T_{\perp,\cC} \equiv \max_{\vec n_{T,\cC}}  \frac{\sum_{i\in\cC} |{\vec
      q}_{\perp i}\cdot {\vec n_{T,\cC}}|}{Q_{\perp,\cC}}\,, 
  \qquad\quad \tau_{\perp,\cC} \equiv 1 - T_{\perp,\cC}\,,
\end{equation}%
a {\bf central thrust minor}
\begin{equation}
  \label{eq:Tm-C}
  T_{m,\cC} \equiv \frac{1}{Q_{\perp,\cC}} \sum_{i \in \cC}
  |q_{xi}|\,,\qquad\quad
%  T_{m,\cE} = T_{m,\cC} + \cE_{\barcC}\,,
\end{equation}
and a {\bf central three-jet resolution threshold}, $y_{23,\cC}$ defined by
the algorithm of section~\ref{sec:dir-glob} applied only to the final state
particles in $\cC$ (but maintaining the ``beam'' distance,
eq.~(\ref{eq:diB_hh})).

Finally, we define ``exponentially suppressed'' variants of the
event-shapes of sec.~\ref{sec:dir-glob} by adding to the
central event-shapes a power of $\cE_\barcC$ which makes the
event-shape continuously global~\cite{Banfi:2004nk}. We obtain the
{\bf exponentially suppressed transverse thrust, thrust minor and
three-jet resolution},
\begin{equation}
  \label{eq:tauperp-E}
  \tau_{\perp,\cE} \equiv \tau_{\perp,\cC} + \cE_{\barcC}\,,
\end{equation}
\begin{equation}
  \label{eq:Tm-CE}
  T_{m,\cE} = T_{m,\cC} + \cE_{\barcC}\,,
\end{equation}
\begin{equation}
  \label{eq:y23-E}
  y_{23,\cE} \equiv y_{23,\cC} + \cE_\barcC^2\,.
\end{equation}

Additionally, one can consider event-shapes which are more naturally
defined using particles only in a restricted region, like jet-masses
and broadenings.
Given a central transverse thrust axis $\vec n_{T,\cC}$, one can
separate the central region $\cC$ into an up part $\cC_U$ consisting
of all particles in $\cC$ with $\vec p_{\perp} \cdot \vec n_{T,\cC} >
0$ and a down part $\cC_D$ with $\vec p_{\perp} \cdot \vec n_{T,\cC} <
0$ respectively.
One then defines, in analogy with $\ee$ \cite{Clavelli}, the
normalised squared invariant masses of the two regions
\begin{equation}
  \label{eq:mass-XC}
  \rho_{X,\cC} \equiv \frac{1}{Q_{\perp,\cC}^2}
  \left(\sum_{i\in \cC_X} q_{i}\right)^2\,,\qquad X = U, D\,,
\end{equation}
from which one can obtain a (non-global) {\bf central sum of masses and
heavy-mass},
\begin{equation}
  \label{eq:mass-C-sum-heavy}
  \rho_{S,\cC} \equiv \rho_{U,\cC} + \rho_{D,\cC}\,,\qquad\quad
  \rho_{H,\cC} \equiv \max\{\rho_{U,\cC}, \rho_{D,\cC}\}\,,
\end{equation}
and the corresponding global event-shapes, the {\bf
exponentially-suppressed sum of masses and heavy-mass}
\begin{equation}
  \label{eq:mass-E-sum-heavy}
  \rho_{S,\cE} \equiv \rho_{S,\cC} + \cE_{\barcC}\,,\qquad\quad
  \rho_{H,\cE} \equiv \rho_{H,\cC} + \cE_{\barcC}\,.
\end{equation}

With the same division into up and down regions as for the jet masses,
one can define jet broadenings. To do so in a boost-invariant manner,
one first introduces rapidities and azimuthal angles of axes for the
up and down regions,
\begin{equation}
  \label{eq:broadening-axes}
  \eta_{X,\cC} \equiv \frac{\sum_{i\in \cC_X} q_{\perp i}
    \eta_i}{\sum_{i\in \cC_X} 
    q_{\perp i}}\,,\qquad
  \phi_{X,\cC} \equiv \frac{\sum_{i\in \cC_X} q_{\perp i}
    \phi_i}{\sum_{i\in \cC_X} 
    q_{\perp i}}\,,\qquad
   X = U, D\,,
\end{equation}
and defines broadenings for the two regions,
\begin{equation}
  \label{eq:BX-C}
  B_{X,\cC} \equiv \frac{1}{2Q_{\perp,\cC}} \sum_{i\in\cC_X}
  q_{\perp i}\sqrt{(\eta_i-\eta_{X,\cC})^2 + (\phi_i - \phi_{X,\cC})^2}\,,
  \qquad X = U, D\,,
\end{equation}
from which one can obtain {\bf central total and wide-jet broadenings},
\begin{equation}
  \label{eq:B-C-total-wide}
  B_{T,\cC} \equiv B_{U,\cC} + B_{D,\cC}\,,\qquad\quad
  B_{W,\cC} \equiv \max\{B_{U,\cC}, B_{D,\cC}\}\,.
\end{equation}
Adding the forward term one obtains the global {\bf
  exponentially-suppressed total and wide-jet broadenings},
\begin{equation}
  \label{eq:BTWE}
  B_{T,\cE} \equiv B_{T,\cC} + \cE_{\barcC}\,,\qquad
  B_{W,\cE} \equiv B_{W,\cC} + \cE_{\barcC}\,.
\end{equation}
We note that an observable that effectively has exponentially
suppressed forward behaviour has also been studied in
\cite{Stewart:2009yx}.

%---------------------------------------------------------------------------
\subsection{Observables with recoil term (indirectly global observables)}
\label{sec:rec-obs}
Because of transverse momentum conservation, if radiation is emitted
in the forward region $\bar \cC$, recoil effects will cause the vector
sum of the transverse momenta in the complementary, central region
$\cC$ to be non-vanishing. It is then possible to exploit this effect
to make observables (continuously) global, despite the fact that only
a central subset of particles in the event effectively enters the
definition of the event shapes. To do so, we add to the central
event-shapes a suitable power of a recoil term, the two-dimensional
vector sum of the transverse momenta in $\cC$,
\begin{equation}
  \label{eq:recoil-term}
  \cR_{\perp,\cC} \equiv \frac{1}{Q_{\perp,\cC}} \left|\sum_{i \in \cC} {\vec
        q}_{\perp i}\right|\,,
\end{equation}

We obtain than {\bf recoil enhanced transverse thrust, thrust minor,
three-jet resolution, sum- and heavy-jet masses, total and wide
broadenings}
\begin{subequations}
\begin{align}
  \label{eq:tauperp-R}
  \tau_{\perp,\cR} &\,\equiv\, \tau_{\perp,\cC} + \cR_{\perp,\cC}\,,
  \\
  \label{eq:Tm-R}
  T_{m,\cR} &\,\equiv\, T_{m,\cC} + \cR_{\perp,\cC}\,,
  \\
  \label{eq:y23-R}
  y_{23,\cR} &\,\equiv\, y_{23,\cC} + \cR_{\perp,\cC}^2\,,
  \\
  \label{eq:mass-R-sum-heavy}
  \rho_{S,\cR} &\,\equiv\, \rho_{S,\cC} + \cR_{\perp,\cC}\,,\;\,\qquad\quad
  \rho_{H,\cR} \,\equiv\, \rho_{H,\cC} + \cR_{\perp,\cC}\,,
  \\
  \label{eq:BTWR}
  B_{T,\cR} &\,\equiv\, B_{T,\cC} + \cR_{\perp,\cC}\,,\qquad\quad
  B_{W,\cR} \,\equiv\, B_{W,\cC} + \cR_{\perp,\cC}\,.
\end{align}
\end{subequations}

%----------------------------------------------------------------------
\subsection{Particles versus jets as inputs} 
\label{sec:part-v-jets-as-input}

The event shapes discussed so far have all been defined in terms of
the particles in the event. 
The experiments don't measure particles directly. They do, however,
have methods such as the combination of information from
electromagnetic and hadronic calorimeters into ``Topoclusters''
(ATLAS~\cite{Aad:2009wy}) and, with tracking, ``particle flow''
(CMS~\cite{Ball:2007zza}), that provide inputs to jet algorithms that
are quite close to particles.
These same inputs would probably also be well suited to event-shape
studies.

In uses of event shapes to cut on event topology in
beyond-standard-model searches, as well as in the study of
ref.~\cite{Dissertori:2008es}, it is not particles but instead jets
that have been used as inputs.
The jets are usually defined through an angular resolution parameter
$R$ (as in eq.~(\ref{eq:dij_hh})) and a transverse momentum cutoff,
which we will denote $p_{t0}$.
One of the interests of using jets is that the $p_{t0}$ cutoff
eliminates much of the contamination from the underlying event, which
can easily contribute $\order{100\GeV}$ of transverse momentum to the
rapidity region covered by LHC detectors.

From the point of view of resummation, the use of jets as inputs poses
two main problems.
One comes from the presence of the new scale $p_{t0}$ in the
problem: in terms of the parameters $a$ and $b_{1,2}$ defining the
event-shape's sensitivity to radiation along the incoming legs
(cf.\
table~\ref{tab:evshp-param} and section~\ref{sec:preCAESAR}), this
new scale causes separate regions of event shape value to each involve
different logarithmic structure: for cases with $b_{1,2}>0$ the
potentially different regions are $v \gg (p_{t0}/Q)^a$, $(p_{t0}/Q)^a
\gg v \gg (p_{t0}/Q)^{a+b_{1,2}}$ and $v \ll (p_{t0}/Q)^{a+b_{1,2}}$.
The first of these regions may be within the scope of \caesar if
$p_{t0}/Q$ is sufficiently small.

A second problem is that of globalness. Emissions collinear to any
outgoing hard parton will be clustered together with its emitter to
form a jet. The observable's sensitivity to these emissions will then
depend on the jet recombination scheme. 
In the $E$-scheme, the current default at the Tevatron and LHC, the jet
four-momentum is constructed by simply adding the four momenta of its
constituents. Therefore, all observables defined using transverse
momenta will get no sensitivity to emissions inside each of the two
hard jets, and will therefore be non-global.
This statement is true for any recombination scheme that adds
three-momenta vectorially.
For variables with sensitivity to longitudinal degrees of freedom,
globalness can only be assessed on a case by case basis. For instance
if one considers any global version of the total and heavy-jet mass
(with exponentially suppressed or recoil term), in the $E$-scheme the
mass of each central hard jet will enter the hemisphere central jet mass
$\rho_{X,\cC}$ in eq.~(\ref{eq:mass-XC}). Therefore one obtains the
same result for the central component of the event shape as would have
been obtained using hadrons as inputs (modulo the fact that the jet
clustering may affect which particles are considered central).

One alternative to the use of jets as inputs, in order to avoid the
globalness issue, is the following: use as inputs the particles that
are inside the two hardest jets, together with all the remaining jet
momenta.
Note that this does not eliminate the issue of the extra scales
related to $p_{t0}$, though it does maintain the reduced sensitivity
to underlying event that comes from the use of jets.

%======================================================================
\section{Structure of the perturbative calculation}
\label{sec:pert-struct}

Typically one wishes to consider  event shapes only for events that
are sufficiently hard, requiring for example at least one jet above some minimum
transverse momentum threshold $p_{t,\min}$ and in some central
rapidity region. 
We will denote this kind of hardness selection cut by a function
$\cH(q_1,\ldots,q_N)$ of the $N$ particles in the event;
$\cH(q_1,\ldots,q_N)$ is equal to $1$ for events that pass the cuts
and $0$ otherwise. 
One can then define the cross section for events
that pass the cuts,
\begin{equation}
  \label{eq:sigmacut}
  \sigma = \sum_{N} \int d\Phi_N\,
  \frac{d\sigma_N}{d\Phi_N}\,
  \cH(q_1,\ldots,q_N)\,,
\end{equation}
where $d\sigma_N/d\Phi_N$ is the differential cross section for
producing $N$ particles in some configuration $\Phi_N$. 
One can determine $\sigma$ perturbatively as long as $\cH$
corresponds to an infrared and collinear (IRC) safe selection
procedure.

One also defines the partial integrated cross section $\Sigma(v)$ for
events that pass the cut and for which additionally the event shape
observable $V(q_1,\ldots,q_N)$ is smaller than some value $v$,
\begin{equation}
  \label{eq:SigmaIntcut}
  \Sigma(v) = \sum_{N} \int d\Phi_N
  \frac{d\sigma_N}{d\Phi_N} \,\Theta(v - V(q_1,\ldots,q_N))\,
  \cH(q_1,\ldots,q_N)\,.
\end{equation}
The differential normalised distribution for the event shape is then
given by
\begin{equation}
  \label{eq:diffdist}
  \frac{1}{\sigma} \frac{d\Sigma(v)}{dv}\,.
\end{equation}
Perturbatively we will write $\sigma$ and $\Sigma(v)$ as
expansions in the number of powers of the coupling that they contain,
\begin{equation}
  \sigma    = \sigma_{0} + \sigma_{1} + \sigma_{2} + \ldots\\
\end{equation}
where $\sigma_{0}$ is the leading order (LO) result,
$\sigma_{1}$ the NLO result, etc.;
$\sigma_{i}$ is proportional to $\as^{2+i}$. We have chosen not to
extract the powers of $\as$ from the $\sigma_{i}$ coefficients,
because the scale of $\as$ may depend on the kinematics of the events
over which one has integrated.

The expansion for $\Sigma(v)$ is similar
\begin{equation}
  \label{eq:Sigma-expansion}
  \Sigma(v) = \Sigma_{0}(v) + \Sigma_{1}(v) +
  \Sigma_{2}(v) + \ldots \,,
\end{equation}
with the property that $\Sigma_{0}(v) \equiv \sigma_{0}$
because the observable vanishes at $\order{\as^2}$.
$\Sigma_{1}(v)$ looks like a NLO term in
eq.~(\ref{eq:Sigma-expansion}), but it is usually determined from the
LO $\as^{3}$ term for the differential cross section of $v$,
\begin{equation}
  \label{eq:Sigma-from-diff}
%  \Sigma_{1}(v) = \sigma_{1} - \int_{v'>v}\!\!\!\!\! dv'\, \frac{d\Sigma_{1}}{dv'}\,.
  \Sigma_{1}(v) = \sigma_{1} + \bar \Sigma_1(v), \qquad 
\bar \Sigma_1(v) = - \int_v dv'\, \frac{d\Sigma_{1}(v')}{dv'}\,.
\end{equation}
The quantity $\bar \Sigma_{2}(v)$ is similarly determined from the NLO term of the
differential cross section of $v$. In the following we shall never use explicitly
$\Sigma_2$, since $\sigma_2$, the NNLO correction to the dijet cross
section, has yet to be calculated and since its effect would lead to
terms that are beyond our accuracy in differential distributions.

%----------------------------------------------------------------------
\subsection{Resummation}
\label{sec:resummation}

Resummations are relevant in the region of small $v$, where
logarithmically enhanced contributions of soft and collinear origin,
as large as $(\as \ln ^2 v)^n$, appear at all orders in the integrated
cross section $\Sigma(v)$, thus making fixed-order predictions
unreliable. 
There is a large class of observables for which one can write a common
``master'' resummation formula, as was done in~\cite{caesar}, in order
to sum such terms to all orders in $\as$.
In this section we will first examine what the class of observables
is, and then review the broad structure of the resummation. 

\subsubsection{Prerequisites for resummation with \caesar}
\label{sec:preCAESAR}
In order for an observable to be resummed within the \caesar framework,
its functional behaviour in the presence of an arbitrary number of
soft and/or collinear emissions has to satisfy a number of
conditions. These have been extensively discussed in~\cite{caesar},
and are checked automatically by \caesar given a computer subroutine
that computes the value of the an observable given a set of four-momenta.
The conditions are:
\begin{enumerate} 
\item
  a specific functional form for the observable's dependence
  $V(\{{\tilde p}\},k)$ on the momentum of a \emph{single} soft
  emission $k$, collinear to one of the hard ``Born'' partons
  (``legs'') in the event:
\begin{equation}
  \label{eq:parametricform}
  V(\{{\tilde p}\}, k)=
  d_{\ell}\left(\frac{k_t^{(\ell)}}{Q}\right)^{a_\ell}
  e^{-b_\ell\eta^{(\ell)}}\, 
  g_\ell(\phi)\>,
\end{equation}
where $\{{\tilde p}\}$ denote the Born momenta (including recoil
effects) and $k$ is the soft collinear emission; $k_t^{(\ell)}$ and
$\eta^{(\ell)}$ denote respectively its transverse momentum and
rapidity, as measured with respect to the Born parton (`leg') labelled
$\ell$; $\phi$ is the azimuthal angle of the emission with respect to
a suitably defined event plane (when relevant); $g(\phi)$ can be any
function for which $\int d\phi \ln g(\phi)$ is well defined;
$Q$ is a hard scale of
the problem (taken here to be the sum of the transverse
momenta of the two hardest jets).

\item {\em continuous globalness}~\cite{NG,DiscontGlobal}, a
  requirement on the observable's single-emission scaling properties
  in every region of the phase space. First, all the $a_\ell$ have to
  be equal, $a_1 = a_2 = \ldots \equiv a$, and the $d_\ell$ have to be
  all non-zero. Second, the observable's scaling at the boundaries of
  the soft collinear region has to be consistent with
  eq.~(\ref{eq:parametricform}), i.e.~in the soft large-angle region
  we require $V(\{\tilde p\},k)\sim k_t^{a}$ for a fixed angle of $k$,
  whilst for hard emission 
  collinear to leg $\ell$ we must have $V(\{\tilde p\},k)\sim
  k_t^{a+b_\ell}$ at fixed energy for $k$.

\item {\em recursive infrared and collinear safety}, a subtle
  mathematical condition (see~\cite{caesar} for its precise
  formulation) concerning the observable's scaling in the presence of
  multiple soft/collinear emissions.

\end{enumerate}
Table~\ref{tab:evshp-param} summarises the values of the coefficients
$a_\ell$ and $b_\ell$ for the event shapes presented in
section~\ref{sec:event-shape-defin}. We stress that central
observables, like the central transverse thrust
eq.~(\ref{eq:Tperp-C}), defined using only hadron momenta in a
selected rapidity interval, tend to have $d_{1,2}=0$, and 
therefore be non-global.
\comment{but not true for $\tau_{\perp,g}$.}
The exponentially-suppressed term $\cE_\barcC$
in eq.~(\ref{eq:exp-term}) or the recoil term $\cR_{\perp,\cC}$ in
eq.~(\ref{eq:recoil-term}), as explained in sections~\ref{sec:exp-obs}
and~\ref{sec:rec-obs}, are added to central event shapes precisely so
as to make them global. The different powers of $\cE_\barcC$ and
$\cR_{\perp,\cC}$ that appear in the definition of these modified
event shapes (see for instance eqs.~(\ref{eq:tauperp-E}) and
(\ref{eq:y23-E})) are chosen so as to ensure their continuous
globalness. This can be seen by observing that, for each event shape,
the coefficients $a_\ell$ corresponding to different legs are
equal. This above discussion holds for most observables but there may
be exceptions. For example, the central variant of the thrust-minor
$T_{m,\cC}$ is actually a global observable because of an indirect
sensitivity to non-central emissions due to recoil. This is the reason
why the $b_{1,2}$ coefficients for $T_{m,\cE}$ are not those that
usually appear for observables with exponentially-suppressed
components, but are rather those typical of a (linear) recoil term.
\begin{table}
  \centering
  \begin{tabular}{|c|cc|cc|}\hline
                       & $a_{1,2}$ & $b_{1,2}$ & $a_{3,4}$ & $b_{3,4}$ \\\hline
    $\tau_{\perp,g}$   &  1        &    0      &    1      &  1        \\
    $\tau_{\perp,\cE}$ &  1        &    1      &    1      &  1        \\
    $\tau_{\perp,\cR}$ &  1        &    1      &    1      &  0        \\\hline
    $T_{m,g}$          &  1        &    0      &    1      &  0        \\
    ${T_{m,\cE}}$      &  1        &    0      &    1      &  0        \\
    $T_{m,\cR}$        &  1        &    0      &    1      &  0        \\\hline
    $y_{23}$           &  2        &    0      &    2      &  0        \\
    $y_{23,\cE}$       &  2        &    2      &    2      &  0        \\
    $y_{23,\cR}$       &  2        &    0      &    2      &  0        \\\hline
    $B_{T/W,\cE}$      &  1        &    1      &    1      &  0        \\
    $B_{T/W,\cR}$      &  1        &    0      &    1      &  0        \\\hline
    $\rho_{S/H,\cE}$   &  1        &    1      &    1      &  1        \\
    $\rho_{S/H,\cR}$   &  1        &    0      &    1      &  1        \\\hline
    $F_{g}$            &  1        &    0      &    1      &  1        \\\hline
    $\spherog$         &  2        &    0      &    2      &  0        \\\hline
  \end{tabular}
  \caption{ Table of event shapes being considered here and the powers of their
    parametric sensitivity to the transverse momentum ($a_\ell$) and
    collinear angle ($b_\ell$) of an emission along incoming  ($a_{1,2},
    b_{1,2}$) and outgoing ($a_{3,4}, b_{3,4}$) hard partons.
  }
  \label{tab:evshp-param}
\end{table}

Recursive infrared and collinear (rIRC) safety, a detailed discussion of which is
beyond the scope of the present paper, is trivially satisfied for all
observables that we discuss here.

\subsubsection{NLL resummation structure}
\label{sec:NLL_resum}

For global observables, in events with $v \ll 1$, it is possible,
unambiguously, to associate the event kinematics with that of a $2\to
2$ (Born) event. This is because the requirement $v\ll 1$ forces all
radiation to be either soft or collinear.
At perturbative level it is also possible to unambiguously attribute a
partonic subprocess to the event, for example $qq\to qq$ (doing so
requires a flavour infrared and collinear safe procedure, as in \cite{jetflav},
but the result is independent of the choice of the 
procedure).
Here we will use $\momConf$ to label the event's $2\to2$ kinematics and
$\subProc$ to label its $2\to 2$ flavour structure.
Then we can write $\Sigma(v)$ as a sum over partonic subprocesses and
an integral over Born configurations that pass the hard event cuts,
\begin{equation}
  \label{eq:Sigmacut_sum_int}
  \Sigma(v) = \sum_\subProc \Sigma^{(\delta)}(v)\,,\qquad
  \Sigma^{(\delta)}(v) =  \int d\momConf\,
  \frac{d\Sigma^{(\subProc)}(v)}{d\momConf} \cH(\momConf)\,,\
  \quad \qquad (v\ll 1).
\end{equation}
The ambiguities in such a decomposition of $\Sigma(v)$ are suppressed
by powers of $v$.

For observables that satisfy the properties of the previous section,
the result of ref.~\cite{caesar} is that we can write
\begin{equation}
  \label{eq:Sigmacut_through_f}
  \frac{d\Sigma^{(\subProc)}(v)}{d\momConf} =
  \frac{d\sigma_0^{(\subProc)}(v)}{d\momConf}
  \vProb^{(\subProc)}_\momConf(v) (1 + \order{\as})   \qquad (v\ll 1) \,.
\end{equation}
where $d\sigma_0^{(\subProc)}(v)/d\momConf$ is the LO cross section,
differential in the Born configuration, separated into
subprocesses, and understood to have been evaluated with a
factorisation scale $\mu_F \sim Q$.
The function $f^{(\subProc)}_\momConf(v)$ encodes
the resummation, and has the form~\cite{CSS,CTTW}
\begin{equation}
  \label{eq:vProb-general}
  \vProb_{\momConf}^{(\subProc)}(v) = \exp\left[ 
    L g_1^{(\subProc)}(\as L) 
    + g_{2,\momConf}^{(\subProc)}(\as L, \mu_R, \mu_F)
    + \order{\as^n L^{n-1}} 
  \right]\,, \qquad L = \ln \frac1v\,,
\end{equation}
where $\as \equiv \as(\mu_R)$, with $\mu_R$ some renormalisation
scale of order $Q$.

The order-by-order expansion of $\vProb_{\momConf}^{(\subProc)}(v)$
involves terms of the form $\as^n L^{2n}$.
It is because of the property of ``exponentiation'' (a consequence of
rIRC safety and of coherence\footnote{%
  The validity of coherence is brought into question by the findings
  of refs.~\cite{Forshaw:2006fk,Forshaw:2008cq} and we discuss the
  possible implications of this in section~\ref{sec:superl-logar}.
}) that one can write it in the form
eq.~(\ref{eq:vProb-general}), whose leading-logarithmic (LL)
contribution in the exponent, $L g_1^{(\subProc)}(\as L) $, resums
terms $\as^n L^{n+1}$: i.e.\ corrections to the first order $\as L^2$
term involve only powers of $\as L$.
The function $g_{2,\momConf}^{(\subProc)}(\as L, \mu_R, \mu_F)$ resums
``next-to-leading logarithmic'' (NLL) terms in the exponent, $\as^n
L^{n}$, also referred to sometimes as single-logarithmic terms.

The LL function $Lg_1^{(\subProc)}(\as L)$ can be computed
analytically given only the $a_\ell$ and $b_\ell$ values.  It is given
by
\begin{align}
  \label{eq:Lg1-expansion}
  Lg_1^{(\subProc)}(\as L) &= 
  -\sum_\ell 
  \frac{C_\ell^{(\subProc)}  L}{2\pi\beta_0 \lambda b_{\ell}}
  \left((a-2 \lambda)
    \ln\left(1-\frac{2\lambda}{a}\right)
    -(a+b_{\ell}-2\lambda)
    \ln\left(1-\frac{2\lambda}{a+b_{\ell}}\right)
  \right)
  \nonumber
  \\
  &=
  -\sum_\ell \frac{C_\ell^{(\subProc)}}{a(a +
    b_\ell)}\frac{\as L^2}{\pi}  +
  \order{\as^2 L^3}\,,
\end{align} 
where $C_\ell^{(\subProc)}$ is the colour charge ($\CF$ or $\CA$) of
hard parton $\ell$ for the hard-scattering subprocess $\subProc$,
$\lambda = \as \beta_0 L$ and $\beta_0 = (11\CA - 4\TR n_f)/(12\pi)$.
Since the coefficients $a\equiv a_1 = a_2 = \ldots$ and $b_\ell$ do
not depend on the particular momentum configuration $\momConf$ of the
hard partons, $g_1^{(\subProc)}(\as L)$ is also independent of $\momConf$.
Its dependence on the subprocess arises only through the colour
charges of the incoming and outgoing partons.

The NLL function $g_{2,\momConf}^{(\subProc)}(\as L, \mu_R, \mu_F)$
can be decomposed into three types of terms~\cite{caesar}
\begin{equation}
  \label{eq:g2-separated}
  g_{2,\momConf}^{(\subProc)}(\as L, \mu_R, \mu_F) =
  g_{2s,\momConf}^{(\delta)}(\as L, \mu_R) + 
  \sum_{\ell=1}^2\ln\left[
    \frac{q_\ell^{(\subProc)}(x_\ell^{(\momConf)}\!,v^{\frac{1}{a+b_\ell}}\mu_F)}
    {q_\ell^{(\subProc)}(x_\ell^{(\momConf)}\!,\mu_F)}
  \right] + \ln \cF^{(\delta)}(R'(\as L))\,,
\end{equation}
The term $g_{2s}(\as L)$ accounts for NLL corrections associated with
the event kinematics, the particular values of the $d_\ell$ and
$g_\ell(\phi)$ coefficients in eq.~(\ref{eq:parametricform}), the
choice of renormalisation scale and scheme used in $\as$ in $g_1(\as
L)$, as well as the non-trivial colour evolution of large-angle soft
virtual gluon
resummation~\cite{BottsSterman,KS,KOS,Oderda,KidonakisOwens}.

The second term on the right-hand side (RHS) of eq.~(\ref{eq:g2-separated})
involves parton distribution functions (PDF) for the parton
flavours in the initial state of the given subprocess,
$q^{(\subProc)}_\ell(x_\ell^{(\momConf)}\!,\mu_F)$, at a longitudinal
momentum fraction $x_\ell^{(\momConf)}$ for each leg that depends on
the Born kinematics.
This term arises because the PDFs in $d\sigma_0^{(\subProc)}/d\momConf$
in eq.~(\ref{eq:Sigmacut_through_f}) were evaluated at a factorisation
scale $\mu_F \sim Q$.
The presence of a PDF at  scale $\mu_F \sim Q$ implies
that one integrates over all possible incoming collinear emissions, up
to $k_t \sim Q$.
However the requirement that the event shape be small, $V(k) \lesssim
v$, restricts collinear emissions to have $k_t
\lesssim v^{1/(a+b_\ell)}Q$.
Thus the PDFs should actually be evaluated at a factorisation scale
$\sim v^{1/(a+b_\ell)} Q \sim v^{1/(a+b_\ell)}\mu_F$ (as occurs also
in Drell-Yan transverse-momentum
resummations~\cite{CSS,KodairaTrentadue,CollinsSoper}).
The ratio of PDFs in eq.~(\ref{eq:g2-separated}) serves to
replace $q^{(\subProc)}_\ell(x_\ell^{(\momConf)},\mu_F)$ as used in
the Born cross section with a PDF at the correct factorisation scale.

The third term on the RHS of eq.~(\ref{eq:g2-separated}) accounts for
the NLL corrections associated with the presence of multiple soft and
collinear emissions, when each has $V(k)\sim v$ and they are all
widely separated in rapidity.  It is a function of
\begin{equation}
  \label{eq:Rprime}
  R'(\as L) \equiv -\,\partial_L L g_1(\as L)\,,
\end{equation}
and is known analytically for some observables (e.g. $\tau_{\perp,g}$),
while in all other cases \caesar can compute it numerically via a
suitable Monte Carlo procedure.
$\cF^{(\subProc)}(R')$ sometimes depends on the underlying scattering channel
$\subProc$, but not (for the observables studied here) on the hard
momentum configuration.

The behaviour of $\cF(R')$ with increasing $R'$ (decreasing $v$) is a
characteristic feature of each event shape. It depends on whether
multiple emissions tend to increase or decrease the value of the event
shape.  
In the first case, for a fixed value $v$, $\cF(R')$ has to
account for an extra suppression of emissions so as to keep the event
shape's value less than $v$, \ie $\cF(R') < 1$.  
For the special case of $V(k_1, k_2,\ldots) =
\max(V(k_1),V(k_2),\ldots)$ then $\cF(R')\equiv 1$ (for example the
$y_{23}$ jet resolution threshold~\cite{y3} in the $e^+e^-$ Cambridge
jet algorithm~\cite{Cam}).
Conversely if the
contributions of multiple emissions tend to cancel, the function
$\cF(R')$ has to compensate the excessive suppression given by the LL
function $L g_1(\as L)$, therefore $\cF(R') > 1$. 

This last case appears most
dramatically when it is a cancellation between multiple emissions, and not
a direct veto on real emissions, that is the dominant effect that
keeps the event shape small. 
In this case the LL function $L g_1(\as L)$ (whose functional form
depends only on the effect of single emission) no longer accounts for
the dominant contribution to the distribution. Furthermore, no NLL function 
such as $\cF(R')$ can fully compensate for this.
This inconsistency reveals itself through a divergence of $\cF(R')$ at
a given critical value $R'_c$, which can be inferred from
considerations on the cancellation mechanism, as explained in
refs.~\cite{Banfi:2004nk,caesar}.
Such a divergence is present, for example, for $T_{m,\cC}$ and
$T_{m,\cE}$ and occurs at $R'_c=C_T/(C_1+C_2)$, where $C_1$ and $C_2$
are the colour charges of incoming partons and $C_T$ the total colour
charge of the hard parton system.
It will prevent us from obtaining sensible NLL resummed results
for these observables.
In the case of recoil observables there is also a divergence, but at
larger $R'_c$ (smaller $v$), for example at $R'_c = 2C_T/(C_1+C_2)$
for $T_{m,\cR}$ and $B_{T/W,\cR}$.
The effect on the corresponding differential distributions will be
discussed when presenting matched results.

%......................................................................
\subsubsection{NNLL$_\Sigma$ accuracy}
\label{sec:NNLL_Sigma}

As well as discussing the LL, NLL, etc. accuracy of resummation in the
exponent of eq.~(\ref{eq:vProb-general}), one can also discuss the
accuracy in the order by order expansion of $\Sigma$ itself. 
In this way of counting logarithms, ``LL$_\Sigma$'' terms involve powers
$\as^n L^{2n}$, NLL$_\Sigma$ involve $\as^n L^{2n-1}$, etc.
A NLL resummation in the exponent automatically guarantees
NLL$_\Sigma$ accuracy.
However it is also possible (if not entirely straightforward), given
the information at our disposal, to obtain NNLL$_\Sigma$ accuracy.
To see how, observe that the terms that we
neglect in eqs.~(\ref{eq:Sigmacut_through_f}) and
(\ref{eq:vProb-general}) are an overall $\as$ correction without
logarithms, as well as terms $\as^n L^{n-1}$ in the exponent, starting
at $\as^2 L$.
The latter, if they multiply $\as^n L^{2n}$ when expanding the exponent,
lead at most to $\as^{n+2} L^{2n+1} \sim \as^n L^{2n-3}$, i.e.\ they are
NNNLL$_{\Sigma}$. We can therefore ignore them.
As for the overall $\as$ correction, when multiplied by the double
logarithms, it gives us terms $\as^{n+1} L^{2n} \sim \as^n L^{2n-2}$, which are
NNLL$_{\Sigma}$ and therefore cannot be neglected.
This means that we need to determine the coefficient of the pure
$\order{\as}$ term.

To do so, let us define the NLL resummed cross section as
\begin{equation}
  \label{eq:Sigmar_def}
  \frac{d\Sigma_r^{(\subProc)}(v)}{d\momConf} \equiv
  \frac{d\sigma_0^{(\subProc)}(v)}{d\momConf}
  \vProb^{(\subProc)}_{\NLL,\momConf}(v)  \,.
\end{equation}
with $\vProb^{(\subProc)}_{\NLL,\momConf(v)}$ containing \emph{only}
the LL and NLL resummation terms, 
\begin{equation}
  \label{eq:vProb-NLL}
  \vProb_{\NLL,\momConf}^{(\subProc)}(v) \equiv \exp\left[ 
    L g_1^{(\subProc)}(\as L) 
    + g_{2,\momConf}^{(\subProc)}(\as L, \mu_R, \mu_F)
  \right]\,.
\end{equation}
Then we can determine the coefficient $C_{1,\momConf}^{(\subProc)}$ in
terms of the first order expansion of the exact and resummed
distributions, and in particular their difference as $v\to0$,
\begin{equation}
  \label{eq:C1-def}
  \as C_{1,\momConf}^{(\subProc)}
  \equiv
  \lim_{v\to0} \left(\frac{d\Sigma_1^{(\subProc)}(v)}{d\momConf} -
    \frac{d\Sigma_{r,1}^{(\subProc)}(v)}{d\momConf} \right)
  \Bigg/
  \frac{d\sigma_0^{(\subProc)}}{d\momConf} 
  \,.
\end{equation}
The $C_{1,\momConf}^{(\subProc)}$ constant involves many
contributions, including parts that cancel the $\mu_R$ and $\mu_F$
dependence present in the Born cross section, parts that are sensitive
the observable's exact behaviour with respect to soft large-angle
emission and hard collinear splitting and parts related to the exact
structure of the 1-loop $2\to2$ scattering diagram.

Now we can write the NNLL$_\Sigma$  resummed distribution as
\begin{equation}
  \label{eq:NLLL_Sigma_differential}
  \frac{d\Sigma_r^{(\subProc)}(v)}{d\momConf} (1 + \as
  C_{1,\momConf}^{(\subProc)})\,. 
\end{equation}
The fact that we may multiply
$(1 + \as C_{1,\momConf}^{(\subProc)})$ and
$\frac{d\Sigma_r^{(\subProc)}(v)}{d\momConf}$ in order to get
NNLL$_\Sigma$ accuracy is a consequence of the property that
soft-collinear virtual corrections, which give powers of $\as L^{2}$,
affect neither the flavour, the momentum nor the colour involved in
the hard scattering or the PDFs and therefore straightforwardly
multiply all the more complicated contributions that are present in
$C_{1,\momConf}^{(\subProc)}$. 

In practice, it is not feasible to (numerically) determine the exact
first order distribution for $v$ fully differentially in the Born
configurations.
Considering instead quantities $\Sigma^{(\subProc)}(v)$,
$\Sigma_r^{(\subProc)}(v)$ (and their order-by-order expansions),
integrated over configurations that pass the event cuts, as in
eq.~(\ref{eq:Sigmacut_sum_int}), one can define a $C_1^{(\subProc)}$
coefficient averaged over Born momentum configurations,
\begin{equation}
  \label{eq:C1_subProc}
  \VEV{\as C_{1}^{(\subProc)}} \equiv \lim_{v\to 0}
  \frac{\Sigma_{1}^{(\subProc)}(v) -
    \Sigma_{r,1}^{(\subProc)}(v)}{\sigma_0^{(\subProc)}}
  =
  \frac{1}{\sigma_0^{(\subProc)}}\int d\momConf 
  \frac{d\sigma_0^{(\subProc)}}{d\momConf}
  \as C_{1,\momConf}^{(\subProc)} 
\,. 
\end{equation}
Writing 
\begin{equation}
  \label{eq:still_NNLL_Sigma}
  (1 + \VEV{\as C_{1}^{(\subProc)}}) \Sigma_{r}^{(\subProc)}(v)\,,
\end{equation}
gives a distribution that is still correct to NNLL$_\Sigma$ accuracy,
because the LL, $\exp(L g_1^{(\subProc)}(\as L))$, component of
$\Sigma_{r}^{(\subProc)}(v)$ is independent of the momentum
configuration.

In contrast, if one considers $C_1$ averaged additionally over
subprocesses 
\begin{equation}
  \label{eq:C1_fullav}
  \VEV{\as C_{1}} \equiv 
  \lim_{v\to 0}   \frac{\Sigma_{1}(v) - \Sigma_{r,1}(v)}{\sigma_0}
  = 
  \frac{1}{\sigma_0} \sum_\subProc 
  \VEV{\as C_{1}^{(\subProc)} } \sigma_0^{(\subProc)}\,,
\end{equation}
then 
\begin{equation}
  \label{eq:not_NNLL_Sigma}
  (1 + \VEV{\as C_{1}}) \Sigma_{r}(v)\,,
\end{equation}
is not accurate to NNLL$_\Sigma$, because
\begin{equation}
  \label{eq:why_not_NNLL_Sigma}
  \sum_\subProc
  \VEV{\as C_{1}^{(\subProc)} }
   \sigma_0^{(\subProc)} \exp(L
    g_1^{(\subProc)}(\as L))
  \ne
  \left(
    \frac{1}{\sigma_0}\sum_\subProc 
    \VEV{\as C_{1}^{(\subProc)} }
    \sigma_0^{(\subProc)}
  \right)
  \left(
    \sum_\subProc \sigma_0^{(\subProc)}
    \exp(L g_1^{(\subProc)}(\as L))
  \right),
\end{equation}
since the coefficient of the double logarithms in $L
g_1^{(\subProc)}(\as L)$ does depend on the subprocess, through the
colour charges of the hard partons.

%----------------------------------------------------------------------
\subsection{Matching of NLL to NLO}
\label{sec:matchNLLNLO}

While the resummation of logarithms is necessary in the region where
event-shape values are small and their logarithms large, the region of
large values of $V$ is dominated by events with three or more well
separated jets. %partons.
Those types of events are described more reliably by fixed order
calculations. It has therefore become standard to match resummed
calculations to next-to-leading order (NLO) to have a reliable
prediction over a larger range of values of $V$.
In this section we will present the formulae we use to perform the
matching.

In the following we will denote with $f(v) = \Sigma (v)/\sigma$ the
integrated event-shape fraction, where $\Sigma(v)$ and $\sigma$ are
defined in eqs.~\eqref{eq:SigmaIntcut} and ~\eqref{eq:sigmacut}
respectively.
After a NLL+NLO matching this quantity should satisfy the following requirements
\begin{enumerate}
\item it should respect the physical constraints that, when the event shape
  reaches its maximum value $v_{\max}$, we have $f(v_{\max})=1$
  exactly and $\frac{df(v)}{dv}\big|_{v=v_{\max}}\!\!\!\!=0$;
%$df(v)/dv=0$ for $v=v_{\max}$.
\item its expansion up to relative $\cO{\as^2}$ should reproduce the
  exact NLO result for the corresponding differential distribution;
\item one should obtain NNLL$_\Sigma$ accuracy, i.e.\ all logarithms
  $\cO{\as^n L^{m}}$ with $m \ge 2n-2$ should be correctly accounted
  for, which implies that the matching formula should reduce to
  eq.~(\ref{eq:still_NNLL_Sigma}) in the limit of small $v$.
  Preferably this should be the case without having to go through the
  tedious procedure of manually determining $C_1^{(\subProc)}$
  separately for each event shape.
\end{enumerate}
There are various matching procedures that satisfy these
requirements and therefore formally have the same accuracy.  We
consider here the so-called log-$R$~\cite{CTTW} and multiplicative~\cite{DISresum}
matching schemes, adapted to hadron-hadron
collisions. 
In particular both need to be modified in accordance with the need,
section~\ref{sec:NNLL_Sigma}, to have the $\order{\as}$ constant
$C_1^{(\subProc)}$ term multiply the resummation separately for each
subprocess.
Actually, what matters is not so much the subprocess but the colour
charges of the incoming and the outgoing Born partons.
Therefore we can consider all subprocesses $qq\to qq$, $qq'\to qq'$,
$q\bar q \to q\bar q$, etc., with the same incoming and the same
outgoing colour charges as belonging to a single colour channel $a =
qq\to qq$. The other colour channels are $qg \to qg$, $q\bar q \to
gg$, $gg \to q\bar q$ and $gg \to gg$.%
\footnote{ We study here only  observables whose double
  logarithms depend only on the total colour charge of the two incoming
  and two outgoing partons, so that we do not to distinguish incoming
  partons $1$ and $2$ (or outgoing partons $3$ and $4$).  This means that
  for the matching only the colour structure is relevant, therefore this
  colour labelling does not distinguish quarks from anti-quarks or quarks
  of different flavour. A given colour channel $a$ is then in general a
  sum over multiple partonic channels $\subProc$.}

We denote by $\Sigma_{r, i}^{(a)}(V)$ the expansion of the resummed
cross section corresponding to a specific colour channel and by
$\Sigma_{i}^{(a)}(V)$ the corresponding exact fixed order
prediction. In analogy with eq.~(\ref{eq:Sigmacut_sum_int}), where the
index $a$ is omitted a sum over all possible colour channels is
understood. The index $i$ denotes the order in $\as$ of the expansion
(relative to the Born cross section).

We have obtained fixed order cross-sections
using the code \nlojet~\cite{NLOJET}. The publicly available version computes
cross-sections summed over the flavour of outgoing partons. We
therefore extended it so as to have access to the flavour of
both incoming and outgoing partons in the calculation of $\sigma_0$,
$\sigma_1$ and $\Sigma_1(v)$, though not for $\bar \Sigma_2(v)$
since its colour-channel separation  is not needed for NNLL$_\Sigma$
accuracy.
To assign events with more than two outgoing partons to a definite
$2\to 2$ colour channel we used the exclusive flavour-$k_t$ algorithm
of~\cite{jetflav} to cluster events to a $2 \to 2$ topology while
keeping track of flavour in an infrared safe manner.
During the clustering procedure, quarks of different flavour might end
up in the same jet, giving rise to multi-flavoured jets, i.e.\ jets whose flavour
does not correspond to any QCD parton.
These events, which do not correspond to any Born $2\to 2$ processes
and have vanishing weights for $v\to 0$, will be labelled as having
$a=\mathrm{other}$. % Note that

The matching equations are defined in terms of the following resummed
distribution (and its fixed order expansions $\tilde
\Sigma_{r,1}^{(a)}(v),\tilde \Sigma_{r,2}^{(a)}(v)$),
\begin{equation}
  \label{eq:Sigmacut_resummedtilde}
  \tilde \Sigma_r^{(a)}(v) =\sum_{\subProc \in a} \int d\momConf\,
  \frac{d\sigma_0^{(\subProc)}}{d\momConf}\,\cH(p_{3},p_{4}) \,  
  \,\tilde \vProb_{\momConf}^{(\subProc)}(v) \,,
\end{equation}
where $\tilde \vProb_{\momConf}^{(\subProc)}(v)$ is 
the resummed probability
$\vProb_{\momConf}^{(\subProc)}(v)$ (eq.~(\ref{eq:vProb-NLL}))
with $L$ replaced by~\cite{CTTW,DISresum}
\begin{equation}
  \tilde L \equiv \frac{1}{p} \ln \left(\left(\frac{1}{x_V v}\right)^p
    -\left(\frac{1}{x_V v_{\rm max}}\right)^p +1\right),\qquad x_V = X
    \cdot X_V\,, 
  \label{eq:Ltilde}
\end{equation}
where $v_{\rm max}$ is the maximum kinematically allowed value of the
event shape, so that $\tilde L(v=v_{\rm max}) = 0$. We take the values
of $v_{\rm max}$ from the NLO calculation, which is sensible since we
want the differential distributions to reproduce the NLO result at
high $v$.
The factors $x_V$ and $p$ modify the definition of the
logarithm that one is resumming.
The main effect of $x_V$ is to modify the logarithm at small values of
$V$, and it will therefore affect subleading logarithmic terms (the
change at NLL is cancelled via a suitable compensatory term in $g_2(\as
L)$). The main effect of $p$ on the contrary is to modify $L$ at large
values of $V$, it will therefore mainly affect power suppressed
terms. Our default values for $x_V$, $X_V$ are given by (see Appendix A
of ref.~\cite{Banfi:2004nk}) are fixed by setting $X=1$ and 
\begin{equation}
\label{eq:Xv}
\ln X_V = -\frac{1}{n}\sum_{\ell = 1}^n \left(\ln d_\ell + \int
\frac{d\phi}{2\pi} \ln g_\ell(\phi)\right)\,. 
\end{equation}
In the same way as one varies renormalization and factorization scales
around a central value by a factor of 2, we will probe the $x_V$
dependence through a variation of $X$ in the range $1/2 \le X \le 2$. This
will provide an estimate of the error associated with unknown NNLL
contributions to the resummation.
In principle, one can also vary the power $p$ around the value 1 (as a
probe of terms that are suppressed by powers of $v$), though for
simplicity in the following we just fix $p=1$ and therefore do not
include any uncertainty related to its variation.

We now introduce the log-R matching formula
\begin{equation}
       f(v) = \frac{\tilde f(v)}{\tilde f(v_{\rm max})}\,,
    \label{eq:second-logR-hh-alt}
\end{equation}
with 
\begin{multline}
    \label{eq:second-logR-hh-alt2}
       \tilde f(v) = \frac{1}{\sigma_{0}+\sigma_{1}}\left\{\sum_{a \ne \mathrm{other}}
      \tilde \Sigma_{r}^{(a)}(v)
      \exp\left[\frac{\Sigma_{1}^{(a)}(v)-\tilde \Sigma_{r, 1}^{(a)}(v)}{\sigma_{0}^{(a)}} 
      \right]
      \times \right.\\ \left.\times
      \exp\left[\frac{\bar\Sigma_{2}(v)-\tilde \Sigma_{r, 2}(v)}{\sigma_0}
        - \frac{1}{\sigma_0} \sum_{a\neq\text{other}}
          \frac{(\Sigma_{1}^{(a)}(v))^2-(\tilde \Sigma_{r, 1}^{(a)}(v))^2}{2\sigma_{0}^{(a)}}
      \right]
      +\Sigma_{1}^{(\mathrm{other})}(v)\right\}\>, 
\end{multline}
where $\bar \Sigma_2(v)$ has been introduced after
eq.~\eqref{eq:Sigma-from-diff}.  It is straightforward to verify that
with this matching equation $f(v)$ satisfies all three requirements listed at
the beginning of this section.

The alternative, multiplicative matching (mod-R) scheme that we use is 
%% \begin{multline}
%%   \label{eq:second-logR-hh-alt}
%%   f(V) = \frac{1}{\sigma_0+\sigma_1}\left\{\sum_{a \ne \mathrm{other}}
%%     \sigma^r_a(V)
%%     \left[1 + \frac{\sigma^e_{1,a}(V)-\sigma^r_{1,a}(V)}{\sigma_{0,a}} 
%%     \right]
%%     % 
%%     \times \right.\\ \left.\times
%%     \left[1 + \frac{\bar\sigma^e_2(V)-\sigma^r_2(V)}{\sigma_0}
%%       - \frac{1}{\sigma_0} \sum_{a\neq\text{other}}
%%       \sigma^r_{1,a}(V)
%%       \frac{\sigma^e_{1,a}(V)-\sigma^r_{1,a}(V)}{\sigma_{0,a}} 
%%     \right]
%%     %% 
%%     +\sigma^e_{1,\mathrm{other}}(V)\right\}\>.  
%% \end{multline}
%% Or perhaps better:
%% \begin{multline}
%%   \label{eq:second-mult-hh}
%%   f(V) = \frac{1}{\sigma_0+\sigma_1}\left\{\sum_{a \ne \mathrm{other}}
%%     \sigma^r_a(V)
%%     \left[1 + \frac{\sigma^e_{1,a}(V)-\sigma^r_{1,a}(V)}{\sigma_{0,a}} 
%%     \right]
%%     % 
%%     + \right.\\ \left.+
%%     \sigma^r(V)\left[\frac{\bar\sigma^e_2(V)-\sigma^r_2(V)}{\sigma_0}
%%       - \frac{1}{\sigma_0} \sum_{a\neq\text{other}}
%%       \sigma^r_{1,a}(V)
%%       \frac{\sigma^e_{1,a}(V)-\sigma^r_{1,a}(V)}{\sigma_{0,a}} 
%%     \right]
%%     %% 
%%     +\sigma^e_{1,\mathrm{other}}(V)\right\}\>.  
%% \end{multline}
%% which can equivalently be rewritten as 
\begin{multline}
  \label{eq:second-mult-hh-alt}
  f(v) = \frac{1}{\sigma_0+\sigma_1}\left\{\sum_{a \ne \mathrm{other}}
    [\tilde \Sigma_{r}^{(a)}(v)]^{Z} (\sigma_{0}^{(a)})^{1-Z}
    \left[1 + \frac{\Sigma_{1}^{(a)}(v) - Z \tilde \Sigma_{r,1}^{(a)}(v)}{\sigma_{0}^{(a)}} 
    \right.
    +\frac{\bar\Sigma_2(v)- Z \tilde \Sigma_{r,2}(v)}{\sigma_0}
     \right.\\ \left.\left.
      - \frac{1}{\sigma_0} \sum_{a'\neq\text{other}}
      Z\tilde \Sigma_{r,1}^{(a')}(v)
      \frac{\Sigma_{1}^{(a')}(v) 
        - \frac{Z+1}{2} 
        %+((Z-1)/2-Z)
        \tilde \Sigma_{r,1}^{(a')}(v)}{\sigma_{0}^{(a')}} 
    \right]
    +\Sigma_{1}^{(\mathrm{other})}(v)\right\}\>,  
\end{multline}
where $Z = \left(1-\frac{v}{v_{\max}}\right)$.  This matching equation
has the same matching accuracy as eq.~\eqref{eq:second-logR-hh-alt},
so that using both matching procedures provides an additional way of
estimating the uncertainty in the matched distributions.

In both matching formulae, $\tilde \Sigma_{r,1}(v)$ and $\tilde
\Sigma_{r,2}(v)$ require the calculation of the order $\as$ and
$\as^2$ expansions of the ratios of PDFs at different scales that
appear in eq.~(\ref{eq:g2-separated}).
These have been obtained using \hoppet~\cite{Salam:2008qg}.

In the following we will present results for normalised differential 
distributions
\begin{equation}
\label{eq:diffdist2}
\frac{1}{\sigma}\frac{d\sigma(v)}{dv} =\frac{d f(v)}{dv}\,.
\end{equation}
Notice that two-loop corrections to $\sigma_2$, currently unknown, are
not needed for a second order matching, as they do not contribute to
the differential distribution within the target accuracy.

%......................................................................
\subsection{Coherence-violating (super-leading) logarithms}
\label{sec:superl-logar}

One of the assumptions that enters into the derivation of the
generalised resummations of \cite{BSZ03} is ``coherence''
\cite{Coherence}, the property that real emissions and virtual
corrections at large angles are independent of the structure of real
emissions that have occurred at small angles (with respect to any of
the incoming and outgoing legs).
Physically this can be understood as arising because a large-angle
emission (or virtual correction) sees only the sum of colour charges
of a bunch of collinear partons and that sum of colour charges is
conserved under collinear splitting.\footnote{For initial-state
  splittings, large-angle emission sees the difference in colour
  charges between incoming and outgoing partons that are collinear to
  an incoming direction.}

The assumption of coherence is challenged by the results of
ref.~\cite{Forshaw:2006fk,Forshaw:2008cq}, which found ``super-leading logarithms'' (SLL),
terms that go as $\as^4 L^5$, when calculating the probability of
there being no soft radiation (above scale $Q e^{-L}$) in a finite
patch of rapidity and azimuth.
Based on coherence, one would have expected only terms $\as^n L^m$
with $m\le n$ for such an observable.
Therefore, one might also call the terms of \cite{Forshaw:2006fk,Forshaw:2008cq}
``coherence-violating logarithms'' (CVL), a name that is suitable also
in the case of observables whose leading-logarithmic structure
involves double logarithmic terms $\as^n L^{2n}$ (for which $\as^4 L^5$
is not super-leading).

The interpretation of the result in ref.~\cite{Forshaw:2006fk,Forshaw:2008cq} is that one specific class of
(soft) single logarithmic virtual correction, ``Coulomb-gluon
exchange,'' can be affected by small-angle (collinear) initial-state gluon
emission, independently of how small that angle is.
This is because in the calculation of \cite{Forshaw:2006fk,Forshaw:2008cq} Coulomb
gluons are exchanged either between two 
incoming partons or between two outgoing partons but not between one
incoming and one outgoing parton (whereas other classes of soft contribution treat incoming
and outgoing partons on an equal footing); real initial-state splittings,
however small in angle, lead to a redistribution in colour between
incoming and outgoing states and therefore Coulomb-gluon exchange
cares about them (but not about the corresponding 
collinear virtual initial-state corrections).
This means that the coefficient of the Coulomb single logarithms
$\as^n L^n$ is proportional to the probability of soft-collinear
initial state emission, $\as^m L^{2m}$ and hence one obtains terms
$\as^{n+m} L^{n+2m}$, which are super-leading with respect to the
expected $\as^n L^n$.

The calculation of the impact of this effect requires that one follow
through the soft colour evolution of the $2\to2$ scattering
\cite{BottsSterman,KS,KOS,Oderda,KidonakisOwens}, for which the Coulomb-exchange
terms provide imaginary contributions.
For the purposes of our discussion here it is not necessary to enter
into the full detail of the soft colour evolution. Rather it suffices
to be aware, following~\cite{Forshaw:2008cq}, that for the case of
vetoing emissions into a finite patch (gap) the lowest order
coherence-violating terms are contributions with structures such as
\begin{subequations}
    \label{eq:CVL_gap}
  \begin{align}
    \mathrm{CVL}_\gap &\sim C \as^4 \int
    \frac{dk_{t1}^{(v)}}{k_{t1}}\;
    \frac{d k_{t2}^{(r/v)}}{k_{t2}} \frac{d\theta_2^{(r/v)}}{\theta_2}
    \Theta(1-\theta_2)\Theta(Q \theta_2 - k_{t2}) \;
    \frac{dk_{t3}^{(v)}}{k_{t3}}\;
    \frac{dk_{t4}^{(v)}}{k_{t4}}\;\cdot
    \nonumber \\
    &\quad \quad\cdot \Theta(Q-k_{t1}) \Theta(k_{t1} - k_{t2})
    \Theta(k_{t2} - k_{t3}) \Theta(k_{t3} - k_{t4})
    \Theta(k_{t4} - Q e^{-L})\\
    &= \frac{2 C}{5!}  \as^4 L^5 + \order{\as^4 L^4}\,,
  \end{align}
\end{subequations}
where we have shown only one of the two orderings given in
\cite{Forshaw:2008cq} (the other  gives either the same number or fewer
logarithms, depending on the observable).
In the integration measures, we have labelled each momentum
with $(v)$ if it can only be virtual, and $(r/v)$ if we are
considering the difference between real and virtual cases. 
Gluon $2$,
the collinear, possibly real gluon, can have an angle corresponding to
anywhere outside the gap region, down to the smallest kinematically
allowed angles $\theta_2 \sim k_{t2}/Q$.
Gluons $1$, $3$ and $4$ have only transverse momentum integrations
because they are either Coulomb exchange gluons or the virtual
counterparts of large-angle soft-gluon emission. (In the other
ordering it is gluon $1$ that is collinear and possibly real).
The integral for gluon $4$ is limited to be above $Q
e^{-L}$ because below that scale the observable places no constraint
on real emissions and so all real and virtual effects should cancel,
by virtue of unitarity.
Finally, the constant $C$ depends on the kinematics of the hard
scattering and the definition of the gap region.

The extension to the event-shapes case involves restricting the
$(r/v)$ integration for gluon 2 to regions of phase-space that are
consistent with the the real gluon's contribution to the event shape
being $\lesssim e^{-L}$. Eq.~(\ref{eq:CVL_gap}) therefore becomes
\begin{multline}
  \label{eq:CVL_event-shape}
  \mathrm{CVL}_\evshp \sim C \as^4 \int \frac{dk_{t1}^{(v)}}{k_{t1}}\;
  \frac{d k_{t2}^{(r/v)}}{k_{t2}}
  \frac{d\theta_2^{(r/v)}}{\theta_2}
  \Theta(1-\theta_2)\Theta(Q \theta_2 - k_{t2})
  \Theta(Q e^{-L/a} - k_{t2} \theta_2^{b/a})
  \;
  \frac{dk_{t3}^{(v)}}{k_{t3}}\;
  \frac{dk_{t4}^{(v)}}{k_{t4}}\;\cdot
  \\
  \cdot \Theta(Q-k_{t1})
  \Theta(k_{t1} - k_{t2})
  \Theta(k_{t2} - k_{t3})
  \Theta(k_{t3} - k_{t4})
  \Theta(k_{t4} - Q e^{-L/a})\,,
\end{multline}
where the coefficients $a$ and $b$ are those that appear in
eq.~(\ref{eq:parametricform}) for the incoming legs, $\ell=1,2$, for simplicity
we have neglected the $d_l$ and $g_\ell$ factors there, and the
constant $C$ may differ somewhat from that in eq.~(\ref{eq:CVL_gap})
(since there it could depend on the gap definition).
Three cases arise:
\begin{subequations}
  \label{eq:CVL-evshp-3cases}
  \begin{align}
    b < 0 \quad & \to\quad  \mathrm{CVL}_\evshp \sim C \as^4 L\\
    b = 0 \quad & \to\quad  \mathrm{CVL}_\evshp \sim C \as^4 L^2\\
    b > 0 \quad & \to\quad \mathrm{CVL}_\evshp \sim C \as^4 L^5
  \end{align}
\end{subequations}
where for the cases $b\le 0$ the number of powers of $L$ is that
obtained when relaxing the constraint $Q e^{-L/a} > k_{t2}
\theta_2^{b/a}$ to become $Q e^{-L/a} \gtrsim k_{t2} \theta_2^{b/a}$
(consistent with the fact that we have ignored factors of $d_\ell$,
$g_{\ell}$).
In words, the CVL contributions only appear with a large number of
logarithms when the collinear gluon, $2$, if real, is allowed to be
harder than the virtual Coulomb gluons ($3$, $4$).\footnote{ Note that
  for the
  exponentially-suppressed observables, with $b=a$, we expect the
  coefficient of the CVL to be significantly suppressed as compared to
  the gap case, because of the way in which the event-shape constraint
  restricts the phase-space integration region.}
For observables with $b < 0$ the one logarithm arises from the
integration over $k_{t1}$, while the $Q e^{-L/a} \gtrsim k_{t2}
\theta_2^{b/a}$ constraint forces $k_{2}$ to be at large angles with
$k_{t2} \sim Qe^{-L/a}$, with the knock-on effect that  $k_{t3}$ and
$k_{t4}$ should also be $\sim Qe^{-L/a}$.
For $b = 0$, we instead have just $Q e^{-L/a} \gtrsim k_{t2}$  for
gluon 2, and an extra logarithm then arises from the integration of
$\theta_2$ in the collinear region.

The expectation for yet higher order terms is that for $b\le 0$ (all
observables of this paper except the exponentially-suppressed ones)
the results in eq.~(\ref{eq:CVL-evshp-3cases}) could be multiplied by
additional powers of $\as L$ giving at worst a series $\as^n L^{n-2}$,
which is subleading both with respect to our NLL accuracy in the
exponent and to our NNLL$_\Sigma$ accuracy in its expansion.\footnote{%
  This result involves the assumption that there must be at least two
  large-angle or Coulomb virtual gluons softer than the collinear
  gluon.
  While this is the case for the contributions found in the gap case
  \cite{Forshaw:2006fk,Forshaw:2008cq} we do not show here that it
  will always necessarily be the case for event-shapes.
  If there could instead be coherence-violating contributions with
  just one large-angle or Coulomb virtual gluon that is softer than the
  collinear one, one might expect terms up to $\as^n L^{n-1}$, which,
  however, are still subleading relative to the calculations of this
  paper.  }
For $b> 0$ (the exponentially suppressed observables) one would obtain
terms $\as^n L^{2n-3}$. It is not clear how they would fit into the
exponential resummation, except that they would certainly destroy NLL
accuracy in the exponent; in the expansion of the resummation they
would represent terms NNNLL$_\Sigma$ and therefore be subleading with respect
to our accuracy.

One caveat with regard to the above discussion is that the results of
\cite{Forshaw:2006fk,Forshaw:2008cq} have been obtained in a
strongly ordered eikonal approximation, with the assumption that
the ``strong ordering'' parameter is transverse momentum.
With different assumptions,  the results change. For example, if
the correct ordering parameter were energy, then
eq.~(\ref{eq:CVL_gap}) would be modified in such a way as to give an
infinite result.
If instead one considered emission-time (or virtuality) ordering,%
\footnote{%
  One gets the same result for the ordering based on two different
  considerations. 
  Physically, the time scale for the collinear emission to take place
  is $(1/k_{t2}) \cdot (\omega_2/k_{t2}) \sim 1/(k_{t2}\theta_2)$, where
  $\omega_2$ is the energy of gluon 2, to be compared with $1/k_{t}$
  for a large-angle virtual gluon exchange.
  In terms of the diagrammatic structure, ordering is in part related
  to the virtualities of propagators and for a hard scattering of
  partons with energy $E$, the squared propagator virtuality induced
  by soft and collinear gluon emission is $\sim E \omega_2 \theta_2^2
  \simeq E k_{t2}\theta_2$ to be compared with $E k_t$ for large-angle
  virtual gluon exchange.}
which leads to
$k_{t1} \gg k_{t2} \theta_2 \gg k_{t3}$, then the contribution of
eq.~(\ref{eq:CVL_gap}) would be halved,
\begin{equation}
  \label{eq:CVL_gap_t_ordered}
  \mathrm{CVL}_\gap^{(\text{T-ordered})} 
  \sim 
  \frac{C}{5!}  \as^4 L^5\,,
\end{equation}
and the corresponding result for the event shapes case would become
\begin{subequations}
  \label{eq:CVL_evshp_t_ordered}
  \begin{align}
    b < a \quad & \to\quad  \mathrm{CVL}_\evshp^{(\text{T-ordered})}  \sim C \as^4 L\\
    b = a \quad & \to\quad  \mathrm{CVL}_\evshp^{(\text{T-ordered})}  \sim C \as^4 L^2\\
    b > a \quad & \to\quad \mathrm{CVL}_\evshp^{(\text{T-ordered})}  \sim C \as^4 L^5
  \end{align}
\end{subequations}
In this case, for all the observables being discussed in this paper, the
CVL terms would be subleading relative to our accuracy.

To conclude: given today's knowledge it is not clear whether
coherence-violating terms matter at our accuracy for event-shape
resummations.
The critical issue is that of the appropriate ordering parameter
($k_t$, time or virtuality ordering, or some other ordering). 
The correct ordering needs to be derived (by going beyond the eikonal
approximation), unless of course there exists some yet-to-be found
contribution that cancels the CVL terms.
If CVL terms do exist and $k_t$ ordering is correct, then they will
invalidate our statement of NLL accuracy in the exponent for the
exponentially suppressed class of observables, though not our
statement of NNLL$_\Sigma$ accuracy in the expansion of the distribution.
In practice we have reason to believe that their numerical impact will
still be small: partly because the CVL terms were already not very
large in \cite{Forshaw:2006fk}; and partly because the large colour
factors multiplying the double logarithms of our resummation force the
majority of events to be in a region where the logarithms are not
actually all that large (a reflection of this will appear in
section~\ref{sec:naive-exp}, where we will see that naive
exponentiation of the NLO calculation is not too different from the
full NLO+NLL result, even though it misses classes of LL terms in the
exponent and LL$_\Sigma$ terms in its expansion).

%----------------------------------------------------------------------
\section{Perturbative results}
\label{sec:pert-res}

In this section, we shall consider numerical results both for Tevatron and LHC
collision scenarios.
We will start by presenting the event selection cuts that we use.
We shall show results for NLL+NLO matched calculations for a range of
observables.
We will pay particular attention to the estimation of uncertainties on
our predictions, and comparisons to separate pure NLO and NLL
calculations.
We will also compare our results to Monte Carlo parton-shower results,
with and without tree-level matrix element matching.

%......................................................................
\subsection{Event selection cuts}
\label{sec:cuts}

The Tevatron scenarios involve $p\bar p$ collisions at
centre-of-mass energy $\sqrt{s} = 1.96\TeV$.
Events are clustered with the SISCone jet algorithm
\cite{Salam:2007xv} (similar to the MidPoint
algorithm~\cite{RunII-jet-physics} that is in widespread use at the
Tevatron, but infrared safe), with a jet radius $R=0.7$ and a split--merge overlap
threshold $f=0.75$.
The two hardest (highest-$p_t$) jets in the event should have
rapidities $|y| < 0.7$. 
Events are accepted for a low-$p_t$ sample if the hardest jet has $p_t
> 50\GeV$, while they are accepted for a high-$p_t$ sample if the
hardest jet has $p_t > 200\GeV$.
As concerns the event shapes, the central region is defined by
$\eta_\cC = 1$.

The LHC scenarios  involve $p p$ collisions at a centre-of-mass
energy $\sqrt{s} = 14\TeV$.\footnote{%
  The LHC will initially run at
  centre-of-mass energies that are below $\sqrt{s}=14\TeV$,
  though the exact energy of collisions is subject to uncertainty and
  will vary over the course of the initial runs.
  Given that the generation of the NLO results for a single combination of
  collider energy and event-selection cuts requires many CPU-years of
  computing time, we have decided to remain with $\sqrt{s}=14\TeV$ as
  our default choice for the time being.
  The general picture as it applies to other centre-of-mass energies
  can be largely understood by interpolation between the Tevatron
  and $14\TeV$ LHC results.
} %
Events are clustered with the $k_t$ jet algorithm
\cite{KtHH,Kt-EllisSoper}, with a jet radius $R=0.7$.
The two hardest jets in the event should have rapidities $|y| < 1$.
Events are accepted for a low-$p_t$ sample if the hardest jet has $p_t
> 200\GeV$, while they are accepted for a high-$p_t$ sample if the
hardest jet has $p_t > 1\TeV$.
As concerns the event shapes, the central region is defined by
$\eta_\cC = 1.5$.
The larger choice than at the Tevatron reflects the LHC detectors'
larger overall rapidity coverage.

The cross sections for the different selections are given in
table~\ref{tab:cross-sections}. 
These, and all other NLO calculations presented here, have been
carried out with \nlojet 3.0~\cite{NLOJET} (modified to provide access to
parton flavour information up to $\order{\as^3}$), with CTEQ6M Parton
Distribution Functions (PDFs) \cite{Pumplin:2002vw} and FastJet
2.3~\cite{FastJet} for the jet clustering.
The renormalisation and factorisation scales have central values
$\mu_R, \mu_F = p_t \equiv (p_{t1}+p_{t2})/2$, where $p_{t1}$ and
$p_{t2}$ are the transverse momenta of the hardest and second hardest
jet respectively.
The quoted errors correspond to the uncertainties due to scale
variation $p_t/2 < \mu_R,\mu_F < 2 p_t$, with $\mu_F/2 <
\mu_R < 2 \mu_F$.
Given the cross sections in table~\ref{tab:cross-sections} one easily
concludes that with available (anticipated) luminosities at the
Tevatron (LHC) there will be large event samples on which to study
event shapes.
One also observes that NLO corrections are larger than one is used to
seeing for (say) inclusive jet cross sections.
This is a consequence of our selection based on the value of $p_{t1}$.
A selection based on the average $p_t$ of the two hardest jets would
instead have given K-factors rather similar to those for the inclusive
cross section.%
\footnote{The choice of a cut on $p_{t1}$ was originally motivated by
  the observation in the context of
  HERA~\cite{Chekanov:2001fw,Aktas:2003ja,KK} that identical
  simultaneous cuts on $p_{t1}$ and $p_{t2}$ led to poor convergence
  of the perturbative series, for reasons discussed
  in~\cite{symm-cuts,Banfi:2003jj}.
  Cutting on $p_{t1}$ was intended as a way of avoiding this problem,
  but, as we see here, seems to introduce issues of its own.
  Note that it is probably not advisable to introduce a staggered cut
  on $p_{t1}$ and $p_{t2}$, e.g.\ $p_{t1}> 50\GeV$ and $p_{t2}>
  40\GeV$, because that introduces an extra small parameter in the
  problem, related to the difference between the two $p_t$ cuts.
}

Table~\ref{tab:cross-sections} also shows the breakdown into the 3
main partonic scattering channels (as calculated at LO). At each of
the colliders, for the lower $p_t$ cut, channels involving gluons are
dominant, while for the higher $p_t$ cut channels involving quarks
play a bigger role. This difference between low and high-$p_t$ samples
will be clearly visible in the final results.

\renewcommand\arraystretch{1.3}

\begin{table}
  \centering
  \begin{tabular}{l|cc|ccc}
    & LO & NLO & $qq\to qq$ & $qg\to qg$ & $gg\to gg$\\\hline
    Tevatron, $p_{t1} > 50\GeV$  & $60^{+22}_{-15} \nb$ & $116^{+28}_{-21} \nb$ 
    & 10\% & 43\%& 45\%
    \\
    Tevatron, $p_{t1} > 200\GeV$  & $59^{+25}_{-16} \pb$ & $101^{+27}_{-22} \pb$
    & 41\% & 43\% & 12\%
    \\
    $14\TeV$ LHC, $p_{t1} > 200\GeV$  & $13.3^{+3.4}_{-2.5}\nb$ & 
    $23.8^{+3.9}_{-3.2} \nb$
    & 7\% & 40\% & 50\%
    \\
    $14\TeV$ LHC, $p_{t1} > 1\TeV$  & $6.4^{+2.0}_{-1.4} \pb$ & 
    $10.5^{+2.2}_{-2.0} \pb$
    & 31\% & 51\% & 17\%
  \end{tabular}
  \caption{Cross section for events that pass the selections cuts
    described in the text. The uncertainty is that due to scale
    variation with the choice $ p_t/2 \le \mu_R,\mu_F \le 2 \,p_t$,
    with $\mu_F/2 \le \mu_R \le 2\,\mu_F$,
    where $p_t$ is the average of the transverse momenta of the two
    hardest jets. 
    Also shown is the breakdown (at LO) into the main scattering
    channels; $q$ denotes both quarks and antiquarks, and channels
    that contribute negligibly, such as $gg\to q\bar q$, are not
    shown.}
  \label{tab:cross-sections}
\end{table}

\renewcommand\arraystretch{1.0}

\subsection{Resummed results and uncertainty studies}
\label{sec:res+errors}
Here we present resummed results for the global thrust
minor, $T_{m,g}$ together with a study of its perturbative
uncertainties at the Tevatron for the high-$p_t$ sample ($p_{t1} >
200$ GeV).

\begin{figure}[htbp]
  \centering
  \includegraphics[width=0.7\tw]{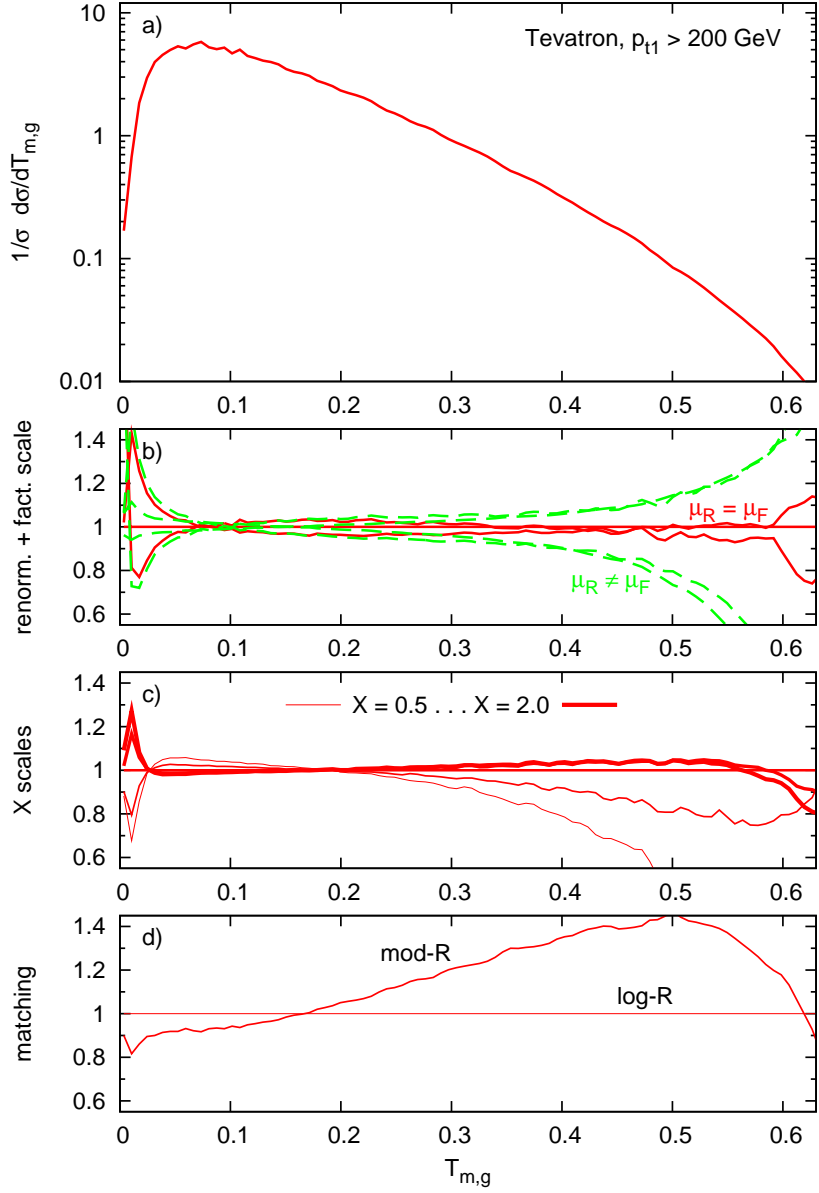}
  \caption{a) NLL+NLO resummed matched distribution for the directly
    global thrust minor at the Tevatron ($p\bar p$,
    $\sqrt{s}=1960\GeV$) with $p_t > 200\GeV$; b) renormalisation and
    factorisation scale uncertainties for $x_R, x_F = 0.5,1.0,2.0$,
    varied separately, with the condition $0.5 \le x_F/x_R \le 2$; c)
    effect of varying $X=0.5,0.7,1.0,1.5,2.0$ in eq.~\eqref{eq:Ltilde}; d) effect
    of changing the matching scheme.}
  \label{fig:uncertainties}
\end{figure}

Fig.~\ref{fig:uncertainties}a illustrates the NLL+NLO matched
distribution obtained in the log-R matching scheme with $X=1$.
The renormalisation and factorisation scales are chosen,
event-by-event, to be $\mu_F=\mu_R=p_t = (p_{t1}+p_{t2})/2$ in both
the resummation and the NLO calculation.
The distribution has a peak at small event-shape values that is
characteristic of all resummed (and physical) event-shape
distributions.

The uncertainty on the prediction is almost as important as the result
itself, especially as it will allow us to gauge the significance of
any disagreements that we will see with other predictive methods.

The most widely used form of uncertainty estimate is the variation of
renormalization and factorization scales. The solid (red) curves in
fig.~\ref{fig:uncertainties}b illustrate the effect of varying these
scales simultaneously, showing the ratio of results with
$\mu_F=\mu_R=p_t/2$ and $\mu_F=\mu_R=2p_t$ to the default
result. Except at very small event-shape values or at very large ones,
where the distribution vanishes, one sees that the impact of symmetric
scale variation is only about $5\%$.
Asymmetric scale variations are shown by the dashed (green) curves,
corresponding to $\mu_F=\{p_t/2, 2p_t\}$ while keeping $\mu_R=p_t$, and
$\mu_R=\{p_t/2, 2p_t\}$ while keeping $\mu_F=p_t$.
For moderate and large values of the event shape they have a
significantly larger impact than symmetric scale variations, of the
order of $10\%$ for moderate $T_{m,g}$. 
This highlights the importance of considering both symmetric and
asymmetric variations.

Fig.~\ref{fig:uncertainties}c shows the impact of varying $X$ in
eq.~(\ref{eq:Ltilde}), with the line thickness increasing from $X=0.5$
to $X=2$. As discussed in Sec.~\ref{sec:matchNLLNLO}, this variation
can be used to estimate the effect of higher order logarithms not
included in our NLL resummation. We find that for moderate and large
values of $T_{m,g}$ the effect is similar in size to the asymmetric
renormalization/factorization scale variation. Closer to the peak of
the distribution (where the bulk of events sits), the impact of the
$X$-scale variation is mildly larger. We also note that the
variation is quite asymmetric: smaller $X$ values distort the central
distribution much more than larger values.

Finally, in Fig.~\ref{fig:uncertainties}d we estimate uncertainties
that arise from the details of the matching procedure.
In particular we show the ratio of the mod-R matched distribution to
the log-R (see eqs.~\eqref{eq:second-logR-hh-alt},
\eqref{eq:second-logR-hh-alt2} and \eqref{eq:second-mult-hh-alt}). It
is clear that at large values of $T_{m,g}$, the difference between the
two matched distributions is large, with differences of up to 45\% for
$T_{m,g}\sim 0.5$. These very large discrepancies occur however only
in the tail of the distribution, where few events are present.
Comparison to NLO at high $T_{m,g}$ (not shown) indicates that of the
two matching 
schemes, log-R matching is the one with smaller higher order terms
(i.e.\ its NLO+NLL result is closer to NLO) at large $T_{m,g}$. In the
following we will therefore
use log-R matching as our default. 

The above findings are representative of the results for the
other event shapes considered here, both at the Tevatron and at the
LHC and for the low- and high-$p_t$ samples
(further NLO+NLL results are shown in
Sec.~\ref{sec:nll+nlo-many-obs} and on the website associated
with this article~\cite{website}).
In particular {\it symmetric} renormalization and factorization
scale variation (as is currently done in many phenomenological
studies) systematically  underestimates the true size of theoretical
uncertainties. 
While detailed error-estimate studies such as variation of X-scale and
matching procedure are possible only for specific (resummed)
calculations, an {\it asymmetric} $\mu_R$ and $\mu_F$ variation can be
carried out for generic observables.
This is perhaps most relevant for multi-scale observables, where the
scale of $\as$ is a priori not clear. 

%----------------------------------------------------------------------
\subsection{Comparison of resummed, NLO and matched results}
\label{sec:resNLOmat}

In this section we compare various levels of fixed order calculations
(LO, NLO), pure resummed ones (NLL) and matched ones (NLO+NLL) at the
Tevatron for the high-$p_t$ sample.
Because NLL+NLO resummations are rarely available (e.g. they are
currently not available for non-global observables), we discuss in
particular the extent to which NLO alone can be used to describe event
shape distributions. As in the previous section we will use $T_{m,g}$
to illustrate our findings, but results are fairly independent of the
specific event shape, the collider and the details of the hard cuts.

\begin{figure}[tbph]
  \centering
  \setlength{\unitlength}{\textwidth}
  \includegraphics[width=0.48\tw]{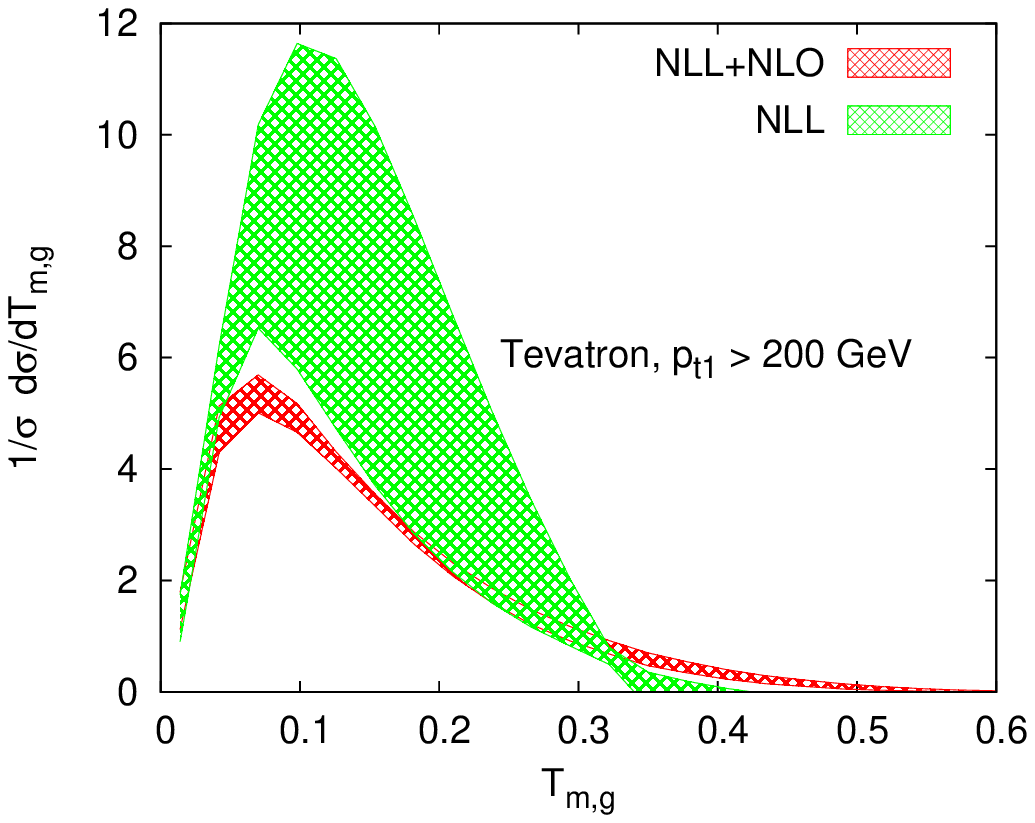}
   \hfill
   % NB: nmat-v-logR is a misnomer, since we have the full 
   %     set of uncertainties that are included (included two 
   %     matching schemes).
  %\framebox(0.48,0.38){b) pure resummed v. matched}
  \includegraphics[width=0.48\tw]{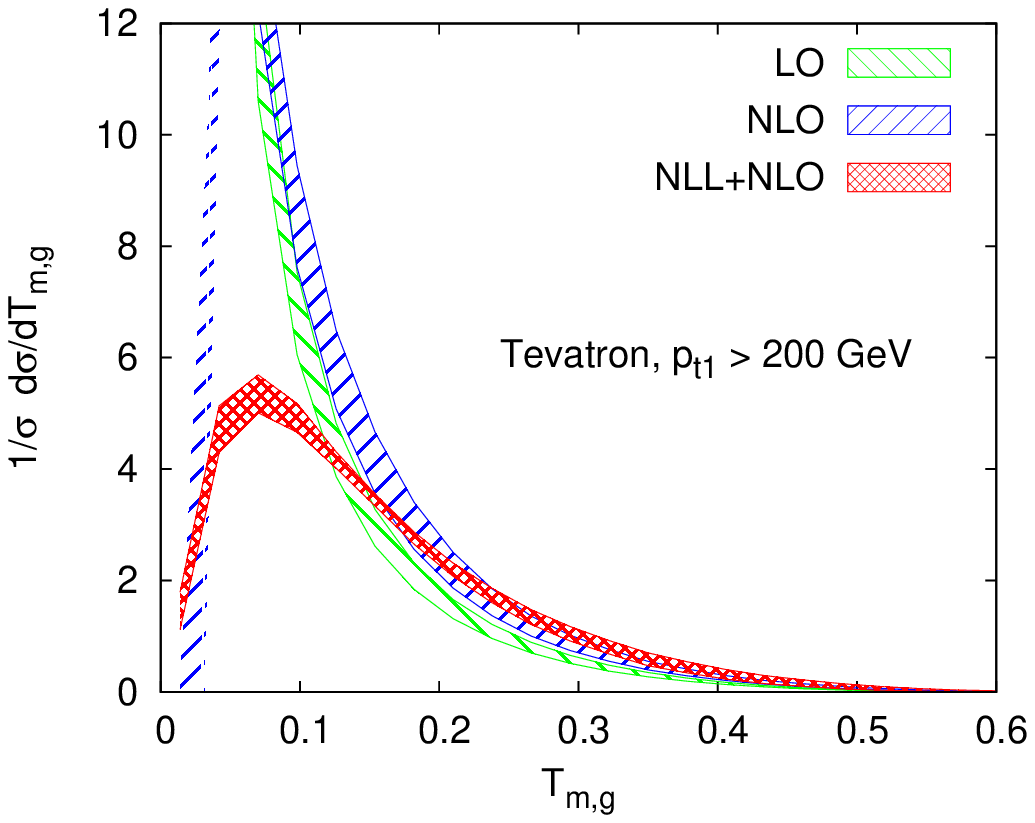}
  \caption{The distribution for the representative observable
    $T_{m,g}$, comparing pure resummation (NLL, left) and pure fixed
    order (LO and NLO, right) with the matched prediction (NLL+NLO).
    For the matched resummed result, the band corresponds to the
    span of all the uncertainties shown in
    fig.~\ref{fig:uncertainties}, while in the fixed order
    calculations it corresponds to the (asymmetric) variations of just the
    renormalisation and factorisation scales; for the pure
    resummed result the band corresponds to the renormalisation,
    factorisation and $X$-scale uncertainties. See text for more details.}
  \label{fig:match-v-nlo-v-resum}
\end{figure}

Fig.~\ref{fig:match-v-nlo-v-resum} shows the result for the log-R
matched $T_{m,g}$ distribution compared to pure resummation (left) and pure
NLO, and LO for reference (right). 
For the matched resummed result, the band corresponds to the span of
all the uncertainties shown in fig.~\ref{fig:uncertainties}, while in
the fixed order calculations it corresponds to the variations of just
the renormalisation and factorisation scales ($ p_t/2 \le \mu_R,\mu_F \le
2 \,p_t$,
    with $\mu_F/2 \le \mu_R \le 2\,\mu_F$);
for the pure resummed result the band corresponds to the
renormalisation, factorisation and $X$-scale uncertainties.

As expected, the matched distribution agrees with the NLO results at
large values of $T_{m,g}$.  However for the pure NLL resummation
without any coefficient function obtained from
eq.~\eqref{eq:Sigmacut_resummedtilde}, the level of agreement with
NLO+NLL is quite poor even at fairly small values of $T_{m,g}$. For
example, the position of the peak of the distribution is not all that
well predicted (at $T_{m,g}\sim 0.09-0.11$ rather than at $T_{m,g}\sim
0.08$). As far as the height of the peak is concerned, both NLL and
NLO+NLL distributions are normalized to one, however the NLL
distribution becomes negative at $T_{m,g}> 0.35$, and this negative
tail causes the distribution to be far too high at low $T_{m,g}$. It
is on the other hand reassuring that these large differences with the
matched distribution are reflected in the very large uncertainty band
of the NLL distribution.

As far as the fixed-order results are concerned,
Fig.~\ref{fig:match-v-nlo-v-resum}b, they are as expected divergent at
small $T_{m,g}$. The LO distribution  essentially never agrees with
the matched distribution, while the NLO does within uncertainties for
$T_{m,g} \gtrsim 0.2$. It is on the other hand evident that scale
uncertainties of the NLO results at small $T_{m,g}$ underestimate the
size of higher order corrections not included in the fixed order
calculations.

Altogether, figure Fig.~\ref{fig:match-v-nlo-v-resum} highlights how
neither NLO nor resummation alone can provide a sensible prediction,
while the combination of NLO and resummation gives significantly
reduced scale-dependence compared to either on its own. Furthermore
the matching procedure gives the general shape that is associated with
the resummation, while maintaining the large-$v$ behaviour of the NLO
prediction.
We finally note that for the event shapes presented here a LO+NLL
resummation agrees well neither with the LO, nor with the
NLO+NLL matched result.

\begin{figure}
  \centering
  \includegraphics[width=0.49\tw]{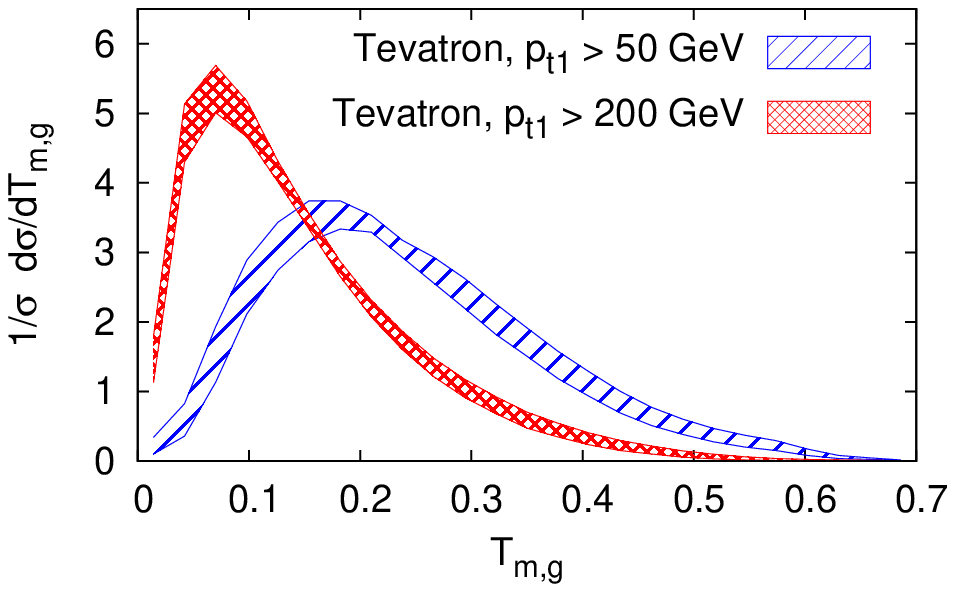}\hfill
  \includegraphics[width=0.49\tw]{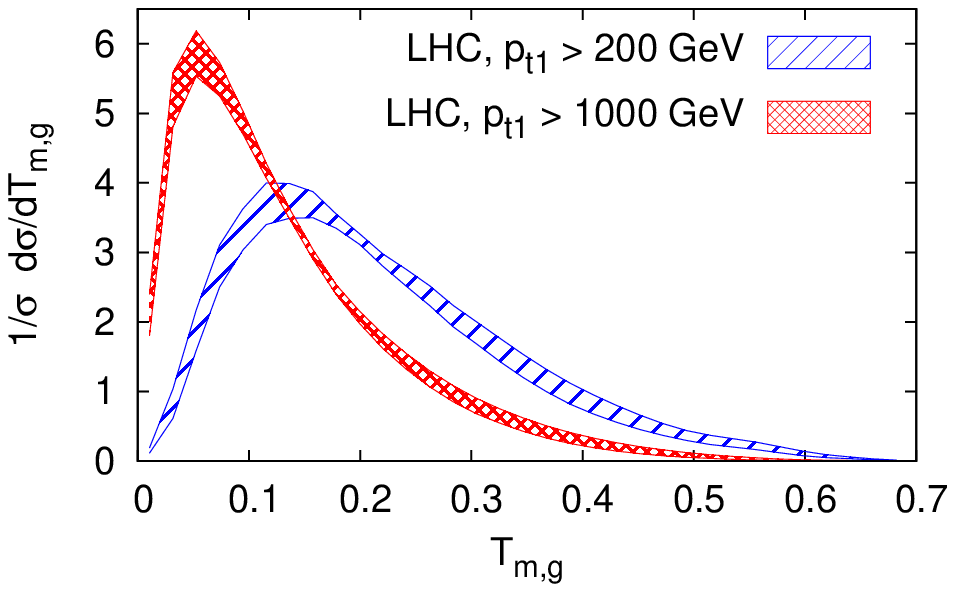}
  \caption{The NLO+NLL matched distribution for the directly global
    thrust minor, $T_{m,g}$ shown for the Tevatron (left) and
    $\sqrt{s}=14\TeV$ LHC
    (right) with two transverse momentum cuts for the event
    selection. 
  }
  \label{fig:different-energies}
\end{figure}

%----------------------------------------------------------------------
\subsection{NLL+NLO matched results for a range of observables}
\label{sec:nll+nlo-many-obs}

In the previous section we established that contrary to NLO or NLL alone,
NLO+NLL provides robust theoretical predictions for event shapes
distributions over a large range of the event-shape values.
This section contains the bulk of results of the present work: we
discuss NLL+NLO resummed distributions for a number of event shapes
variables, both at the Tevatron and at the LHC and for both, low- and
high-$p_t$ samples.

\begin{figure}
  \centering
  \includegraphics[width=\tw]{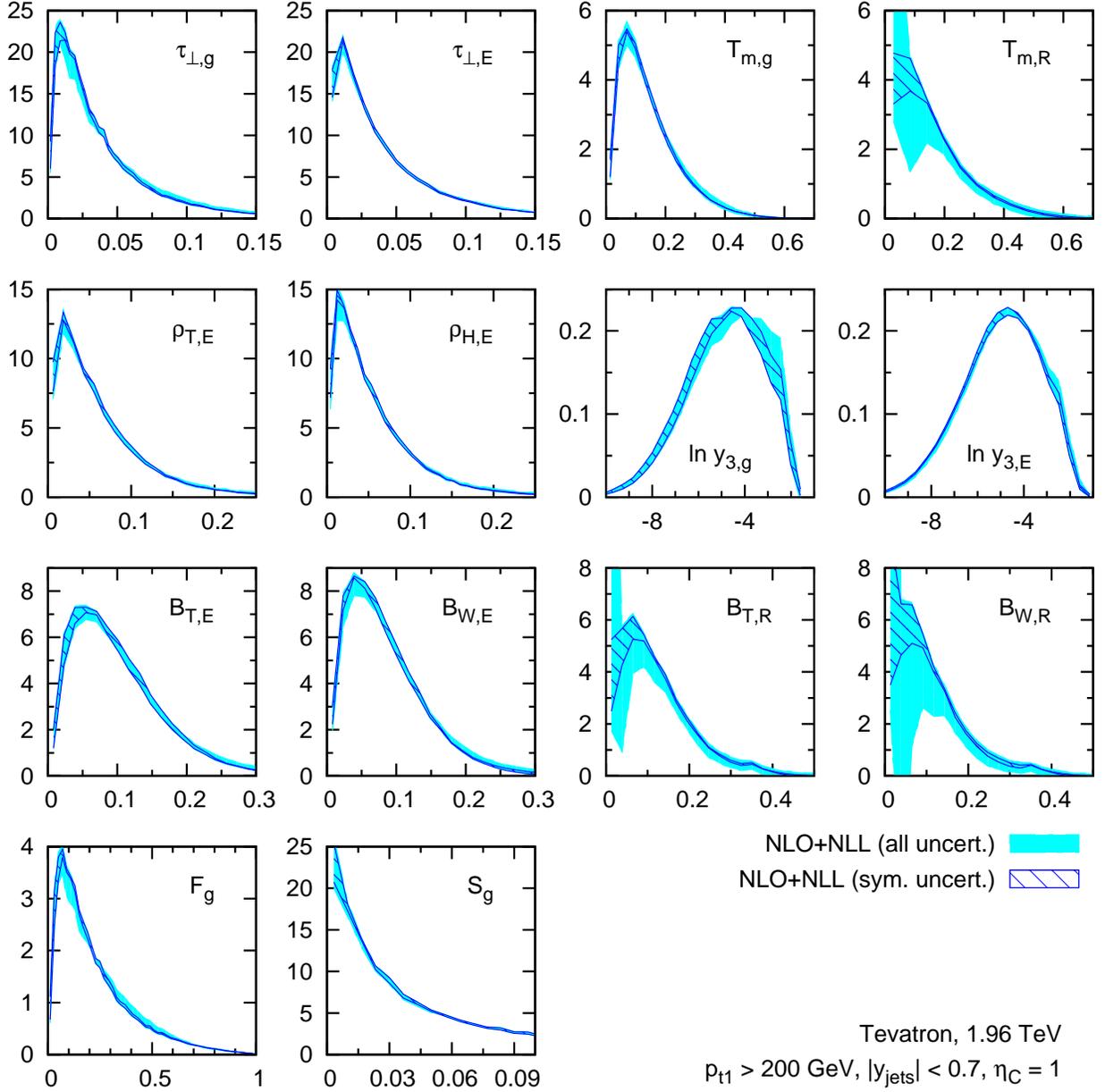}\hfill
  \caption{The normalised NLO+NLL matched distributions, $\frac1\sigma
    \frac{d\sigma}{dv}$, for a range of
    event-shape 
    observables at the Tevatron with $p_{t1} >
    200\GeV$. For further details, see text.}
  \label{fig:nllnlo-tev200}
\end{figure}

\begin{figure}
  \centering
  \includegraphics[width=\tw]{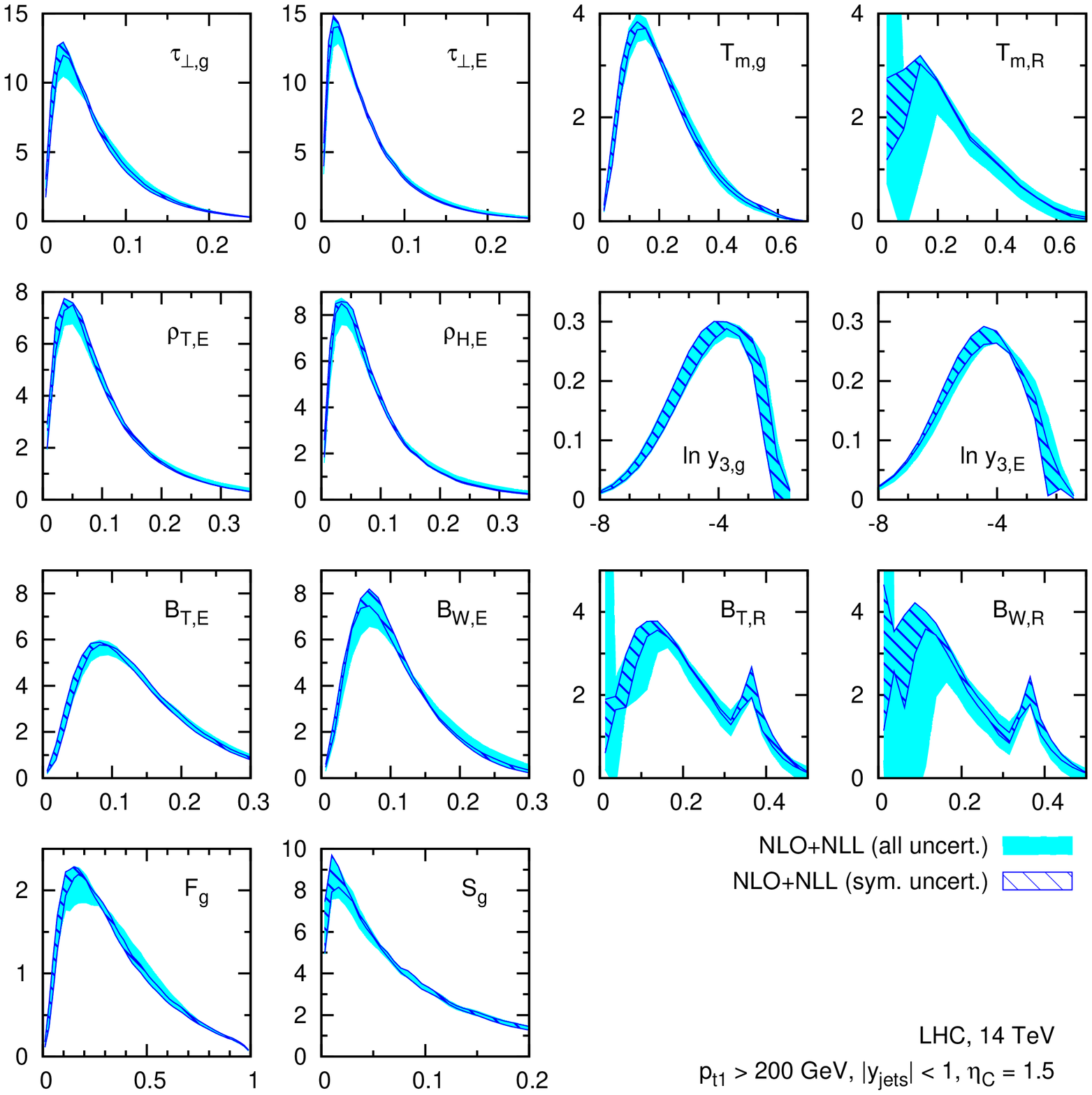}\hfill
  \caption{The normalised NLO+NLL matched distribution,  $\frac1\sigma
    \frac{d\sigma}{dv}$,  for a range of event-shape
    observables at the $\sqrt{s}=14\TeV$ LHC with $p_{t1} >
    200\GeV$. For further details, see text.}
  \label{fig:nllnlo-lhc200}
\end{figure}

We start by looking at the effect of changing the hard selection cuts
and the collider (Tevatron vs LHC) for the same observable discussed
previously, $T_{m,g}$, Fig.~\ref{fig:different-energies}. There is a
striking similarity between the Tevatron (left) and the LHC plot
(right), both for the low-$p_t$ (Tevatron, $p_{t1} > 50$ GeV and LHC,
$p_{t1} > 200$ GeV) and for the high-$p_t$ (Tevatron, $p_{t1} > 200$
GeV and LHC, $p_{t1}
> 1000$ GeV) samples. We also notice that low-$p_t$ curves are broader
and peaked at a higher value of $T_{m,g}$. This is a consequence of
the higher prevalence of gluons in both the initial and final states
of the hard scattering.

Because of this similarly between low- and high-$p_t$ samples at the
two colliders, we examine results for a large range of
observables, as defined in Sec.~\ref{sec:event-shape-defin}, just for the
high-$p_t$ cuts at the Tevatron, Fig.~\ref{fig:nllnlo-tev200}, and for
the low-$p_t$ cuts at the LHC, Fig.~\ref{fig:nllnlo-lhc200}.
For each observable, we give two uncertainty bands, one corresponding
to a symmetric scale variation (hatched, dark blue) and one defined in
terms of {\it all} theoretical uncertainties as discussed in
Sec.~\ref{sec:res+errors} (solid, light blue).
Comparing Fig.~\ref{fig:nllnlo-tev200} and
Fig.~\ref{fig:nllnlo-lhc200} we see that, as observed earlier for
$T_{m,g}$, the peaks of the distributions are further to the right and the
distributions are broader for the LHC (low $p_t$) than for the
Tevatron (high $p_t$).
Looking at specific observables we see that, as already remarked in
the case of $T_{m,g}$ for all observables the {\it symmetric} scale
variation uncertainties are considerably smaller than the {\it full}
uncertainties, and we stress that only the latter are really
indicative of the size of all kinds of neglected higher order terms.

Some final remarks concerns the NLO+NLL results for $T_{m,\cR}$,
$B_{T,\cR}$ and $B_{W,\cR}$.
As discussed in \cite{Banfi:2004nk} and at the end of
Sec.~\ref{sec:NLL_resum}, recoil variables are more difficult to resum
than other variables, because in the \caesar approach, the NLL term
$g_2(\as L)$ has an unphysical divergence at small values of the
observable.
This difficulty is reflected in the substantially larger uncertainty
bands for these observables than for the directly global variants and
those with an exponentially-suppressed forward term.
Among the recoil variables, the thrust minor and broadenings were the
only ones for which an even partially acceptable result could be
obtained.
In order to obtain results for recoil variables of similar quality to
those for the other observables requires a resummation of initial
state emissions in appropriate Fourier transform variables, as done
e.g. for the Drell-Yan $p_t$ resummation~\cite{Parisi:1979se}, 
mixed with a Sudakov type resummation, as was done for the DIS broadening
\cite{Dasgupta:2001eq}.
This is beyond the scope of \caesar.
Another characteristic to be commented on is the spike for $B_{T,\cR}$
and $B_{W,\cR}$ near $0.37$.
We believe this could be related to a Sudakov shoulder
\cite{Catani:1997xc} type phenomenon, and similar (though less
pronounced) artifacts have also been observed in DIS event-shape
distributions.
Again, it is beyond the scope of \caesar to resum the enhanced
higher-order terms associated with these structures.

%----------------------------------------------------------------------
\subsection{Naive exponentiation of NLO}
\label{sec:naive-exp}

In the previous Section we presented full NLO+NLL resummations for a
range of event shapes. Both the NLO Monte Carlo calculation and the
NLL resummation are highly CPU intensive and are usually both run
across many CPUs.
While the NLO part of the calculation is the most computer intensive, this
is to some degree counterbalanced by the fact that many observables
can be computed in the same NLO run. The NLL resummation on the other
hand requires essentially a separate run with \caesar for each
observable. Altogether, a single combination of collider energy and
event-selection cuts requires many CPU-years of computing time.
NLO+NLL resummation also requires that the NLO total cross section and
LO distributions be decomposed into
flavour channels, and this information is not available in the public
version of \nlojet (nor in most other public NLO codes). Furthermore,
\caesar is currently not public, the range of observables that can
be resummed with \caesar is not as broad as one might like (see
Sec.~\ref{sec:preCAESAR}), 
and the matching procedure at hadron colliders is not as
straightforward as in $\ee$, as discussed in Sec.~\ref{sec:matchNLLNLO}.
For the above reasons, it is interesting to explore the possibility of
obtaining predictions with accuracy close to NLO+NLL using publicly
available NLO results only.
For instance the following combination of LO and NLO integrated,
flavour summed distributions, $\Sigma_1(v)$ and $\bar\Sigma_2(v)$, and the
corresponding total cross sections, $\sigma_0$ and $\sigma_1$, as
obtained directly from \nlojet,
\begin{equation}
f(v) = \frac{\tilde f(v)}{\tilde f(v_{\rm max})}\,,\qquad 
\tilde f(v) =
\frac{\sigma_0}{\sigma_0+\sigma_1}\exp\left[{\frac{\Sigma_1(v)}{\sigma_0}+\frac{\bar
\Sigma_2(v)}{\sigma_0}-\frac{1}{2}\left(\frac{\Sigma_1(v)}{\sigma_0}\right)^2}\right]\,,
\label{eq:expNLO}
\end{equation}
has the following properties:
\begin{itemize}
\item $f(v)$ goes to 1 at $v=v_{\rm max}$, without $\cO{\as^3}$
      corrections;
\item the fixed order expansion of the corresponding differential
      distribution, up to relative order $\as^2$ reproduces the
      normalized NLO differential distribution;
\item the formal accuracy is not even LL$_\Sigma$: starting from order
  $\as^3$  the terms $\as^n L^{2n}$ are only correct in the limit in which
  the event sample is dominated by a single colour channel.
\end{itemize}

\begin{figure}
  \centering
  \includegraphics[width=\textwidth]{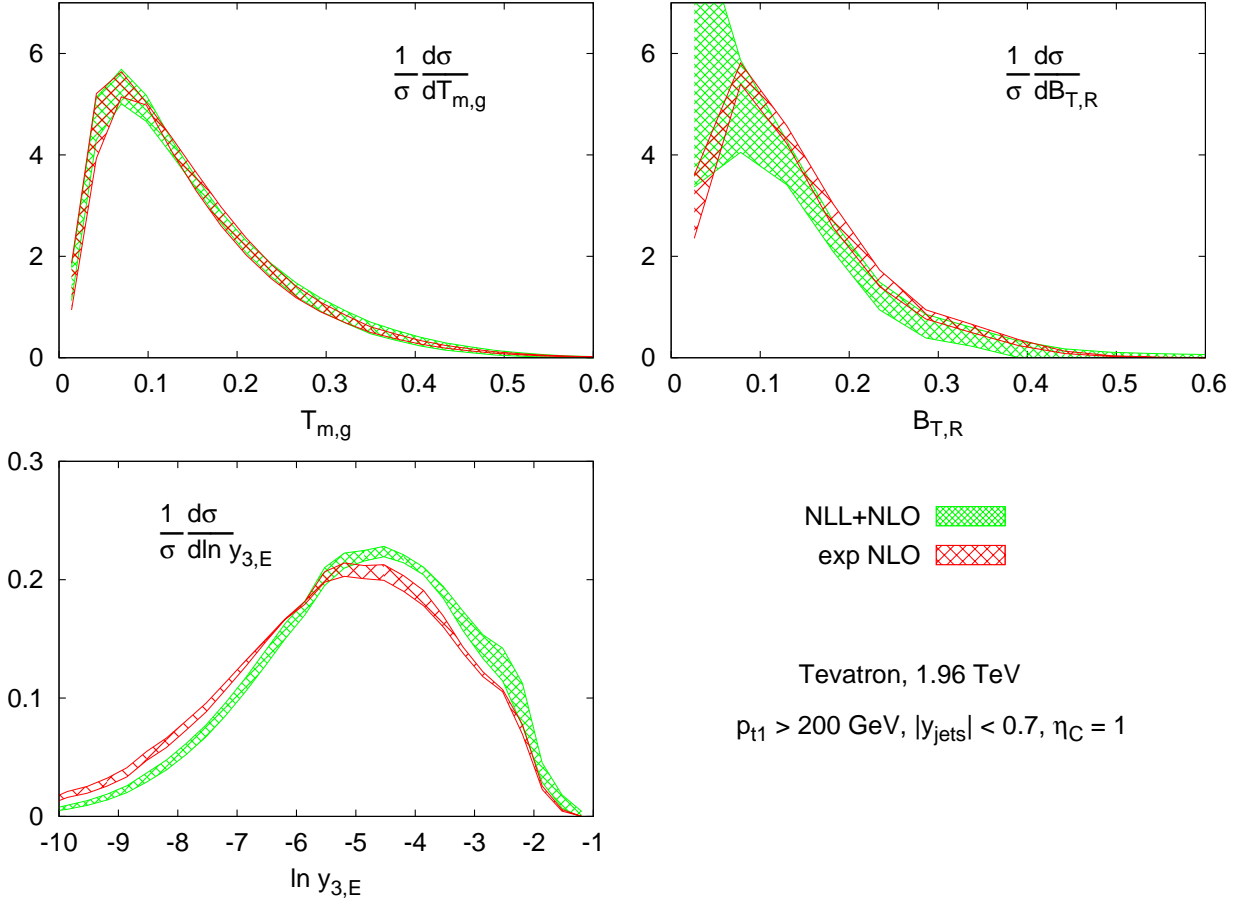}
  \caption{Comparison of the NLO+NLL results with full uncertainties
  and exponentiation of NLO (as defined in eq.~\eqref{eq:expNLO}) with
  asymmetric scale uncertainties for the directly global thrust minor
  (top left), the total broadening with recoil term (top right) and
  for $y_{3,\cE}$ (bottom left). 
    Shown for the Tevatron with a $200\GeV$ cut on $p_{t1}$.} 
  \label{fig:nlo+nll-vs-expnlo}
\end{figure}

Though the method does guarantee any formal resummation accuracy, it
is still instructive to
see how it fares in
practice. We therefore show in fig.~\ref{fig:nlo+nll-vs-expnlo} a
comparison of NLO+NLL matched results with full uncertainties
  and the naive exponentiation of NLO, as defined in
  eq.~\eqref{eq:expNLO} with full scale uncertainties for a
  representative set of observables
for the Tevatron with a 200-GeV jet $p_t$ cut (similar results hold
also at the LHC and for other cuts as well as other observables~\cite{website}).
We see that for the directly global thrust minor (top, left) the
exponentiation result is well-contained in the uncertainty band of the
NLO+NLL matched result, suggesting that the naive exponentiation of NLO
is indeed a quite reasonable procedure (similar results hold in general
for global observables).
The same observation is true also for the total broadening with recoil
term (top, right) but with one important difference. In this case the
NLO+NLL uncertainty band at small values of the observable is
divergent, signaling the breakdown of the resummation (as discussed
in~\cite{Banfi:2004nk} and at the end of Sec.~\ref{sec:NLL_resum}).
The scale uncertainty band of the naive exponentiation does of
course not account for this and is small for all values of the
observable.
Therefore for observables like the recoil event shapes, whose double
logarithms do not fully exponentiate, the naive exponentiation of the
NLO results can only be used as long as one is far from the divergence
of the pure NLL resummation (whose position is one of the pieces of
information provided by \caesar).

Finally, we show in fig.~\ref{fig:nlo+nll-vs-expnlo} (bottom, left) how
well the naive exponentiation does in the case of the three-jet
resolution parameter with exponentially suppressed term. We see that
in this case the naive exponentiation result is {\it not} contained in
the {\it full} uncertainty band of the NLO+NLL resummation. This is
true for the tail of the distribution, where there seems to be too
little radiation suppression and, similarly, for the peak, whose
position is slightly displaced to the left. This softer spectrum is a
feature of other exponentially suppressed observables as well (though
the effect is not as remarkable). 

Altogether it seems that naive exponentiation is a sensible procedure
to extend the range of validity of pure NLO predictions. However,
since it is not guaranteed to provide any formal logarithmic accuracy,
one should rely on full
NLO+NLL predictions for precision studies. In any case, we stress that
before carrying out this exponentiation procedure, one should
understand the basic soft/collinear properties of the observable (be
it with \caesar or in whatever other way).

%----------------------------------------------------------------------
\subsection{Comparison with (matched) parton showers}
\label{sec:partshow}

For most practical applications, it is far more convenient to use
parton-shower Monte Carlo event generators, like Herwig~\cite{Herwig}
or Pythia~\cite{Sjostrand:2006za}, or event generators merged with LO matrix
elements, rather than a full NLL+NLO calculation.
It can however be difficult to estimate the accuracy of these tools
and the reliability of the error estimates that come with them.
The purpose of this section is therefore to compare the NLL+NLO
results with parton-shower based predictions (at parton level, in
order to avoid non-perturbative corrections from hadronisation and the
underlying event).

We will start with Herwig (v6.5) events showered from exact tree-level
matrix elements for $2\to2$, $2\to3$ and $2\to4$ partonic scatterings,
as generated with Alpgen \cite{alpgen}. We use the MLM
prescription~\cite{Alwall:2007fs} to avoid double counting between
emissions generated in the hard $2\to n$ scattering and those
generated by the parton shower.
This combination of parton-shower and tree-level matrix elements is
the standard tool for many Tevatron and LHC predictions.

To gauge uncertainties in the resulting matched samples we shall
simultaneously vary the renormalisation and factorisation scales in
the tree-level matrix elements by a factor of two around their default
settings in the MLM procedure (which are taken as in the CKKW
procedure~\cite{Catani:2001cc}). 
The MLM procedure also involves a separation scale between the region
of phase-space to be accounted for by the tree-level matrix elements
or by the parton shower. 
We take this separation scale to be $0.5 p_{t,\min}$ when we look at
event samples whose hardest jet has $p_t > p_{t,\min}$ and the angular
distance for a jet and a parton to be matched is restricted to be
$\Delta R < 1.05$.
The hard events have been generated with a $p_t$ threshold $0.4
p_{t,\min}$ for all partons (constrained to have $|y|<5$), which must
separated from each other by a distance $\Delta R > 0.7$.
%
  %% Typical output from Alpgen for our ptmin 50 GeV sample
  %%
  %%  Generation cuts for the partonic event sample:                       
  %%       Light jets:                                                     
  %%  ptmin=  20. |etamax|=  5. dR(j-j)>  0.7                              
  %%  
  %% INPUT 0 FOR INCLUSIVE JET SAMPLE, 1 FOR EXCLUSIVE
  %% (SELECT 0 FOR HIGHEST PARTON MULTIPLICITY SAMPLE)
  %% (SELECT 1 OTHERWISE)
  %%  
  %% INPUT ET(CLUS), R(CLUS), ETACLMAX
  %% (SUGGESTED VALUES:  25.  0.7  5.)
  %%  
  %% JET PARAMETERS FOR MATCHING:
  %% ET>  25. ETACLUS<  5. R=  0.7
  %% DR(PARTON-JET)<  1.05
%
In principle the MLM separation scale should also be varied in order to
gauge uncertainties.
However, the generation threshold should also be kept lower than the
separation scale and the production of the $2\to4$ tree-level event
samples with the $0.4 p_{t,\min}$ threshold already turned out to have
very low efficiency and would have become prohibitive with much lower
a threshold.
Therefore we will only show results with a fixed separation scale.
The PDF that  we used was CTEQ5L~\cite{Lai:1999wy}, the default choice
in Alpgen, but we also verified that the effect of switching to CTEQ6M
(as used in our NLO+NLL calculations) was small. 

\begin{figure}
  \centering
  \includegraphics[width=\textwidth]{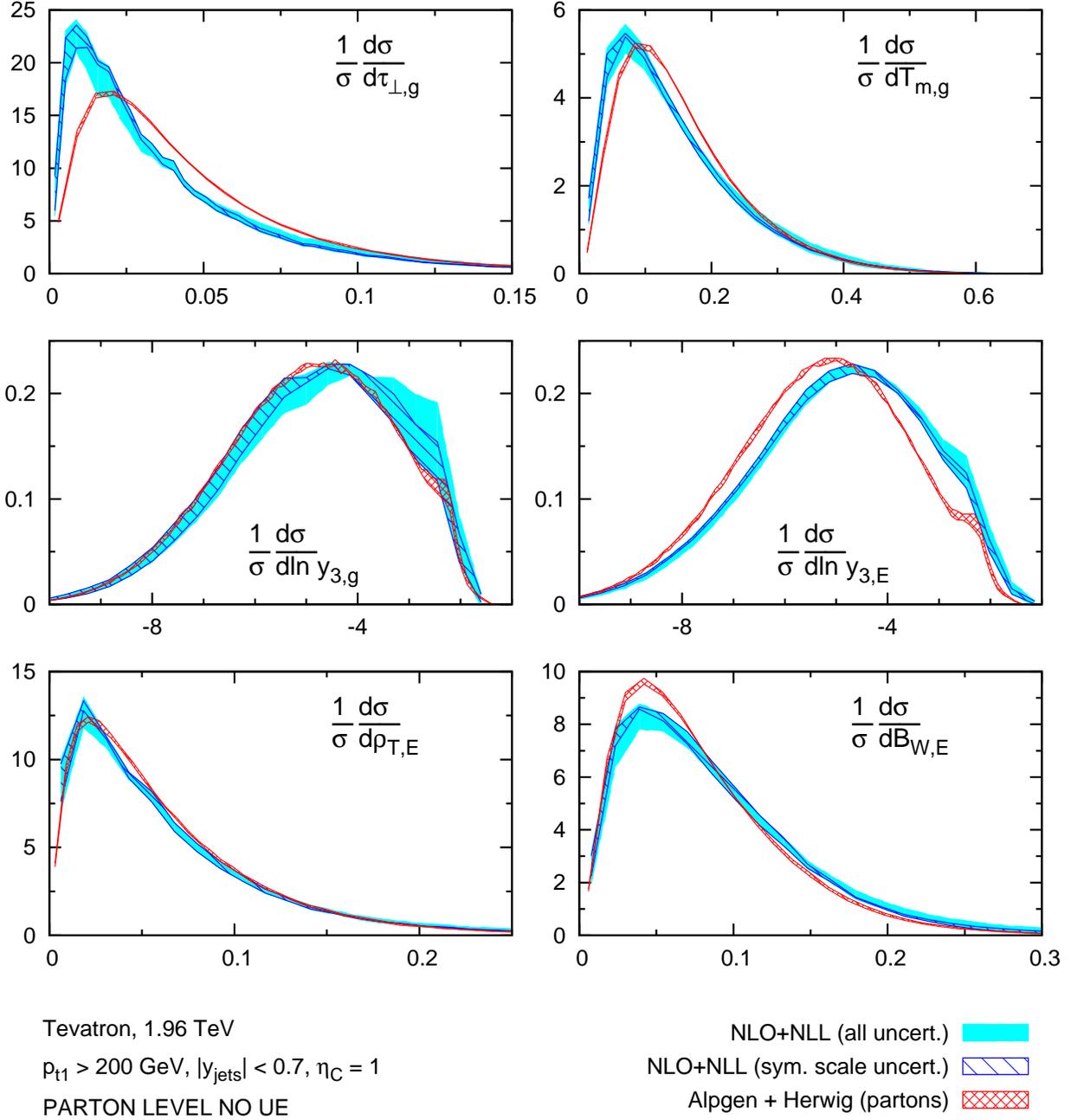}
  \caption{Comparison of the NLO+NLL results with matched
    Alpgen+Herwig results. The latter have just symmetric scale
    variation in their uncertainty bands, so we also include
    subsidiary bands for NLL+NLO with just symmetric scale
    variation. Shown for the Tevatron with a $200\GeV$ cut on $p_{t1}$.}
  \label{fig:nll-v-alp-TEV}
\end{figure}

\begin{figure}
  \centering
  \includegraphics[width=\textwidth]{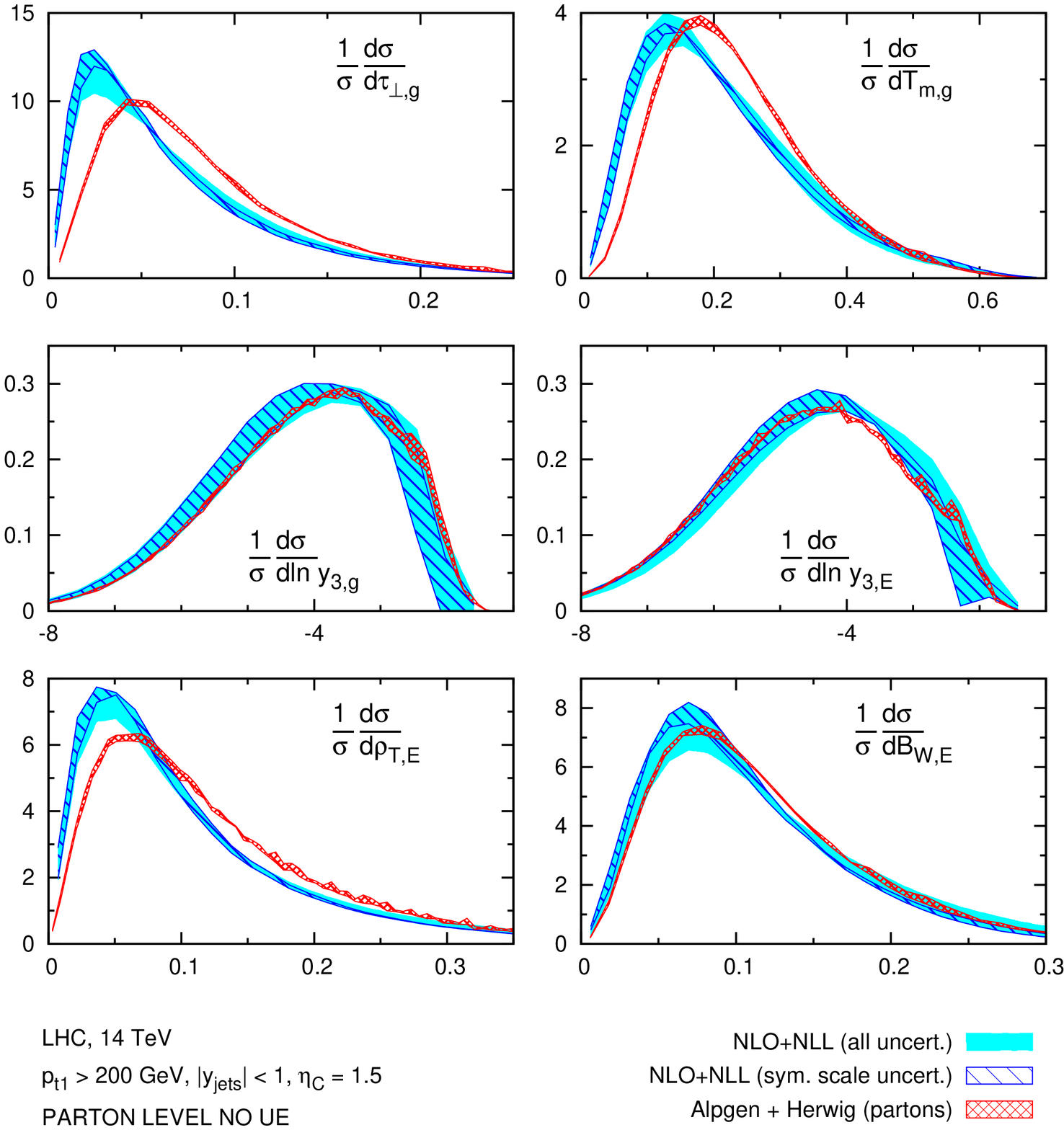}
  \caption{Comparison of the NLO+NLL results with matched
    Alpgen+Herwig results. The latter have just symmetric scale
    variation in their uncertainty bands, so we also include
    subsidiary bands for NLL+NLO with just symmetric scale
    variation. Shown for the $\sqrt{s}=14\TeV$  LHC with a $200\GeV$
    cut $p_{t1}$.}
  \label{fig:nll-v-alp-LHC}
\end{figure}

A comparison of the parton-level Alpgen+Herwig (Tree+PS) results with
the NLL+NLO results is given in figs.~\ref{fig:nll-v-alp-TEV}
and~\ref{fig:nll-v-alp-LHC}, for $p\bar p$ collisions at Tevatron
and LHC, with a cut of $200\GeV$ on the transverse momentum of the
hardest jet.
The Alpgen+Herwig results are shown as red cross-hatched bands.
The NLO+NLL results are shown as two bands: a blue hatched band whose
width corresponds to the uncertainty from just the symmetric
variation of renormalisation and factorisation scales; and a cyan,
solid band corresponding to the full set of uncertainties represented
in fig.~\ref{fig:uncertainties}.

Generally, there is reasonable agreement between the Tree+PS and the
NLO+NLL results.
One feature of note is that the Tree+PS uncertainty band is
significantly narrow than the NLO+NLL band (even that with just
symmetric scale variation).
It is not immediately obvious that this truly reflects smaller
uncertainties in the Tree+PS case, which, based as it is on LO
calculations, would be expected to show larger uncertainties than the
NLO+NLL prediction.
We tend instead to interpret this as indicating that symmetric scale
variation does not provide a good estimate of the true uncertainty on
Tree+PS predictions.\footnote{It should be said that the uncertainty
  band on the Tree+PS prediction is considerably larger if one
  considers differential cross sections instead of normalised
  differential distributions.}

There does not seem to be a clear pattern to the cases where there are
significant differences between the two kinds of predictions.
For example, for the $\tau_{\perp,g}$ and the $T_{\min}$ variables the
Tree+PS predictions seem harder than the NLL+NLO results.
Instead, for $y_{3,\cE}$, the Tree+PS results are generally softer.

\begin{figure}
  \centering
  \includegraphics[width=\textwidth]{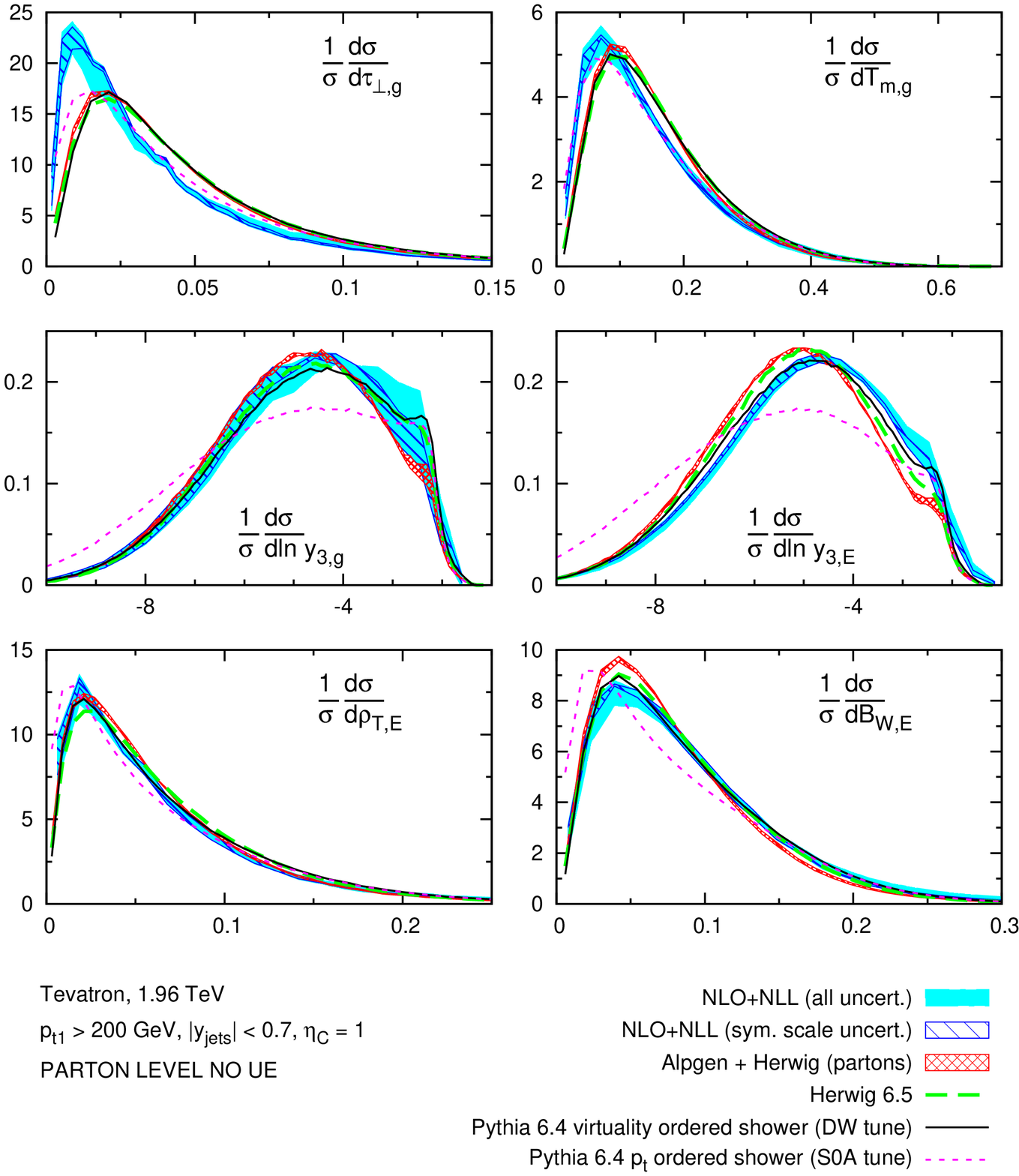}
  \caption{As is fig.~\ref{fig:nll-v-alp-TEV} with, in addition, results
    from plain Herwig and Pythia showers.
    Pythia 6.4 is shown both for the old (virtuality
    ordered) and new (transverse-momentum ordered) showers.
    All results are shown at parton-level, without multiple
    interactions (i.e.\ no underlying event).
    Shown for the Tevatron with a $200\GeV$ cut on $p_{t1}$.  }
  \label{fig:nllnlo+partonshowers-TEV200}
\end{figure}

\begin{figure}
  \centering
  \includegraphics[width=\textwidth]{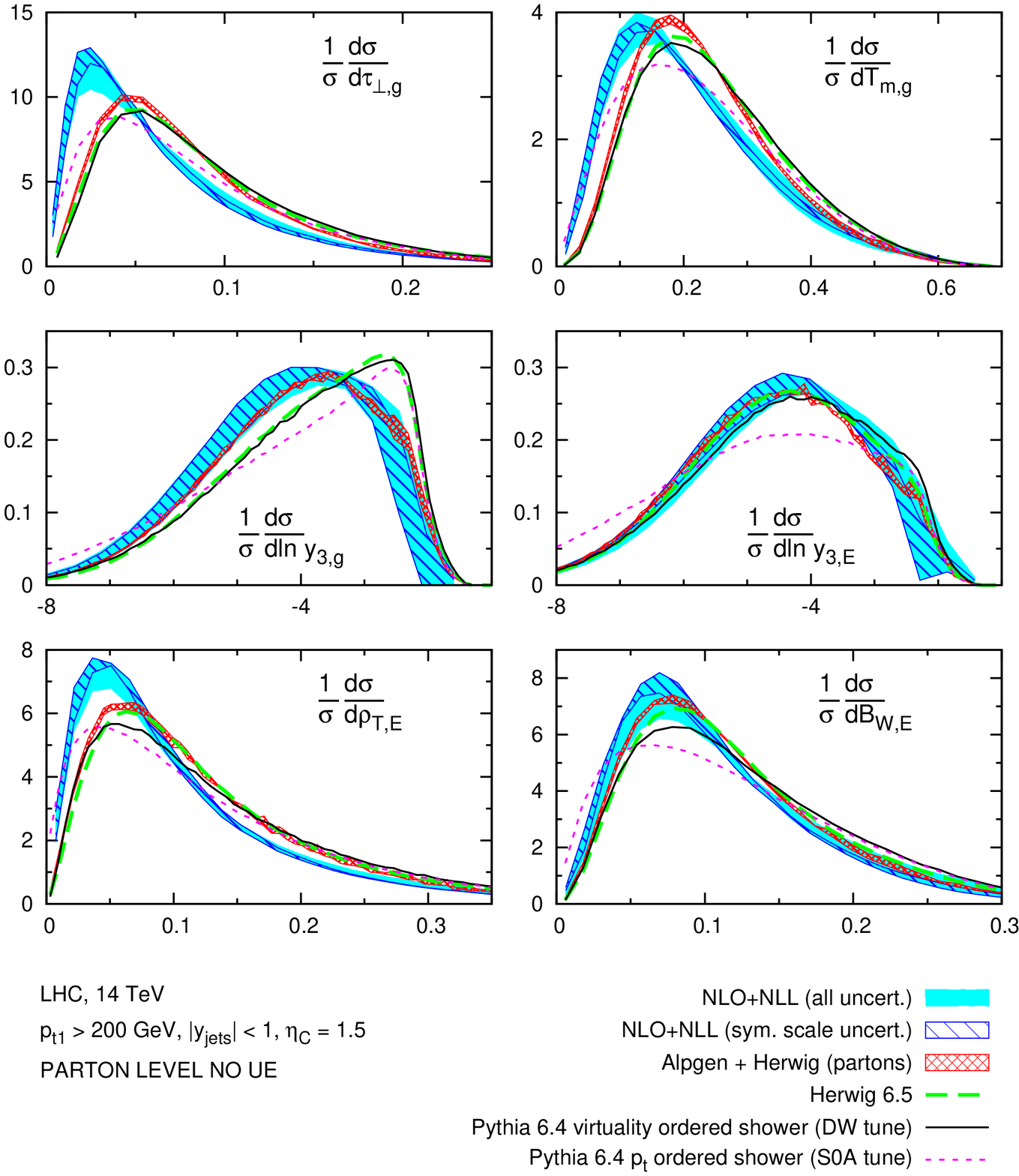}
  \caption{
    As is fig.~\ref{fig:nll-v-alp-LHC} with, in addition, results
    from plain Herwig and Pythia showers.
    Pythia 6.4 is shown both for the old (virtuality
    ordered) and new (transverse-momentum ordered) showers.
    All results are shown at parton-level, without multiple
    interactions (i.e.\ no underlying event).
    Shown for the $\sqrt{s}=14\TeV$ LHC with a $200\GeV$ cut on
    $p_{t1}$.  }
  \label{fig:nllnlo+partonshowers-LHC200}
\end{figure}

We also show in figs.~\ref{fig:nllnlo+partonshowers-TEV200}
and~\ref{fig:nllnlo+partonshowers-LHC200} a comparison between
results obtained from different shower Monte Carlo event generators
with and without matching, for the same subset of the observables at
the Tevatron and the LHC with a minimum $p_t$ on the hardest jet of
$200\GeV$.  Pythia 6.4 is shown both for the old (virtuality ordered)
and new (transverse-momentum ordered) showers.  All results are shown
at parton-level, without multiple interactions (i.e.\ no underlying
event). For reference we show also the result of NLO+NLL
resummation with symmetric scale variation uncertainties (the use of
the full band would complicate the plots excessively). 

In general Herwig's angular-ordered shower and Pythia's virtuality
ordered (old) shower give results that are quite similar (or slightly
harder) to the full matched results, with deviations visible in some
cases, e.g. for $B_{W, \cE}$ at the LHC.

It is perhaps surprising that unmatched, plain parton shower results,
are often harder and sometimes even closer to the NLL+NLO band than
the Tree+PS matched ones, this is particularly evident for $y_{3, g}$ at the
LHC.
This is an unexpected result as the motivation for carrying out
Tree+PS merging is that parton showers are unable to reproduce the
structure of hard large-angle emissions.

What is also evident from figs.~\ref{fig:nllnlo+partonshowers-TEV200}
and~\ref{fig:nllnlo+partonshowers-LHC200} is that there are big
discrepancies between the newer, transverse-momentum ordered shower in
Pythia 6.4 (in the S0A tune) and, essentially, everything else.
These distributions appear to be significantly softer than those from
other parton showers, with the difference most visible in the case of
the $y_3$ variables, and inconsistent with the NLL+NLO calculation.

\begin{figure}[t]
  \centering
  \includegraphics[width=\textwidth]{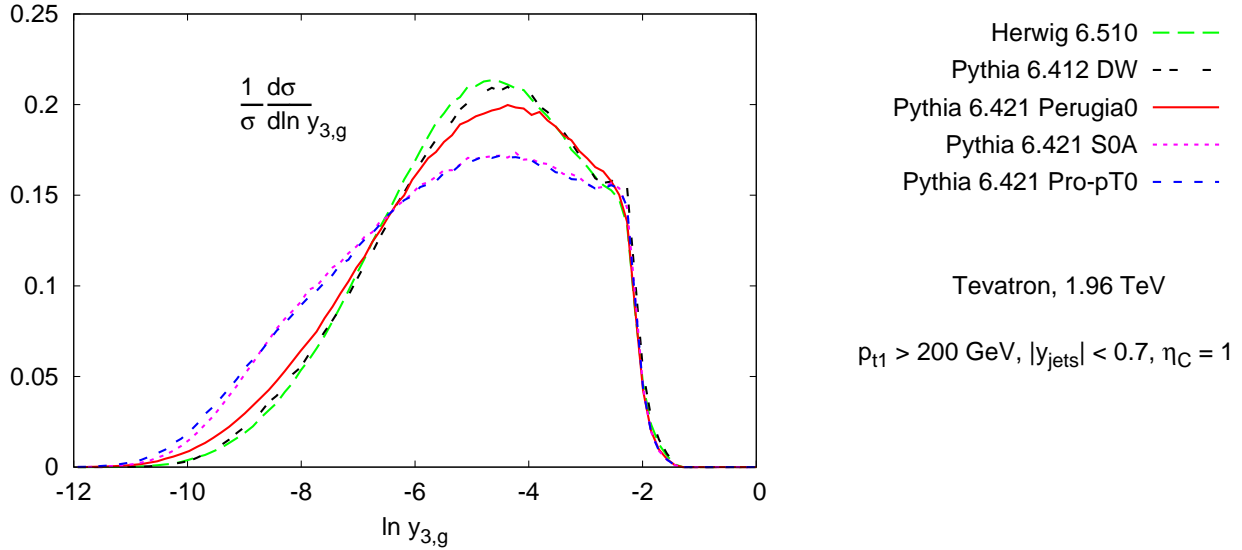}
  \caption{The differential distribution for $y_{3,g}$ at the
  Tevatron, $p_{t1} > 200$ GeV, computed at
    parton level with Herwig and Pythia with both the old shower (DW)
    and the new shower with different tunes.}
  \label{fig:y3g-pytunes}
\end{figure}

It is therefore useful to try to understand whether the origin of the
discrepancies lies in the new shower or in the tuning of the Pythia
parameters. To further probe this issue, we show in
Figure~\ref{fig:y3g-pytunes} a comparison among plain Herwig,
virtuality ordered (DW) Pythia and different tunings of the new
transverse-momentum ordered shower, S0A~\cite{Skands:2007zg}, as used
above, and two more recent tunes, Perugia0~\cite{Skands:2009zm}
Pro-pt0~\cite{Buckley:2009vk},  (shown with 
version 6.421; version 6.412 yields identical results).
Of these the two more recent $p_t$-ordered tunes, Pro-pt0 gives
results very similar to S0A, while Perugia0 is closer (though not
identical) to the Herwig and virtuality-ordered Pythia results.
The conclusion to be drawn from these results is that for
transverse-momentum ordered showers, the shower parameters can have
major implications for the reliability of the results and a consensus
has yet to emerge among current tunes for the choices of these
parameters.

%======================================================================
\section{Non-perturbative effects}
\label{sec:study-np-effects}

\begin{figure}[p]
  \centering
  \includegraphics[width=\tw]{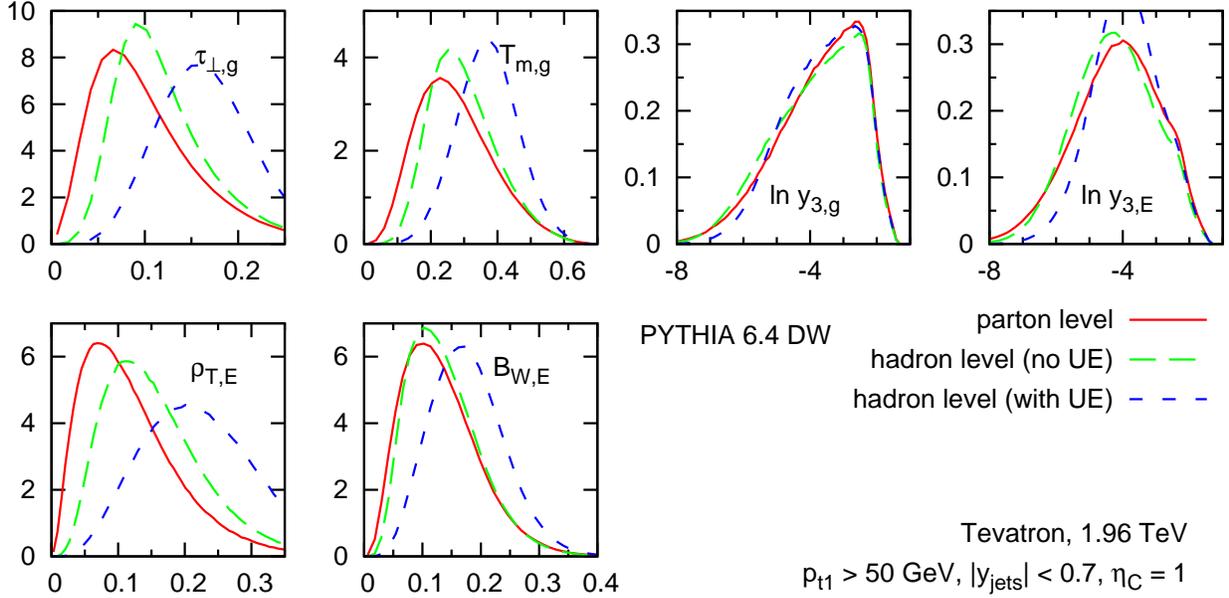}
  \caption{%
    Comparison of parton-level, hadron-level without UE and
    hadron level with UE, for selected event-shape distributions, as
    obtained with Pythia 6.4 (DW tune).% + Jimmy 4.31 (ATLAS tune, as given in text).
    Shown for the Tevatron with a $50\GeV$ cut on
    $p_{t1}$.  }
  \label{fig:pythia3levels-TEV050}
\end{figure}
\begin{figure}[p]
  \centering
  \includegraphics[width=\tw]{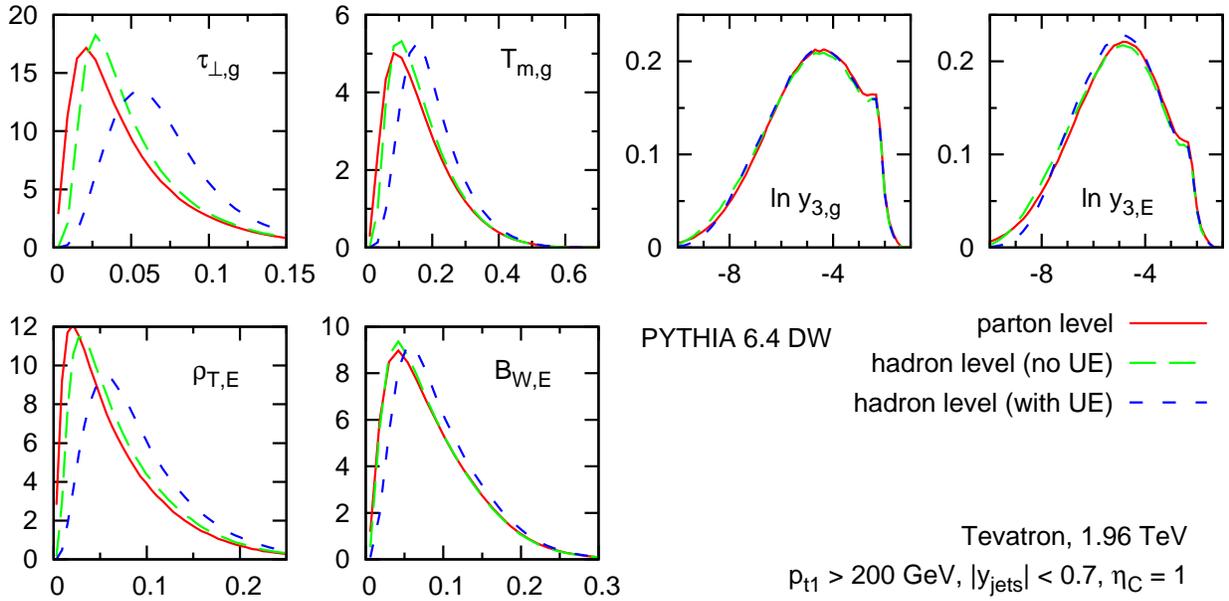}
  \caption{%
    As in fig.~\ref{fig:pythia3levels-TEV050}, but
    for the Tevatron with a $200\GeV$ cut on $p_{t1}$.  
  }
  \label{fig:pythia3levels-TEV200}
\end{figure}

\begin{figure}[p]
  \centering
  \includegraphics[width=\tw]{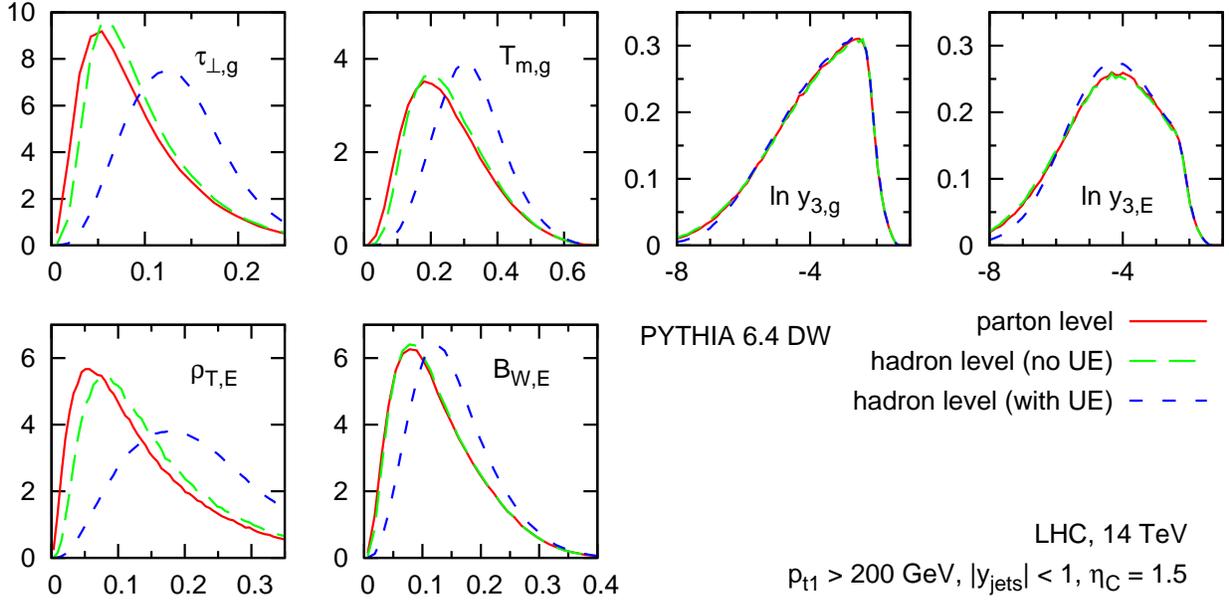}
  \caption{%
    As in fig.~\ref{fig:pythia3levels-TEV050}, but
    for the $\sqrt{s}=14\TeV$ LHC with a $200\GeV$ cut on
    $p_{t1}$.  
  }
  \label{fig:pythia3levels-LHC200}
\end{figure}
\begin{figure}[p]
  \centering
  \includegraphics[width=\tw]{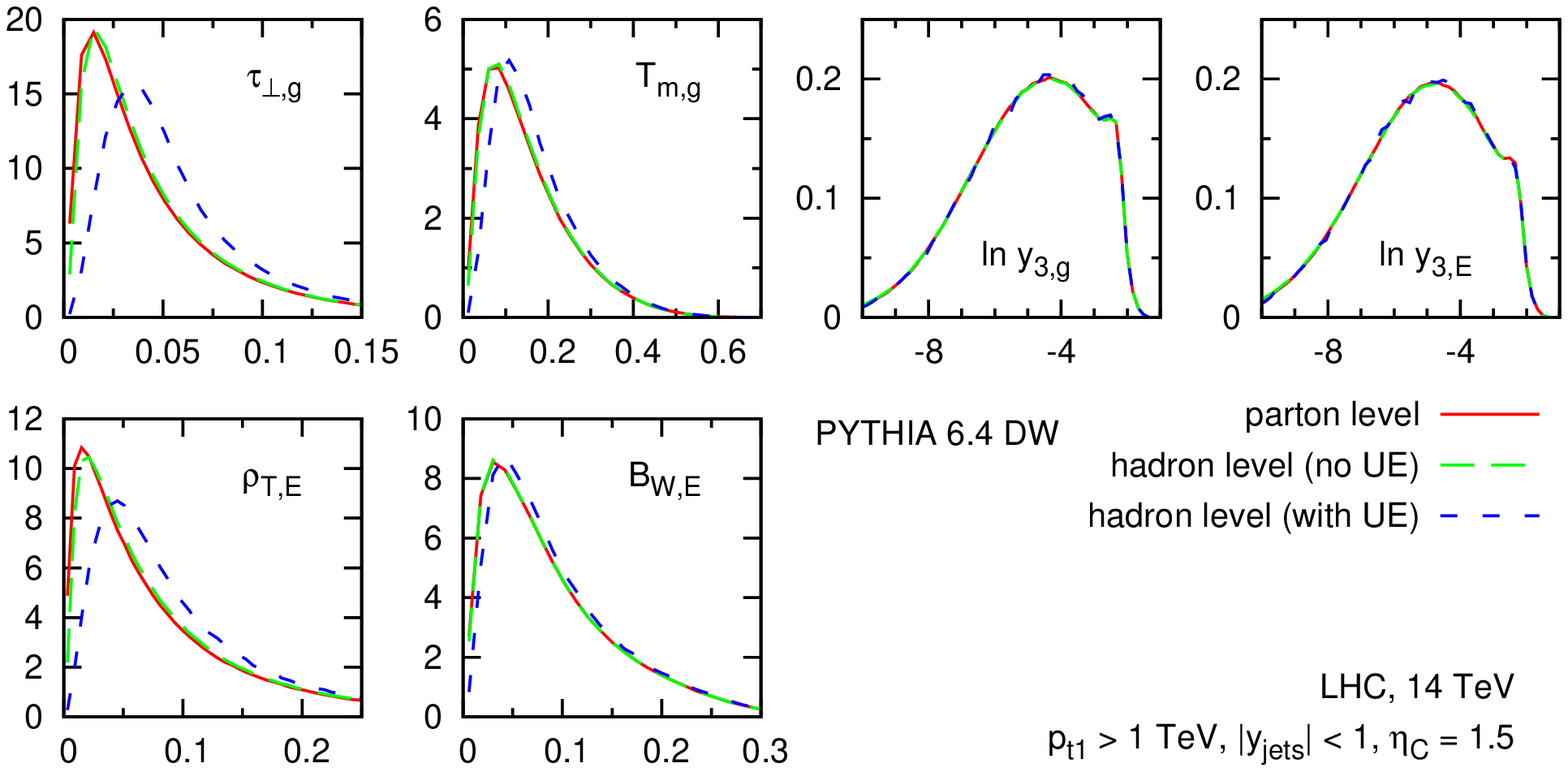}
  \caption{%
    As in fig.~\ref{fig:pythia3levels-TEV050}, but
    for the $\sqrt{s}=14\TeV$ LHC with a $1\TeV$ cut on
    $p_{t1}$.  
  }
  \label{fig:pythia3levels-LHC1000}
\end{figure}

\begin{figure}
  \centering
  \includegraphics[width=\tw]{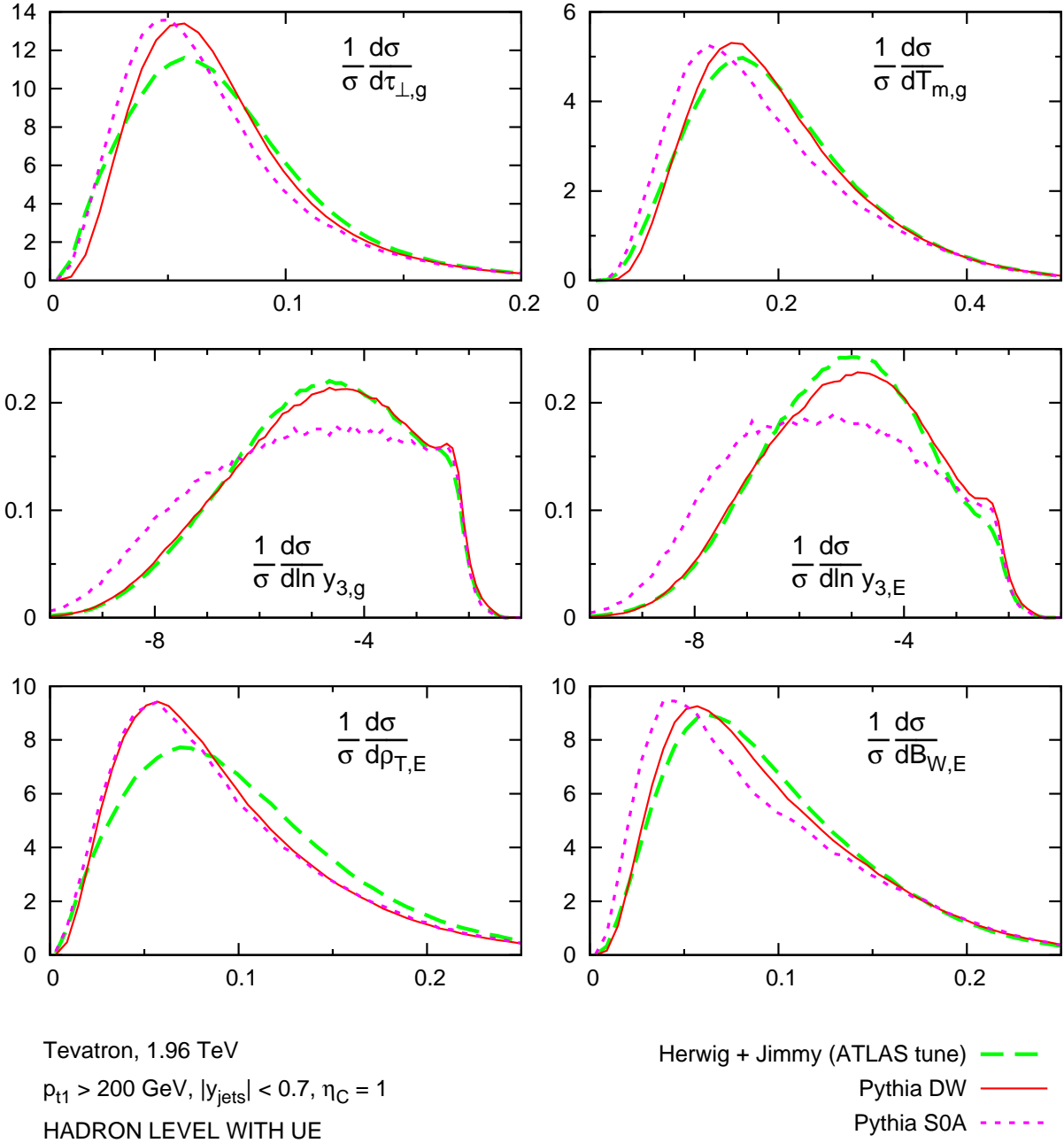}
  \caption{Hadron-level results, including underlying event, for
    selected event-shape distributions, as obtained with Herwig 6.5 +
    Jimmy 4.31 (ATLAS tune, as given in text) and Pythia 6.4 with 2
    tunes, DW (DWT would be identical) and S0A.
    Shown for the Tevatron with a $200\GeV$ cut on
    $p_{t1}$.  }
  \label{fig:UE4tunes-TEV200}
\end{figure}

\begin{figure}
  \centering
  \includegraphics[width=\tw]{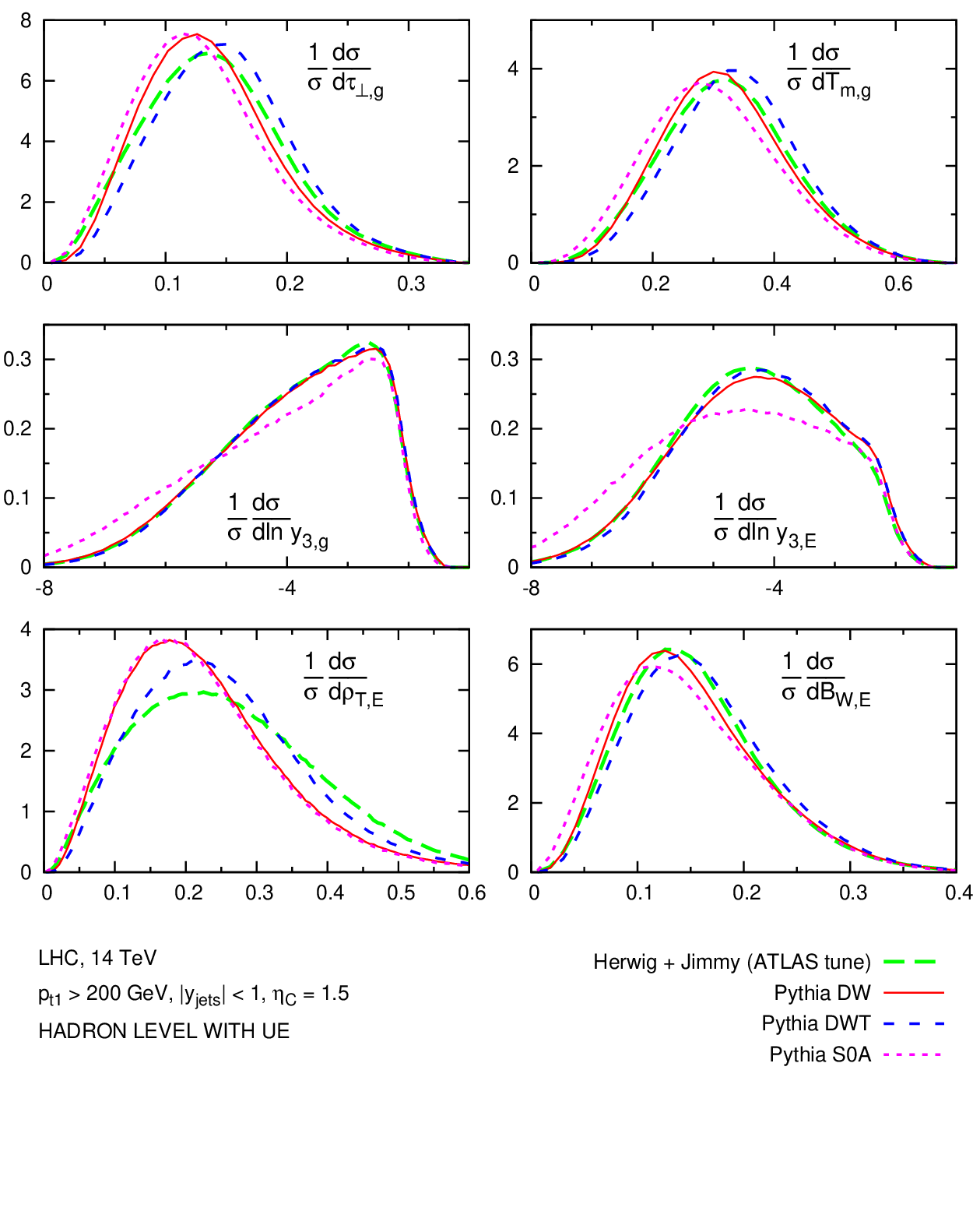}
  \caption{Hadron-level results, including underlying event, for
    selected event-shape distributions, as obtained with Herwig 6.5 +
    Jimmy 4.31 (ATLAS tune, as given in text) and Pythia 6.4 with 3
    tunes, DW, DWT and S0A.
    Shown for the $\sqrt{s}=14\TeV$  LHC with a $200\GeV$ cut on
    $p_{t1}$.  }
  \label{fig:UE4tunes-LHC200}
\end{figure}

So far we discussed only perturbative effects, however, before any
comparison to data can be done, non-perturbative effects have to be
included.
As in $\ee$ annihilation, there are non-perturbative effects due to
hadronisation, i.e.\ related to the transition of partons to
hadrons.
In hadron-hadron collisions there are also interactions between the
two beam-remnants,
the so-called underlying event (UE).
Both effects are suppressed by inverse powers of the hard scale $p_t$
of the high-$p_t$ scattering. For event shapes the dependence is
linear in $1/p_t$~\cite{Banfi:2001aq}, while for jet-resolution
parameters the dependence is even more suppressed, as will become
evident also from the plots presented in this section.\footnote{
  A further potential non-perturbative effect is that due to in-time
  pileup, the additional (usually) soft $pp$ collisions that take
  place during the same beam crossing as the hard $pp$  collision of
  interest. Its impact can largely be eliminated by considering only
  events with a single primary interaction in the beam crossing.
  Given the huge cross-sections for the event selections outlined in
  table~\ref{tab:cross-sections}, this should not pose too much of a
  problem except, possibly, at the higher LHC $p_T$ cut. 
}

At hadron colliders, analytical predictions for non-perturbative
(NP) effects are available only for a limited number of
jet-observbales~\cite{Banfi:2001aq,Dasgupta:2007hr,Dasgupta:2007wa,Dasgupta:2009tm}.
A more general way to estimate those effects is to use event
generators, such as Herwig~\cite{Herwig}or
Pythia~\cite{Sjostrand:2006za}, which can be run at parton- or
hadron-level with or without underlying event. The default Pythia
underlying event model includes multi-parton interactions, while
Herwig needs to be interfaced to Jimmy~\cite{Butterworth:1996zw} to
have a realistic modelling of the underlying event.

In
Figs.~\ref{fig:pythia3levels-TEV050}-\ref{fig:pythia3levels-LHC1000},
we compare parton level and hadron level results without and with
UE for the set of observables discussed previously for the low- and
high-$p_t$ samples at the Tevatron and at the LHC, as obtained with
virtuality ordered Pythia 6.4 (DW tune). 
%Herwig 6.5 + Jimmy 4.31.

As far as hadronisation corrections are concerned, one notices
immediately that for event shapes these effects are quite large at
the Tevatron for the $p_{t1}>50$GeV sample. They systematically shift
the distributions to the right and distort them (mostly squeeze
them). As expected these effects decrease considerably at
$p_{t1}>200$GeV. Going from the Tevatron to the LHC, keeping the
$p_{t1}>200$GeV cut, hadronisation correction are comparable, while
again they decrease when going to the $p_{t1}>1000$GeV sample, where
they are completely negligible.  
Since the majority of events in a sample will have jets with $p_t$
close to the $p_t$-cut, this patten confirms the expected $1/p_{t}$ scaling of
hadronisation corrections.
For $y_3$ distributions hadronisation effects follow the same pattern
but are much smaller, and are already very small at the
Tevatron for $p_{t1}>50$GeV. 

The effect of the underlying event on these distributions is quite
different. For event shape distributions at the Tevatron for
$p_{t1}>50\GeV$, the UE broadens significantly the distributions and
moves them systematically to the right. 
For fixed center of mass (c.o.m.) energy, the UE decreases with
$p_t$, but contrary to the hadronisation corrections, increasing the
c.o.m. energy, at fixed $p_t$ results in an increased UE activity.
%due to the larger phase space available to multiple interactions. 
% 
As for hadronisation corrections,
UE effects are much smaller for $y_3$ distributions than for event
shapes.
This fact means that at sufficiently high $p_t$ one can compare
perturbative predictions directly to data, without additional NP
corrections. This also makes $y_3$ distributions (in particular the
global version) suitable for direct tuning of shower parameters.
We note also that different event shapes have different NP
sensitivities, broadenings seem to have smaller corrections, while
masses and thrust distributions tend to have larger ones. Therefore
the latter seem better suited to study NP effects and to tune models
of hadronisation and underlying event.

To address this last issue further, we show in
Fig.~\ref{fig:UE4tunes-TEV200} how different Monte Carlo showers and
tunes to the same Tevatron data differ from each other for the same
set of observables.
Specifically we use Herwig+Jimmy~\footnote{ 
 The parameters used are PRSOF=0,
 PTJIM$=2.8(s/1800\GeV)^{0.137}\GeV$, an inverse (anti)proton radius
 of JMRAD(73)$=$JMRAD(91)$=1.8\GeV$ and 
 CTEQ6L1~\cite{Pumplin:2002vw} PDFs, as per the ATLAS tune in
 \cite{Albrow:2006rt}.}
% \comment{run
% \verb|./games -dijets -tune hwtune_jimmy_moraes -herwig -out a|
% \verb|./games -dijets -tune hwtune_jimmy_atlas -herwig -out a|
% %
% to check difference,
Pythia's virtuality ordered shower
with the DW  tune  and
Pythia with the $p_t$-ordered shower (S0A tune). 
%
%DW and DWT tunes agree for all observables, but 
It is noticeable how the DW, S0A and Herwig+Jimmy tunes differ (sometimes
substantially), despite the fact that all have been tuned to
Tevatron data.
For S0A in particular this is not really surprising: if 
perturbative predictions already differ substantially, so will 
full results at hadron level. 
However it is nevertheless instructive, because it illustrates to what
extent hadron-level event-shape distributions can help constrain
perturbative aspects of the shower.

Finally, in Fig.~\ref{fig:UE4tunes-LHC200} we show what happens for
the same $p_t$ cut at the LHC. Discrepancies between Herwig and Pythia
survive (but are maybe reduced). In addition to the DW Pythia tune, we
also show DWT (which was identical at Tevatron energies) and see
sizable differences between them.
All this suggests that event shapes have significant scope for tunes of
event generators.

%======================================================================
\section{Multi-jet limit}
\label{sec:multi-jet}

One common use of event shapes is to distinguish two different classes
of multi-jet events: those in which the jets cover phase space quite
uniformly, as in multi-body heavy-particle decays; and those in
which the jets are relatively collimated in few bunches, as induced by
the collinear singularities of massless QCD multi-particle emission.
 %%
 %% search for "sphericity":
 %% ATLAS 0901.0512 (v4): 
 %%   - top & single top, 
 %%   - SUSY, 
 %%   - not for black holes (doesn't work for all # of dim)
 %%
 %% CMS LHCC-2006-021, J. Phys. G: Nucl. Part. Phys. 34 995-1579
 %%   - ttbar (p1211 of physics part)
 %%   - SUSY is inspired by sphericity (but not the same)
 %%   - black holes (p.1436, 1466)
 %%
For example, the ATLAS \cite{Aad:2009wy} and CMS \cite{Ball:2007zza}
discussions of prospective analyses mention the use of event shapes,
most notably the transverse sphericity (see below), in physics studies
that range from $t\bar t$ analyses to searches for supersymmetric
particle decays and black-hole decays (yet other applications include
hidden-valley searches~\cite{Han:2007ae,Strassler:2008fv}).

The purpose of this section is to compare various event shapes'
ability to distinguish characteristically different event topologies.
Firstly we shall examine to what extent current event shapes are able
to distinguish symmetric 3-jet topologies from symmetric multijet
topologies. The main finding will be that they discriminate
principally between 2-jet pencil-like events and symmetric events,
regardless of the number of jets in the latter.
We shall then study the robustness of the identification of symmetric
events: both with respect to parton showering and to the orientation
of the multijet system.
The results from these two studies will then lead us to propose event
shapes that should have enhanced sensitivity to the symmetric multijet
limit.\footnote{
  It is worth noting also the event-shape type observable proposed for
  BSM searches~\cite{Randall:2008rw}, whose aim is not to distinguish
  multijet ``hedgehog'' topologies from dijet topologies, but rather
  to be sensitive to the hadronic structure of BSM signal events with
  large transverse missing energy (e.g. $R$-parity conserving SUSY),
  but without making explicit use of the measurement of missing
  transverse energy.  }

%----------------------------------------------------------------------
\subsection{The transverse sphericity}
\label{sec:spheri}

One event shape that we have not considered so far is the
sphericity. Since it is by far the most widely used for discriminating
symmetric multi-jet topologies, let us briefly examine its properties.
It is defined in terms of the eigenvalues $\lambda_1
\ge \lambda_2$ of the transverse-momentum tensor:
\begin{equation}
  \label{eq:trans-spheri-matrix}
  M_{xy} = \sum_i \left( 
    \begin{array}{cc}
      p_{xi}^2       & p_{xi} p_{yi}\\
      p_{xi} p_{yi}  & p_{yi}^2
    \end{array}
  \right)\,,\qquad\qquad
  \spherig \equiv \frac{2\lambda_2}{\lambda_1 + \lambda_2}\,.
\end{equation}
It has the property that it tends to $1$ for events with circular
symmetry in the transverse plane, and is $0$ for pencil-like events.
However the appearance of a sum of squared momentum components in
$M_{xy}$ makes this observable \emph{collinear unsafe}, as is
the case~\cite{Farhi} for the related variable in $\ee$: for example,
if a hard momentum along the $x$ direction is split into two equal
collinear momenta, then their combined contribution to $\sum_i p_{xi}^2$
will be half that of the original momentum. Therefore collinear
splittings change the sphericity by a factor of order 1.
One consequence of this is that it is impossible to make perturbative
predictions for the sphericity beyond leading order.
Another consequence is that parton showering and hadronisation
significantly alter the value of the observable, limiting its ability
to discriminate different topologies (at least when the input momenta
are particles; often it is jets that are used as inputs).
We shall see this explicitly in
section~\ref{sec:3jets-showering-orientation}. 

%----------------------------------------------------------------------
\subsection{The circular limit}
\label{sec:simple-kinem-stud}

\begin{figure}
  \centering
  \includegraphics[width=0.72\textwidth]{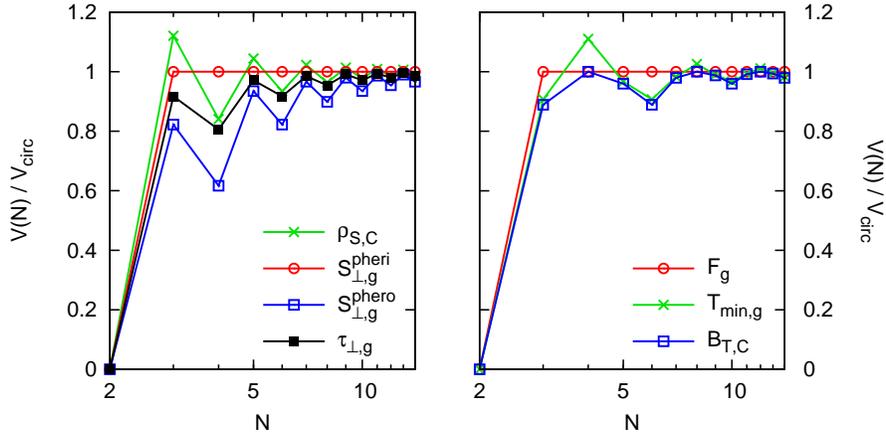}
  \caption{Values of various observables for events with $N$ momenta
    arranged symmetrically in the transverse plane.}
  \label{fig:circular-limit}
\end{figure}

The simplest instructive study that comes to mind for event shapes
intended to distinguish symmetrical multi-jet events from dijet events
is to examine their values $V(N)$ for perfectly symmetrical planar
transverse events with varying numbers  $N$ of momenta, $p_i =
\frac{Q_{\perp,\cC}}{N}(\cos \frac{2\pi i}{N}, \sin \frac{2\pi
  i}{N},0,1)$ for $i = 1\ldots N$.
This is illustrated in
fig.~\ref{fig:circular-limit}. For uniformity the results have
been normalised to the value $V_\mathrm{circ}$ obtained for perfectly
circular planar events ($N\to \infty$), as given in table~\ref{tab:v-infty}.

The only two observables with a monotonic (and trivial) behaviour are
$\spherig$ and $F_g$. The remaining ones have been grouped into the left
and right hand plots according to whether they peak for $n=3$
(thrust-like) or $n=4$ (broadening-like) and one sees that the
perfectly circular limit does not give the largest value for all
observables. Perhaps the most interesting observation from these plots
is the modest difference between the $3$-particle and fully circular
events --- thus they are sensitive to the absence of a unique
preferred transverse direction, but not to the overall degree of
symmetry of the event.

\begin{table}
  \centering
  \begin{tabular}{|c|c|c|c|c|c|c|}\hline &&&&&&  \\[-10pt]
    $\spherig$ & $\spherog$ & $\;\;F_g\;\;$  & $\tau_{\perp,g}$    & $T_{\min,g}$    &
    $\rho_{S,\cC}$ & $B_{T,\cC}$ \\[3pt] \hline  &&&&&&  \\[-10pt]
    1         & 1         & 1    & $1-\frac{2}{\pi}$ & $\frac{2}{\pi}$    &  $\frac12-\frac{2}{\pi^2}$  & $\frac{\pi}{8}$
    \\[4pt]\hline
  \end{tabular}
  \caption{$V_\mathrm{circ}$, the values of various observables in the
    transverse circularly symmetric limit.
    The events have been chosen to be planar --- the variables other
    than $\rho_{S,\cC}$ and $B_{T,\cC}$ are however insensitive to the
    longitudinal components of the momenta.
  }
  \label{tab:v-infty}
\end{table}

%----------------------------------------------------------------------
\subsection{Collinear safety and showered events}
\label{sec:3jets-showering-orientation}

\begin{figure}
  \centering
  \includegraphics[width=0.32\textwidth]{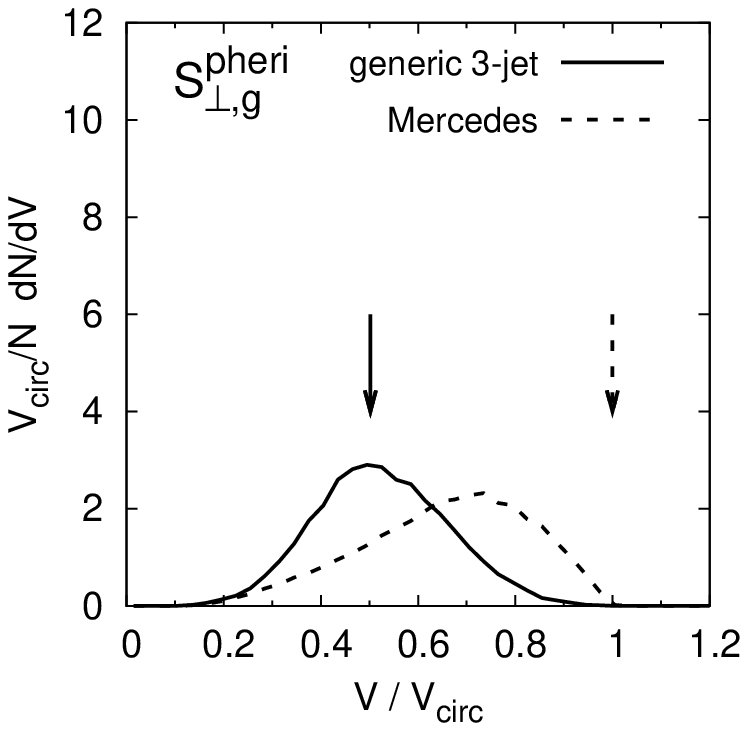}
  \includegraphics[width=0.32\textwidth]{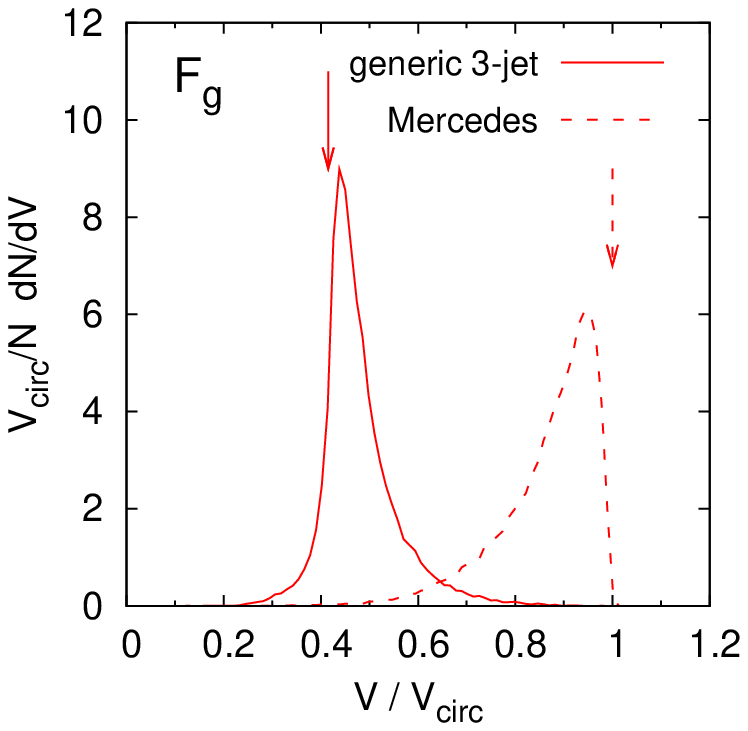}
  \includegraphics[width=0.32\textwidth]{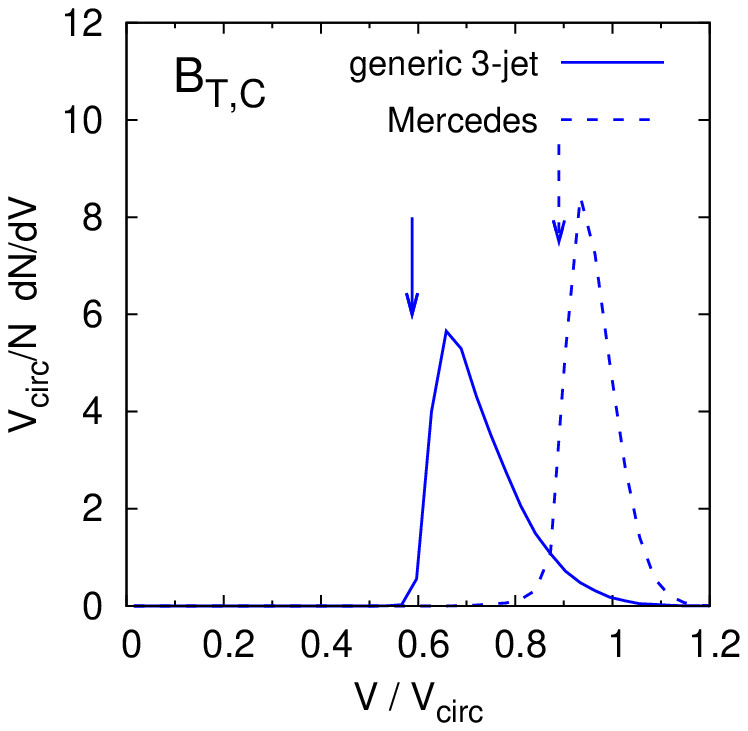}
  \caption{Distribution of event shape values after showering and
    hadronisation for the ``generic'' and Mercedes input 3-parton
    events (for further details of the event generation, see
    text). 
    The arrows indicate the values for the 3-parton events.
    Small overlap between the two distributions and good
    correspondence with the arrows are signs of a good observable.}
  \label{fig:3jet-generic-mercedes}
\end{figure}

Let us now ask the question of how much the collinear unsafety of
$\spherig$ matters in practice. To investigate this we have taken a
number of $2\to 3$ partonic events and showered them with Herwig,
using the ``inclusive'' MLM prescription~\cite{alpgen} to reject
events in which the showering introduces a fourth, harder jet, or
other strong modifications of the
event. Figure~\ref{fig:3jet-generic-mercedes} shows the distribution
of $\spherig$, $F_g$ and $B_{T,\cC}$ for two kinds of input $2\to3$  partonic event,
which are both planar with all particles at rapidity $y=0$:
\begin{center}
  \begin{tabular}{cl|cl}
    \multicolumn{2}{c|}{Event 1 (generic)}     & \multicolumn{2}{c}{Event 2 (Mercedes)}\\\hline
    $p_{t1} = 828$ GeV,&  $\phi_1=0$       & $p_{t1} = 666$ GeV, & $\phi_1=0$ \\
    $p_{t2} = 588$ GeV,&  $\phi_2=3\pi/4$  & $p_{t2} = 666$ GeV, & $\phi_2=2\pi/3$ \\
    $p_{t3} = 588$ GeV,&  $\phi_3=-3\pi/4$ & $p_{t3} = 666$ GeV, & $\phi_3=-2\pi/3$ 
  \end{tabular}
\end{center}
One clearly sees that the collinear unsafe $\spherig$ has much less
discriminating power between the two events than do $F_g$ or $B_{T,\cC}$
(or for that matter any of the other event shapes that were shown in
fig.~\ref{fig:circular-limit}). Furthermore it is clear for the
Mercedes event that the peak at $\spherig=0.7$ has little connection
with the expected 3-parton Mercedes value of $1$. In contrast, the
distributions for the other two observables are peaked close to the
expected values (indicated by the arrows).
%comparison with fig.~\ref{fig:circular-limit} shows reasonable
%agreement for the other two observables.
%
This should of course be of no surprise given the collinear unsafety
of $\spherig$. However, in view of the latter's widespread current use
(albeit with jets, rather than particles, as inputs), we feel that the
point is worth noting.

%----------------------------------------------------------------------
\subsection{Impact of event orientation}

\begin{figure}
  \centering
  \includegraphics[width=0.32\textwidth]{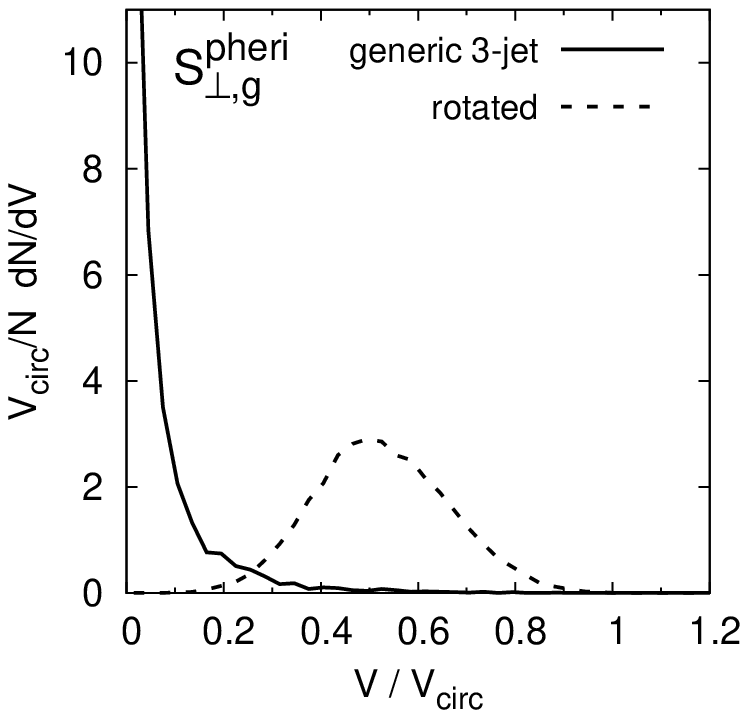}
  \includegraphics[width=0.32\textwidth]{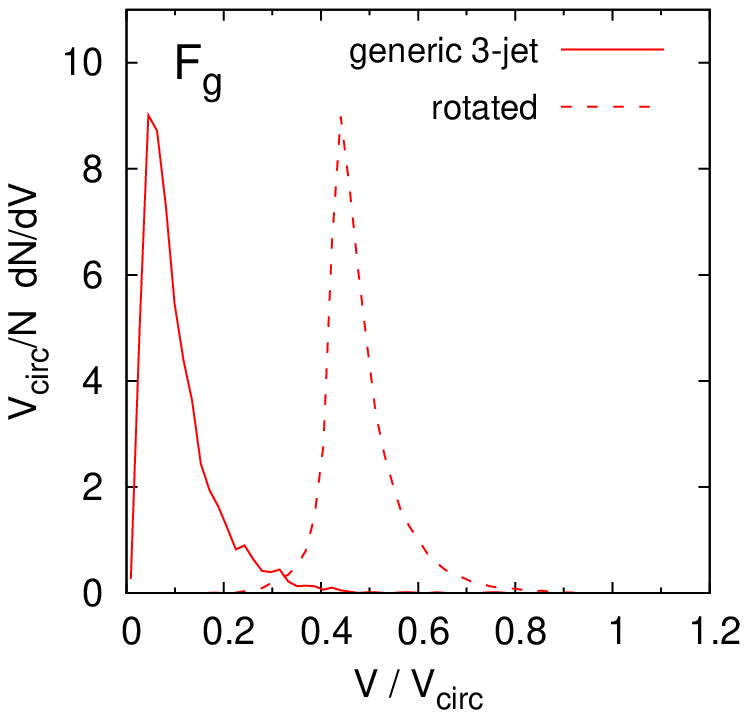}
  \includegraphics[width=0.32\textwidth]{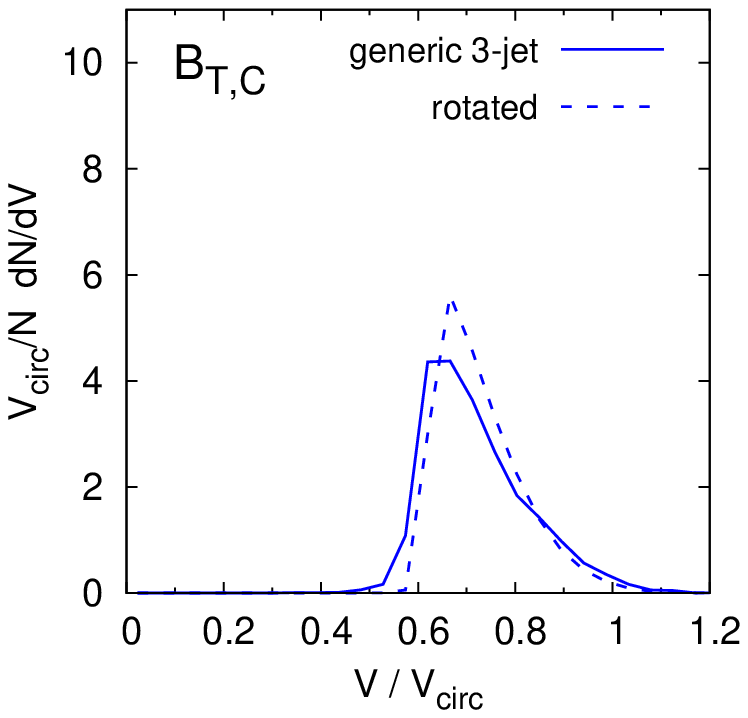}
  \caption{Distribution of event shape values after showering and
    hadronisation for the ``generic'' and rotated generic input 3-parton
    events, as described in the text. 
    Observables that have similar distributions for the two sets of
    events are likely to be more effective for identifying
    massive-object decays.  }
  \label{fig:3jet-generic-rot}
\end{figure}

One of the interests of event-shape studies is in identifying massive
particle decays. Most of the event shapes above have the counterproductive
characteristic that they give very different results for particles
that decay with just transverse components (in the particle's centre
of mass) or with both longitudinal and transverse components.
To illustrate this, we take the generic event given above and rotate
it by $\pi/2$ around the axis of particle $1$, giving
$p_{t2}=p_{t3}\sim416$~GeV, $\phi_{2}=\phi_3=\pi$  and rapidities
$y_{2}=-y_3\simeq 0.88$. We shower it, as explained above, and the
resulting distributions for three event shapes
(normalised to their values in the circular limit) are shown in
fig.~\ref{fig:3jet-generic-rot}.

For $\spherig$ and $F_g$ there is a large difference between
the distributions for the generic and rotated-generic events (and
similarly for \eg $\spherog$, $\tau_{\perp,g}$ and $T_{\min,g}$).
For the broadening in contrast, which we recall involves both the $y$
and $\phi$ dispersions of particles with respect to axes in each of
the two central half-regions, the generic and rotated-generic events
give rather similar distributions. 
A similar phenomenon occurs with the invariant masses of those
regions, in that masses too are sensitive to both directions of
dispersion, though their intrinsic rotational invariance is in part
spoiled when one normalises to $Q_{\perp,\cC}$ as in
eq.~(\ref{eq:mass-XC}).
The rotational invariance is probably in part the origin of the
usefulness of ``cluster-masses'' in the context of hidden-valley
studies~\cite{Strassler:2008fv}.
Note however that masses are significantly more sensitive to
(initial-state) forward semi-hard radiation than are broadenings.

%----------------------------------------------------------------------
\subsection{Increasing sensitivity to the spherical limit}
\label{sec:evshp-spherical}

As is clear from figure~\ref{fig:circular-limit}, none of the event
shapes above are particularly effective at distinguishing truly
spherical events from simpler multi-jet topologies, like symmetric
transverse-planar events. 

What one has in mind when discussing spherical events is that they
have significant ``volume'', symmetrically distributed around the
event. 
One way of obtaining sensitivity to this is to consider the following
matrix, separately for the up and down central regions of an event:
\begin{equation}
  \label{eq:volume}
  M_U = \sum_{i\in \CU} 
      \frac{p_{ti}}{\Qperp} \left(
        \begin{array}{cc}
          \Delta y_{iU}^2                 & \Delta y_{iU} \Delta \phi_{iU}\\
          \Delta y_{iU} \Delta \phi_{iU}  & \Delta \phi_{iU}^2
        \end{array}
        \right)
\end{equation}
where $\Delta y_{iU} = y_i - y_{\CU}$ and $\Delta \phi_{iU} = \phi_i -
\phi_{\CU}$, and similarly for the central down region, $\CD$.
The eigenvalues $\lambda_{U1} > \lambda_{U2}$ of $M_U$ have the
property that $\lambda_{U1}$ is non-zero if there are two non-collinear
particles in the hemisphere, while $\lambda_{U2}$ is non-zero if there
are three non-coplanar particles in the hemisphere. 
The observable
\begin{equation}
  \label{eq:our-spheri-sens-obs}
  S_6 = \min(\lambda_{U2}, \lambda_{D2})\,,
\end{equation}
which we name ``supersphero'',
is therefore non-zero only if there are 3 non-coplanar particles in
each of the hemispheres of the event --- i.e. for events that truly bear
some resemblance to spherical events.

For a perfectly spherical event the two eigenvalues in each hemisphere
are $\lambda_1 = \pi^2/24\simeq 0.411$ and $\lambda_2 =
\pi^2/8-1\simeq 0.234$.\footnote{ One may balance the two eigenvalues more
closely for example by replacing $\Delta y_{iU} \to \Delta y_{iU}(1 +
\Delta y_{iU}^2/6)$ (\ie with the first two terms of the expansion of
$\sinh \Delta y_{iU}$) in eq.~(\ref{eq:volume}), though there is a
significant degree of arbitrariness in this choice.}

The $S_6$ observable, in terms of its use of eigenvalues of a $2\times
2$ matrix, relates of course to the $F_g$-parameter of
eq.~(\ref{eq:fpar}) and to the event shapes studied for boosted
top-quark identification \cite{Thaler:2008ju,Almeida:2008yp}. The
latter's use of a matrix defined in the plane transverse to a jet is
actually quite similar to our use of a matrix defined in a central
half-region.

A detailed study of the $S_6$ observable would benefit from
comparisons of high jet-multiplicity QCD samples and multijet samples
from new-physics scenarios. Such a study is beyond the scope of this
paper, but would, we believe, be of interest.

%\comment{Do we need ref to Almeida, to Thaler?}

%======================================================================
\section{Summary of main results}
\label{sec:ressum}
Given the length of the paper, and the fact that we have addressed quite
a range of issues, we find it useful, before concluding, to summarise here
the main results of the paper. 

There are a number of reason why event shapes provide a powerful
laboratory for studying of a range of aspects of strong-interaction
physics at hadron colliders.
From an experimental point of view, cross sections for the QCD (dijet)
events on which one carries out event-shape studies are very large
both at the Tevatron and at the LHC. This means that high-statistics
event samples are already available at the Tevatron and can be
expected early on at the LHC.
Since event shapes are defined as dimensionless ratios of combinations
of hadron-momenta, and since their differential distributions are also
dimensionless, many experimental uncertainties are reduced.
From a theoretical point of view, one of the attractive characteristics
of event-shape studies is that different variables provide
complementary sensitivities to a broad variety of features of hadronic
events, such as the topology of the final state, the nature of initial
and final-state jet-fragmentation, hadronisation, and the underlying
event.

The above points motivated us to study many hadron-collider event
shapes in a single context.
We exploited the automated NLL resummation procedure
implemented in \caesar to obtain next-to-leading logarithmic resummed
distributions matched to next-to-leading order exact predictions from
\nlojet for a large number of event shapes at the Tevatron and at the
LHC. 

The matching procedure is conceptually simple, but technically
involved, as discussed in detail in Sec.~\ref{sec:matchNLLNLO}.
One issue is that the best logarithmic accuracy achievable once NLO
and NLL predictions are available, namely $\as^n L^{2n-2}$ in the
expansion of the integrated distribution, NNLL$_\Sigma$, can be
obtained only if the NLO code provides full information about the
flavour of all incoming and outgoing partons.
This decomposition into flavour channels is not present in the
publicly available version of \nlojet, so we used the extended version
developed in \cite{Banfi:2007gu} in order to extract this
information. We also needed to use the flavour-$k_t$ jet-algorithm of
ref.~\cite{jetflav} to map the flavour of $2\to3$ events into that of
an underlying $2 \to 2$ Born-like event.
Additionally we needed the order-by-order expansion of PDF evolution,
which was provided by \hoppet \cite{Salam:2008qg}.
The computing effort should also not be neglected: our directory of
resummed results contains $\order{10000}$ files, and we estimate
that several tens of years of CPU time have gone into
the NLO and NLL calculations used here. 
Part of this complexity stems from our choice to consider several
different classes of uncertainties associated with uncalculated
higher-order terms: those from separate variation of renormalisation
and factorisation scales; redefinition of the argument of the
logarithm being resummed ($X$-scale); and two choices of schemes for
combining (matching) the NLO and NLL results.

Among the questions we asked was whether this considerable complexity
is needed. It turned out that the flavour decomposition had only a
modest effect (cf. Appendix~\ref{sec:appA}).
We also found that a simple exponentiation of the NLO result, as
presented in Sec.~\ref{sec:naive-exp}, not even correct to LL or
LL$_\Sigma$ accuracy, comes remarkably close to reproducing most of
the NLO+NLL distributions (albeit not close enough that one would
forgo NLO+NLL if it is available).
One interpretation is that the large amount of radiation that comes
from the 4 Born legs in a $2\to2$ process causes event-shape
distributions to be dominated by regions where the logarithm that is
being resummed is not all that large.
Note, however, that plain (unexponentiated) NLO predictions are very
inadequate substitutes for the full NLL+NLO result and their
uncertainty bands are misleadingly small.

We studied three generic classes of event shapes: the directly global
ones, those with exponentially suppressed terms and those with a
recoil term. The definition of the observables is recaleld in
sections.~\ref{sec:dir-glob}, \ref{sec:exp-obs}, and \ref{sec:rec-obs}
respectively.
While stable numerical results could be obtained for observables
belonging to the first two classes, for the last of these it was
sometimes impossible to obtain numerically sensible results. This is
in part due to cancellations among contributions from multiple
emissions in the recoil term, which cause the resummation provided by
\caesar to have a divergence at small $v$, as explained in
Sec.~\ref{sec:NLL_resum}.
It is also due to structures in the middle of the physical region,
akin to Sudakov shoulders~\cite{Catani:1997xc}, which would require an
additional resummation. Such shoulders are visible e.g. for the
broadenings with recoil term in Figs.~\ref{fig:nllnlo-tev200}
and~\ref{fig:nllnlo-lhc200}.
These observables are also challenging experimentally
because the measurement of the recoil term is affected by
cancellations between large transverse momenta of the two hard jets.

A question mark that hangs over NLL resummations is that of
coherence-violating logarithms (CVL, referred to as super-leading
logarithms (SLLs) in the context of interjet energy flow) \cite{Forshaw:2006fk,Forshaw:2008cq}, terms
potentially starting at $\as^4 L^5$, related to a violation of
coherence, whose validity was a crucial assumption in the
resummations of \cite{caesar}.
There is a risk that this could therefore invalidate our claim of NLL
accuracy for some observables.
We investigated this point in
Sec.~\ref{sec:superl-logar}, and found that the answer depends
critically on the ordering parameter used in the calculation of the
SLL terms. 
If, as in \cite{Forshaw:2006fk,Forshaw:2008cq}, one makes the
\emph{assumption} that the ordering parameter is transverse momentum,
then the claim of NLL accuracy breaks down for our ``exponentially
suppressed'' class of observable (not for the others), while
NNLL$_\Sigma$ remains valid for all observables.
If one instead assumes virtuality ordering, then both NLL and
NNLL$_\Sigma$ accuracies should be valid for all our observables.
This highlights the importance of understanding the question of
ordering for SLLs, which also affects the coefficient of the $\as^4
L^5$ terms in \cite{Forshaw:2006fk,Forshaw:2008cq} and probably
requires that one go beyond the eikonal approximation that was used
there.
Nevertheless, practically we tend to believe that SLLs will not
seriously affect our results, one reason being that we still retain
NNLL$_\Sigma$ accuracy.

Turning to our phenomenological results, a feature common to all
observables is that the shape of the distributions is strongly
influenced by the ratio of quarks to gluons among the incoming
partons. This is because the double-logarithmic Sudakov exponent,
responsible for the position and width of the peak of the distribution
for each underlying subprocess, is determined by the total color
charge of the hard emitting partons. Event samples dominated by gluon
scattering (Tevatron with $p_{t1}> 50$ GeV, LHC with $p_{t1}> 200$
GeV) have broader distributions than those dominated by quark
scattering (Tevatron with $p_{t1}> 200$ GeV, LHC with $p_{t1}> 1000$
GeV). This is evident e.g.\ in Fig.~\ref{fig:different-energies} in the
case of our representative observable $T_{m,g}$ and is discussed in
Sec~\ref{sec:resNLOmat}. We remark that dijet event-shapes at hadron
colliders are the first case in which a change in a kinematical cut
modifies the double logarithmic behaviour of the event-shape
distributions.
This would not be the case for event shapes in hadron-collider
processes such as Drell-Yan production, or  $Z$+jet or $W$+jet.

In the absence of data on the event shapes discussed here, one of the
interesting uses of our NLL+NLO results is to compare them to the
results of two Monte Carlo parton shower programs, this is discussed
in Sec.~\ref{sec:partshow}. We considered Pythia 6.4 and Herwig 6.5
both without and (in the case of Herwig) with matching to multi-parton
tree-level matrix elements (Alpgen, MLM prescription).
%
 % Scores
 % ------
 %
 % Consider 12 var (all 14 minus the two broadenings with recoil)
 %
 %                             TEV 200         LHC 200
 %       -------------------------------------------------
 %       Alpgen                  6 OK             4 OK
 %       Parton showers          9 OK             3 OK
 % 
 %  Lists (OK means one of Herwig/Pythia is "close enough" -- a bit subjective)
 %  - Alp TEV 200:  tauE,TminR,rhoTE, rhoHE,y3g,BTE
 %  - Alp LHC 200:  rhoHE,y3g,y3E,BWE
 %  - PS  TEV 200:  tauE, TminR, rhoTE, rhoHE,y3g,y3E, BTE,BWE,Sg
 %  - PS  LHC 200:  rhoHE, y3E, BWE
 %
The quality of the agreement between plain parton showers and the
resummations depends significantly on the quark/gluon admixture: in
quark-dominated event samples it is often adequate, while in
gluon-dominated samples it is somewhat poorer.
This may be a reflection of the extensive tuning of quark parton
showers carried out with LEP data, while gluon parton showers have
seen fewer constraints.
The importance of tuning parton showers in a context with incoming
beams is highlighted particularly strongly by the results of the newer
$p_t$-ordered shower in Pythia 6.4. In two tunes, S0A and Pro-Pt0, the
agreement both with NLO+NLL and with other showers is quite poor; in
the Perugia0 tune it improves, as can be seen from
Fig.~\ref{fig:y3g-pytunes}. 

One might expect that supplementing parton showers with matching to
multi-parton tree-level events (Tree+PS) should improve the agreement
with NLO+NLL results.
This is the case only for some of the observables.
We also examined the impact of (simultaneous) renormalisation and
factorisation scale variation on the Tree+PS results and found that it
leads to an uncertainty estimate that is far smaller than the actual
differences between Tree+PS and NLO+NLL results, as can be seen in
Figs.~\ref{fig:nll-v-alp-TEV} and~\ref{fig:nll-v-alp-LHC}.\footnote{More
  precisely: it was significant on the 2, 3 and 4-jet differential
  cross sections, but mostly cancelled in the normalised event-shape
  distributions.}
This should not be surprising: in the NLO+NLL calculations
simultaneous scale variation represented only a small part of the full
uncertainties.
Questions that remain open therefore are whether in the Tree+PS approach
uncertainties can be more faithfully estimated if one examines further
``handles'' (independent scale variation, matching scale, etc.), and
whether we would have reached similar conclusions with other matching
schemes (e.g.\ CKKW) and programs.

From a non-perturbative point of view, we estimated both hadronisation
and UE\ corrections using Monte Carlo event generators, as discussed
in Sec.~\ref{sec:study-np-effects}. As expected, hadronisation
corrections decrease when increasing the $p_{t1}$-cut on the
jets. They are fairly negligible with cuts of the order of $200$ GeV
both at the Tevatron and at the LHC, as can be seen in
Figs.~\ref{fig:pythia3levels-TEV200}
and~\ref{fig:pythia3levels-LHC200}.  For lower $p_t$ cuts, they shift
the distributions to the right and, for some observables they squeeze
them, see e.g.\ Figs.~\ref{fig:pythia3levels-TEV050}. 
For jet resolution parameters ($y_{3,g}$ and $y_{3,\cE}$)
hadronisation effects are always small, just a few percent correction
for $p_{t1}>50\GeV$ at the Tevatron, much smaller in all other cases.
These observations are consistent with the experience obtained from
$e^+e^-$ and DIS event-shape studies.

As concerns the UE, there are observables for which
it has a sizable effect even at $p_{t1} > 1\TeV$, most notably for
the thrusts and jet-masses, as can be seen in
Fig.~\ref{fig:pythia3levels-LHC1000}. 
This means that these event shapes are
particularly good for tunes of the UE. 
Jet-resolution parameters are the only observables for which the
UE effects remain consistently small (a few percent for the lower
$p_t$-cut samples, even smaller for the large $p_t$ ones). They are
therefore well suited for tunes of perturbative parameters of showers
and in general for perturbative studies.

Finally, in Sec.~\ref{sec:multi-jet} we examined how well event shapes
can discriminate QCD-like two-jet events from BSM-like multi-jet
events, and how robust this discriminating power is with respect to
parton shower (radiative) corrections. 
In general we find that
event-shapes discriminate well between events with two or more than
two jets, but they do not discriminate well between three or any large
number of jets: the value of event-shapes does not even increase
monotonically with the number of jets for symmetric events, see
Fig.~\ref{fig:circular-limit} in Sec.~\ref{sec:simple-kinem-stud}. 
On the other hand it is possible to design new event shapes, which
start with six jets in the final state, as is the case for our
``supersphero'' $S_6$ event shape defined in
Sec.~\ref{sec:evshp-spherical}. 
We believe these might be particularly promising for extracting
new-physics signals that involve relatively isotropic events with high
jet multiplicity.
Other considerations that we examined in
Sec.~\ref{sec:3jets-showering-orientation} include how well event
shapes retain their discriminatory power after parton showering (the
collinear-unsafe, but widely used transverse sphericity, whose
definition is recalled in Sec.~\ref{sec:spheri} is particularly poor
in this respect); and also their sensitivity not just to transverse
event structure, but also to longitudinal event structure (the
broadenings do well at treating both on an equal footing).

%======================================================================
\section{Conclusions}
\label{sec:short-conclusions}

In this article we have shown the first NLO+NLL (NNLL$_\Sigma$)
predictions, with full 
uncertainty bands, for hadronic observables at $pp$ and $p\bar p$
colliders.
We opted to make these predictions for event shapes in the context of
dijet production, bringing
together calculations with \caesar and a specially adapted version of
\nlojet, despite the fact that the NLO+NLL matching is technically
more challenging than for event shapes in other hadron-collider
processes such as Drell-Yan \cite{Stewart:2009yx} or W/Z+jet
\cite{KoutZ0} production.

Several properties of the dijet process motivated our choice: it
involves both initial and final-state partons; it offers the freedom
to vary the proportion of quarks and gluons involved in the Born
process, through the cut on the hard jets; when that cut is placed at
moderate $p_t$, dijet production involves a substantial $gg\to gg$
scattering component, offering the most accessible example of a
gluon-dominated process; and the cross sections imply large event
samples.

Comparisons of our results with parton-shower Monte Carlo predictions
revealed adequate agreement for historic showers (Herwig 6.5,
virtuality-ordered Pythia 6.4) in quark-dominated cases, while the
showers were generally too hard in gluon-dominated processes.
Some common tunes of the newer, $p_t$-ordered shower in Pythia 6.4
fared noticeably worse than the historic showers.
We also examined one framework for matching to multi-parton tree-level
matrix elements (MLM matching of Alpgen+Herwig 6.5). Though it led to
some improvements, it was
not immediately sufficient to bring about systematic agreement with
the NLO+NLL results.
These findings illustrate how event shapes can provide substantial
input to the quest of understanding perturbative QCD at hadron
colliders.

At hadron level, some event shapes are subject to significant
non-perturbative corrections from hadronisation and the underlying
event.
We saw this to be the case, for example, for the thrusts and jet
masses, while other observables, notably the $y_3$ variants, were
largely unaffected by non-perturbative effects.
Studying a broad range of event shapes, as done here, therefore
provides complementary information on QCD phenomena at hadron
colliders at many different physical scales.

Event shapes are of interest not just for constraining QCD dynamics,
but also for discriminating BSM-like multi-jet topologies from more
QCD-like events. 
There are many interesting questions to ask about event shapes in this
context. 
Some that we addressed here include their robustness to parton
showering (the widely used transverse sphericity fares poorly), their
sensitivity to longitudinal versus transverse event structure and
their behaviour in the high jet multiplicity limit, where new
dedicated event shapes, like the supersphero variable introduced here,
can have particular advantages. 

These first steps of ours in exploring the phenomenology of event
shapes at hadron colliders open a window onto a broad range of
possible new studies, both theoretical and experimental. We 
look forward to their future development.

%======================================================================
\section{Acknowledgments}
We thank Mrinal Dasgupta, G\"unther Dissertori, Pino Marchesini, Lester
Pinera, Peter Skands, Mike Seymour, Matt Strassler, Jesse Thaler and
Matthias Weber for fruitful discussions on this subject.
We thank CERN, Milano-Bicocca University, the LPTHE (UPMC Univ. Paris 6) and
ETH Z\"urich for hospitality while part of this work was carried out.
A.B.\ would like to thank Craig Prescott and the High Performance
Computer center at the University of Florida for the use of computing
facilities in an earlier stage of this work.
G.Z.\ is supported by the British Science and Technology Facilities
Council. 
The work of G.P.S.\ is supported in part by the Agence Nationale de la
Recherche under contract ANR-09-BLAN-0060.

%\newpage 

%======================================================================
\appendix

%======================================================================
%----------------------------------------------------------------------
\section{Cross-checking fixed order and resummation}
\label{sec:appA}

Part of the value of having separate resummed and fixed-order
calculations for event-shape distributions is that they provide
cross-checks as to the validity of each of the approaches. This check
is usually performed by a comparison of the exact fixed order results
$\Sigma_i(v)$ in eq.~(\ref{eq:Sigma-expansion}) with the expansion of the
resummed result $\Sigma_r(v)$ from section~\ref{sec:resummation}. At
small $v$ the two results should differ order-by-order only by terms
suppressed by powers of $v$ or by logarithmically enhanced terms that
are neglected within the resummation accuracy of $\Sigma_r(v)$.

\begin{figure}
  \centering
  \setlength{\unitlength}{\textwidth}
  \includegraphics[width=1.0\textwidth]
  {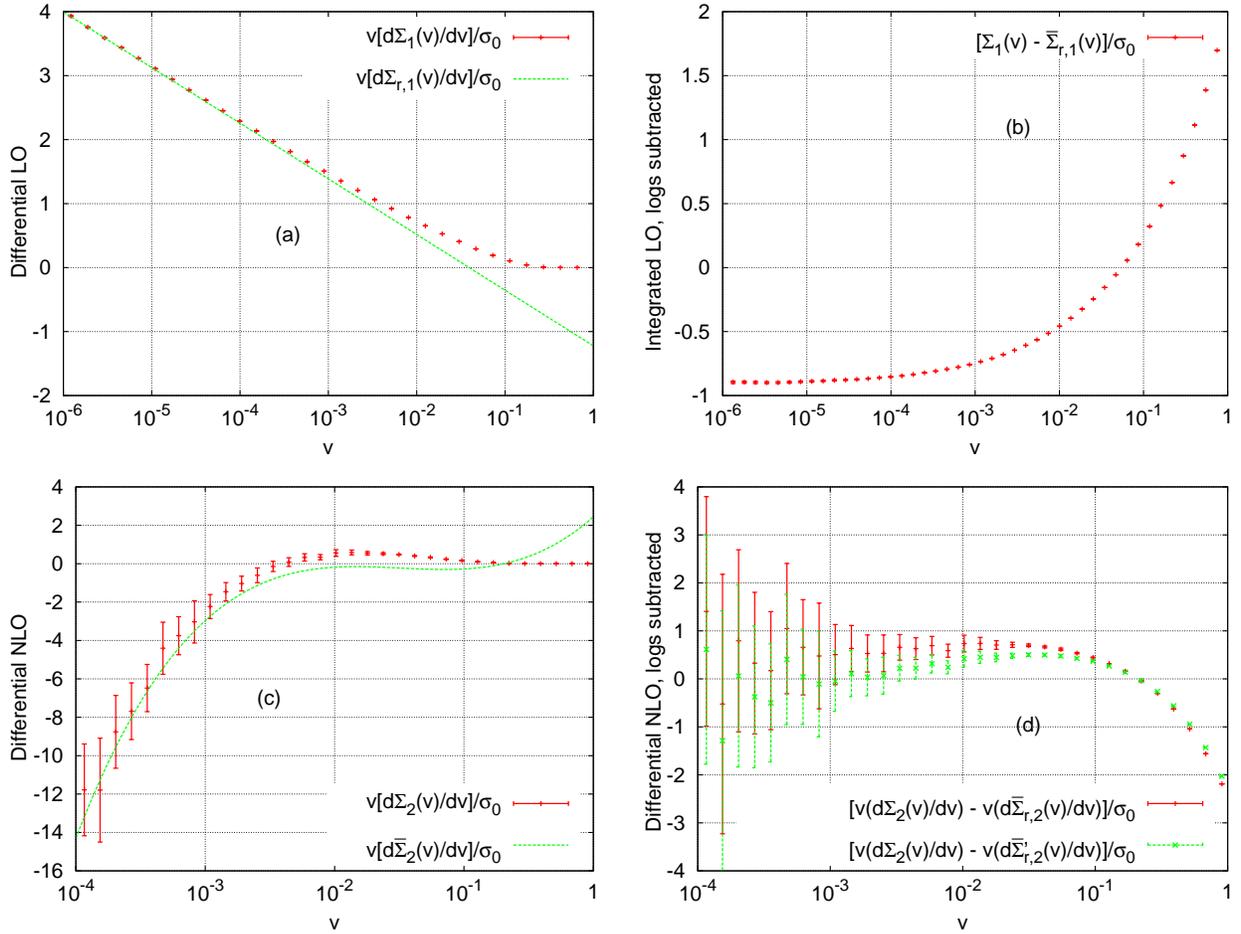}
  \caption{Comparisons of logarithms predicted by the resummation with
    the exact $\order{\as}$ results (a,b) and the $\order{\as^2}$
    results from \nlojet for $\tau_{\perp,g}$.
    Shown for the Tevatron energy and cuts, with $p_{t1} > 200\GeV$.
  \label{fig:logs-correct-othr}
  }
\end{figure}

At order $\as$, the distribution $\Sigma(v)$ of
eq.~(\ref{eq:SigmaIntcut}) has the expansion at small $v$
\begin{equation}
  \label{eq:Sigmar1}
%  \sigma_1(v) = \sum_a \as (H_{12,a} L^2 + H_{11,a} L + C_{1,a})
%  \sigma_{0,a} + \order{v}
  \begin{split}
  \Sigma_{1}(v) & = H_{12} L^2 + H_{11} L + H_{10} 
  = \sum_a (H_{12}^{(a)} L^2 + H_{11}^{(a)} L + H_{10}^{(a)})\,,\quad (v\ll 1)\,,\\
     H_{10}^{(a)} & = \sum_{\subProc \in a} \sigma_0^{(\subProc)} \VEV{\as
    C_1^{(\subProc)}}\,,\qquad L \equiv \ln \frac{1}{v}\,,
  \end{split}
\end{equation}
%where $L \equiv \ln(1/v)$ and $\sigma_0^{(a)}(v)$ is the Born cross
%section for colour channel $a$. 
where $H_{nm}^{(a)}$ is the coefficient of $L^m$, has the dimension of
a cross section and implicitly contains $\as^n$ (notice that
$H_{nm}^{(a)}$ is of order $\as^{n+2}$).
A NLL resummation predicts $H_{12}^{(a)}$ and $H_{11}^{(a)}$, while
$H_{10}^{(a)}$ is obtained from the coefficient $\VEV{\as
  C_1^{(\subProc)}}$ of eq.~(\ref{eq:C1_subProc}) by summing over all
subprocesses $\subProc$ corresponding to the same colour channel $a$,
as indicated in eq.~(\ref{eq:Sigmar1}). The constant $H_{10}^{(a)}$
can be extracted from the exact fixed cross sections $\sigma_1^{(a)}$
and $\bar\Sigma_1^{(a)}(v)$, defined as in
eq.~\eqref{eq:Sigma-from-diff}, as follows:
\begin{equation}
  \label{eq:H10a}
  H_{10}^{(a)} = \sigma_1^{(a)}+
  \lim_{v\to 0}\left[\bar\Sigma_1^{(a)}(v)-
    \left(H_{12}^{(a)} L^2 + H_{11}^{(a)} L\right)\right]\,.
\end{equation}
Fig.~\ref{fig:logs-correct-othr}a shows the prediction for the
differential distribution $v [d\Sigma_{1,r}(v)/dv]$ obtained from
eq.~(\ref{eq:Sigmar1}), compared to the exact result $v
[d\Sigma_1(v)/dv]$ from \nlojet, for the total transverse thrust
$\tau_{\perp,g}$. The two distributions agree at small $v$.

Since fig.~\ref{fig:logs-correct-othr}a contains large logarithms,
a better visual constraint can be obtained by plotting the difference
between $\Sigma_1(v)$ and its logarithmically-enhanced part
$H_{12}L^2+H_{11}L$, which should go to a constant at small $v$, and
indeed does. By performing this exercise separately for each colour
channel one can obtain the $H_{10}^{(a)}$ individually, and can also
verify that $\Sigma_1^{(\mathrm{other})}(v)$ vanishes for small
$v$. From $H_{10}^{(a)}$ one can extract the colour-decomposed average
coefficient constant $\VEV{\as C_1^{(a)}} =
H_{10}^{(a)}/\sigma_0^{(a)}$.  

The coefficients $\VEV{\as C_1^{(a)}}$ obtained in this way are not
precisely the ones that multiply the resummed distribution according
to either of the two matching procedures described in
section~\ref{sec:matchNLLNLO}, because there one resums not logarithms
of $v$ but of a rescaled quantity $X_V v$, eq.~(\ref{eq:Ltilde}).
To get an idea of the size of the $\order{\as}$ term as it is relevant
in the matched resummations, instead of plotting
$\Sigma_1(v)-(H_{12}L^2+H_{11}L)$, in
fig.~\ref{fig:logs-correct-othr}b we plot the difference between
$\Sigma_1(v)$ and the distribution
\begin{equation}
  \label{eq:Sigmabar-r1}
  \bar \Sigma_{r,1}(v) = \bar H_{12} \bar L^2+\bar H_{11} \bar L = 
  H_{12} L^2+ H_{11} L+H'_{10}\,,\qquad
%  H_{12} L^2+ H_{11} L+\bar H_{10}\,,\qquad
  \bar L \equiv \ln\frac{1}{\bar X_V v}\,,
\end{equation}
where $\bar X_V$ is the constant $X_V$ of eq.~(\ref{eq:Xv}) computed
for the reference Born configuration used for the analysis of the
event-shape properties in \caesar (two hard jets in the centre-of-mass
frame with an angle $\theta^*$ with respect to the beam corresponding
to $\cos\theta^*=0.2$).
The constants $\bar H_{1m}$ and $H'_{10}$ are defined in terms of the
$H_{1m}$ so as to give equality between the middle and right-hand
sides of eq.~(\ref{eq:Sigmabar-r1}).
One observes that the difference $\Sigma_1(v)-\bar \Sigma_{r,1}(v)$ in
fig.~\ref{fig:logs-correct-othr}b (normalised to $\sigma_0$) goes, as
expected, to a constant.
That constant should be of order $\as$, whereas numerically it is
$\order{1}$. However, we also know from table~\ref{tab:cross-sections}
that the order $\as$ corrections can come with large coefficients.

Given the $\VEV{\as C_1^{(a)}}$ and the corresponding NLL resummations
$f^{(a)}(v)$, one can predict the NNLL$_\Sigma$ terms in the $\as$ expansion of
$\Sigma(v)$, \ie terms $\as^{n}L^p$ with $n\!-\!2 \le p \le
n$. Specifically, to second order in $\as$, we have
\begin{equation}
  \label{eq:Sigmar2}
  \Sigma_{2}(v) = H_{24}L^4+ H_{23} L^3+H_{22}
  L^2+H_{21}L+H_{20}\,,\qquad (v\ll1)\,,
\end{equation}
and to NNLL$_\Sigma$ accuracy we should control $H_{24}$, $H_{23}$ and
$H_{22}$. 
To see that this is the case we compare $\Sigma_2(v)$ to the
resummation prediction for the modified integrated distribution
\begin{equation}
  \label{eq:Sigmabar-r2}
  \bar\Sigma_{r,2}(v) = \bar H_{24}\bar L^4+ \bar H_{23} \bar L^3+
  \bar H_{22} \bar L^2 = 
  H_{24}L^4+ H_{23} L^3+H_{22} L^2+H'_{21}L+H'_{20}\,,
%  H_{24}L^4+ H_{23} L^3+H_{22} L^2+\bar H_{21}L+\bar H_{20}\,,
\end{equation}
%again with $\bar L \!\equiv\!\ln[1/(\bar X_V v)]$ and
where again the constants $\bar H_{2m}$ and $H'_{2m}$ are defined so
as to given agreement between the middle and right-hand
sides of eq.~(\ref{eq:Sigmabar-r2}).
Fig.~\ref{fig:logs-correct-othr}c shows the exact second-order
differential distribution $v[d \Sigma_2(v)/dv]$,\footnote{
  As mentioned at the beginning of section~\ref{sec:pert-struct},
  there is an unknown overall constant in $\Sigma_2(v)$, which relates
  to the NNLO coefficient of the dijet cross-section. This is
  irrelevant for us here, since we only use the derivative of
  $\Sigma_2(v)$. }
compared to $v[d \bar
\Sigma_{2,r}(v)/dv]$ obtained from eq.~(\ref{eq:Sigmabar-r2}).
Again one sees good agreement, which is more readily verified by
examining the difference between the two distributions,
fig.~\ref{fig:logs-correct-othr}d, which is supposed to be (and is)
flat (the constant results from differentiation of the $\bar H_{21} L$
term in eq.~(\ref{eq:Sigmabar-r2})).
We also include the result that is obtained (lower points with
errorbars) if one does not carry out the colour decomposition for
$\VEV{\as C_1^{(a)}}$, but just computes $\VEV{\as C_1} =
H_{10}/\sigma_0$. This gives rise to a different expansion,
$\bar\Sigma'_{r,2}(v)$, whose coefficient of $L^2$ is different from
that of $\bar\Sigma_{r,2}(v)$. For $\tau_{\perp,g}$ one notices that
the corresponding difference between the exact result $v[d
\Sigma_2(v)/dv]$ and the distribution $v[d\bar\Sigma'_{2,r}(v)/dv]$
exhibits a hint of a slope at small $\tau_{\perp,g}$, indicating a missing
$\as^2 L^2$ term in $\bar\Sigma'_{2,r}(v)$.

\begin{figure}
  \centering
  \setlength{\unitlength}{\textwidth}
  \includegraphics[width=1.0\textwidth]{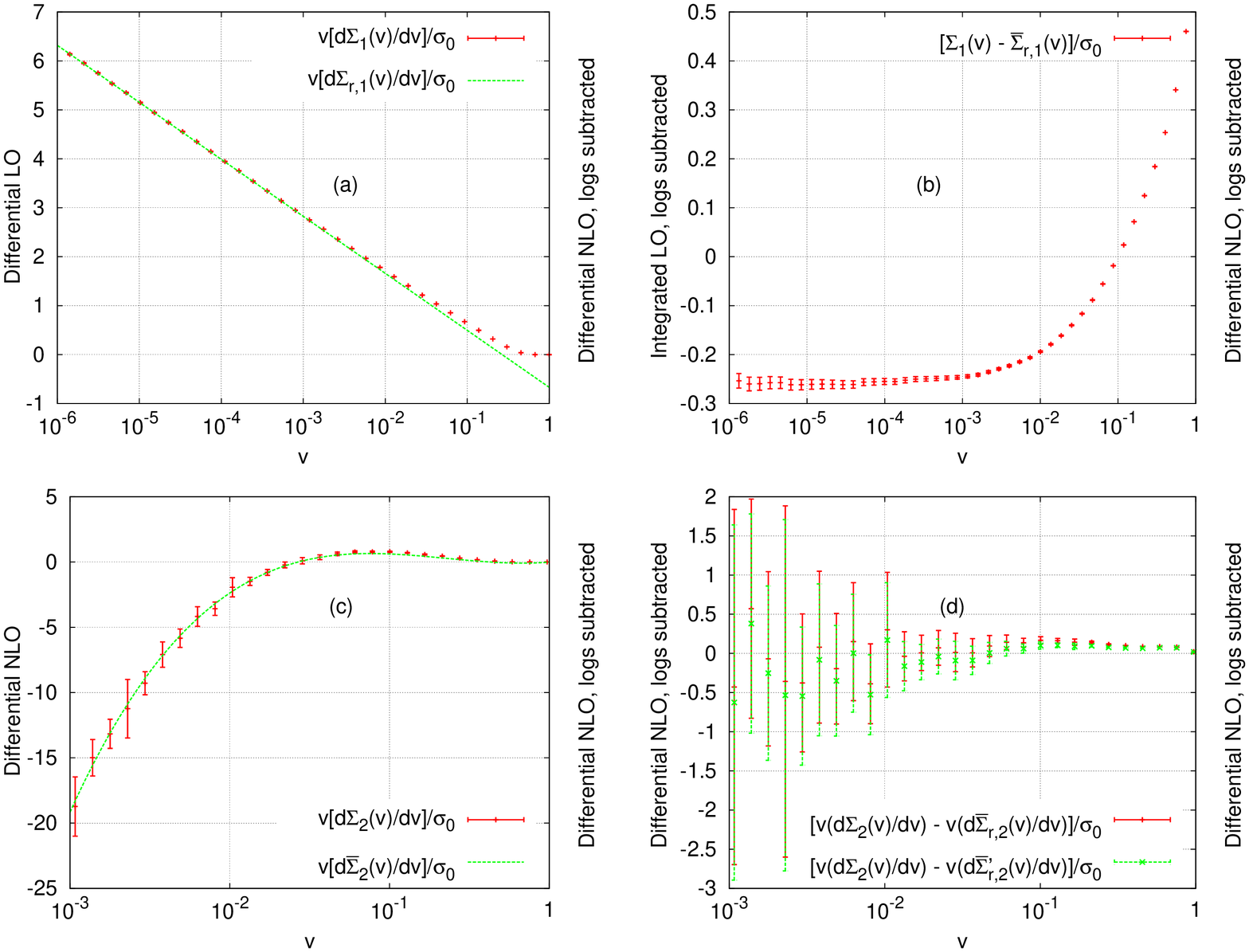}
  \caption{Comparisons of logarithms predicted by the resummation with
    the exact $\order{\as}$ results (a,b) and the $\order{\as^2}$
    results from \nlojet for $T_{m,g}$.
    Shown for the Tevatron energy and cuts, with $p_{t1} > 200\GeV$.
  \label{fig:logs-correct-tmin}
  }
\end{figure}

Fig.~\ref{fig:logs-correct-tmin} shows the same comparison of
fig.~\ref{fig:logs-correct-othr} for the global thrust-minor
$T_{m,g}$.  In this case one is not able, within errorbars, to see any
difference between a resummed prediction containing $\VEV{\as
C_1^{(a)}}$, giving the correct $H_{22}$, and one based on $\VEV{\as
C_1}$, as is evident from Fig.~\ref{fig:logs-correct-tmin}d. This is
possibly due to the fact that the difference between the full
$\cO{\as}$ results and the first order expansion of the resummation,
shown in Fig.~\ref{fig:logs-correct-tmin}a, is small. 

\subsection{Weighted recombination in NLO calculations}
\label{sec:weight-recomb-nlo}

NLO Monte Carlo calculations for multi-jet processes are highly CPU
intensive. 
Consequently, one carries out multiple calculations (runs), spread
across many CPUs, and averages them so as to get the final result.
The correct way of determining the average is to weight each run in
proportion to its number of events.
In practice, however, it is common for the distribution of each run to
contain one or two bins that are ``outliers'', obviously inconsistent
with the distribution as a whole, and which are a consequence of
a handful of real and subtracted NLO events with very large
opposite-sign weights that end up in different bins.
These outliers lead to visible anomalies also in the number-weighted
average and make it almost impossible to use the final distribution
without some (often questionable) prescription to deal with the
outlying bins.

A common alternative to number-weighted averaging is, for each bin of
a run, to choose a weight that is inversely proportional to the square
of the bin's error in that run.
This is an option for example in \nlojet (and is implicit also for the
total cross section in programs like MCFM \cite{MCFM} that use VEGAS).
Since outlier bins tend to have much larger errors than normal bins,
they contribute little to the average, resulting in much smoother
final distributions.
However, the error-weighted averaging procedure introduces a bias,
because there tends to be a correlation between the value in a bin and
its error: for example, in event samples with positive-definite
weights, it is well known that runs with larger bin values also
have larger errors, and the final error-weighted average systematically
undershoots the correct result.

\begin{figure}[t]
  \centering
  \includegraphics[width=1.0\textwidth,height=0.4\textwidth]{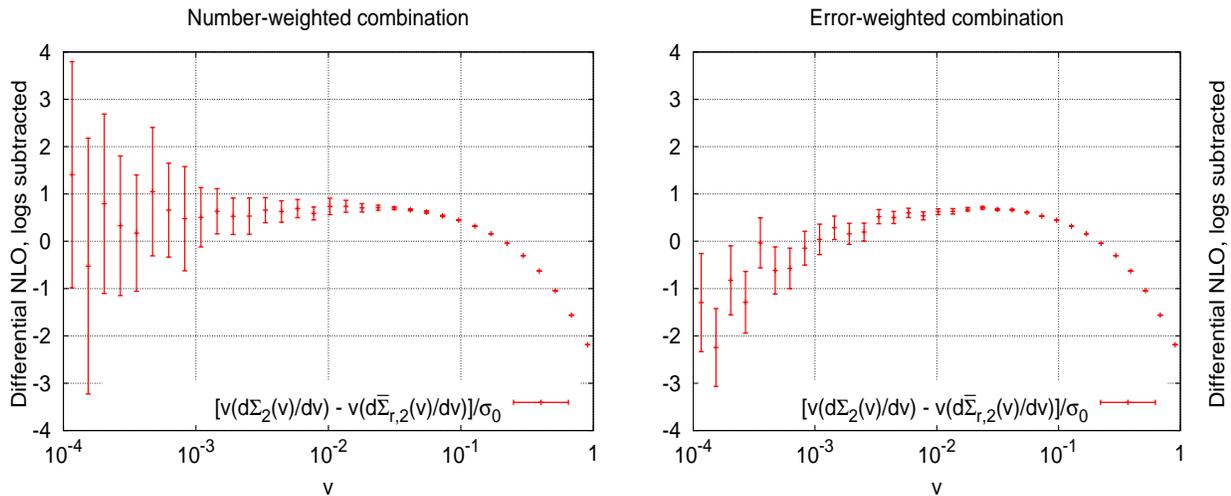}
  \caption{Comparisons of logarithms predicted by the resummation with
    the exact $\order{\as^2}$ results from \nlojet. The left-hand plot
    corresponds to the case with (event) number-based averaging of NLO
    results from separate runs. 
    In the right-hand plot, for each bin of the NLO results, each run
    has been given a weight inversely proportional to the square root
    of the error on that bin in the run.
    The results are for the $\tau_{\perp,g}$ observable, with Tevatron
    energy and cuts and $p_{t1} > 200\GeV$.
    \label{fig:nlo-logs-wgt}
  }
\end{figure}

Fig.~\ref{fig:nlo-logs-wgt} shows the analogue of
fig.~\ref{fig:logs-correct-othr}, comparing event number-based and
error-based weighting. At 
large negative values of $L$ there is a clear slope, i.e.\ the bias in
the error-weighted procedure causes the result to disagree with the
expectations based on resummation.
Only with number-weighted averaging does one obtain results like
fig.~\ref{fig:logs-correct-othr}, which show agreement between the
logarithmic structure of the NLO and resummed calculations.

So as to deal systematically with the issue of outlying bins
figs.~\ref{fig:logs-correct-othr} and \ref{fig:logs-correct-tmin} use
a modified version of the number-based weighting, as
follows. One first determines an  error-based average for a bin
$b^{(w)}$, and a corresponding uncertainty on its contents
$\sigma^{(w)}$ --- this provides an estimate for the correct
value. 
One then carries out the number-based average with the following
modification: for a given bin, one excludes runs whose result is
further than $N \sigma^{(w)}$ from the $b^{(w)}$ (we use $N=100$ for
15 runs; $N$ should scale as the square-root of the number of runs).
This then gives us a final result that is smooth and with a
substantially reduced bias relative to an error-weighted
recombination.

Note that in the phenomenological plots of
sections~\ref{sec:pert-struct} and \ref{sec:pert-res} we have used the
error-based recombination weights. On one hand the bias that it
introduces is modest compared to uncertainties from subleading
effects. On the other, some of our runs used
Rambo~\cite{Kleiss:1985gy} phase space and others the
dipole~\cite{Catani:1996vz} phase space, and this automatically
privileges whichever of the two gives best convergence in a given
phase-space region.

\section{Comment on effect of forward rapidity cut}
\label{sec:appB}

For both generic global event shapes and those with an
exponentially suppressed forward term, in order to satisfy the
globalness requirement needed for the NLL resummation, we included all
particles in the event, including those in the forward/backward
regions. Experimentally however, it is not possible to perform
measurements up to infinite rapidity. At the Tevatron the forward
detector coverage goes up to $y \simeq 3.5$ and, at the LHC, measurements up to
$y \simeq 5$ are viable.
Theoretical arguments suggest that as long as the event-shape's value
is not too small, the effect of not including forward emissions should
be negligible~\cite{KoutZ0}, specifically if $v\gtrsim v_{\min}$, with
$v_{\min}$ given by~\cite{Banfi:2004nk}
\begin{equation}
  \label{eq:vmin}
%  v_{\min} \sim e^{-(a+\min(b_1,b_2)) \eta_{\rm max}}\,,
  v_{\min} \sim e^{-(a+b_{1,2}) \eta_{\rm max}}\,,
\end{equation}
where the $a$ and $b_i$ parameters were discussed in
section~\ref{sec:preCAESAR}.
Examining the pure resummed distributions in~\cite{Banfi:2004nk}, we
came to the conclusion that the result in eq.~(\ref{eq:vmin}) for $v_{\min}$
ensured that the cutoff would usually have an impact only well below
the maximum of the distributions.
Here we supplement this analysis with a numerical study 
that investigates the impact of the rapidity cut in practice.

For this purpose we compute the NLO+NLL prediction
using a rapidity cut on input particles for the NLO part of the
calculation and compare this to the full NLO+NLL without forward
rapidity cuts.
In parallel we carry out an estimate using a Monte Carlo event
generator, since it is straightforward to run it with a rapidity cut.

\begin{figure}[htbp]
  \centering
  \includegraphics[width=\textwidth]{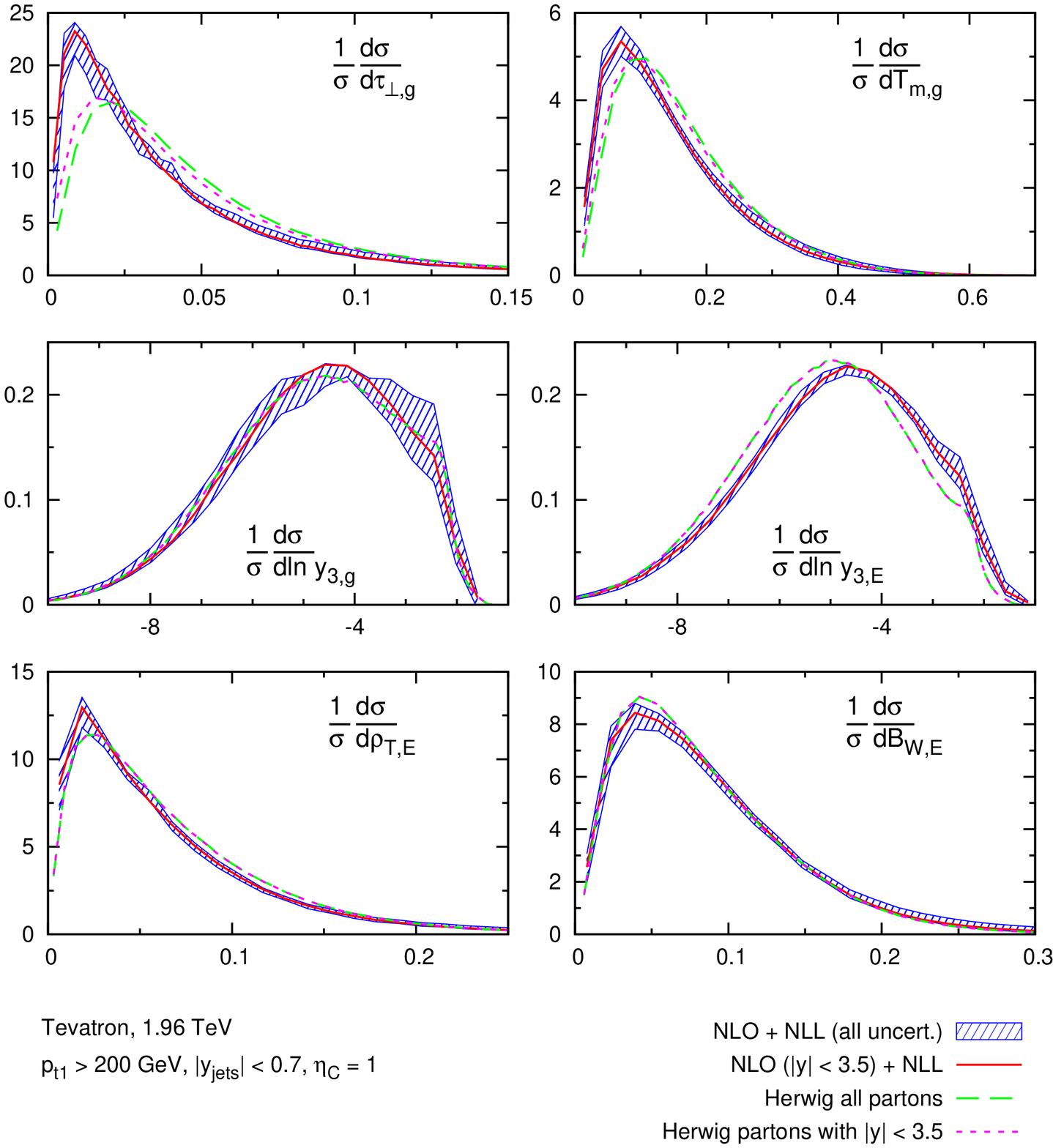}
  \caption{Resummed distributions for a selection of event shapes with and without forward rapidity cut for the high-$p_t$ sample at the Tevatron.}
  \label{fig:ycut-tev200}
\end{figure}

\begin{figure}[htbp]
  \centering
  \includegraphics[width=\textwidth]{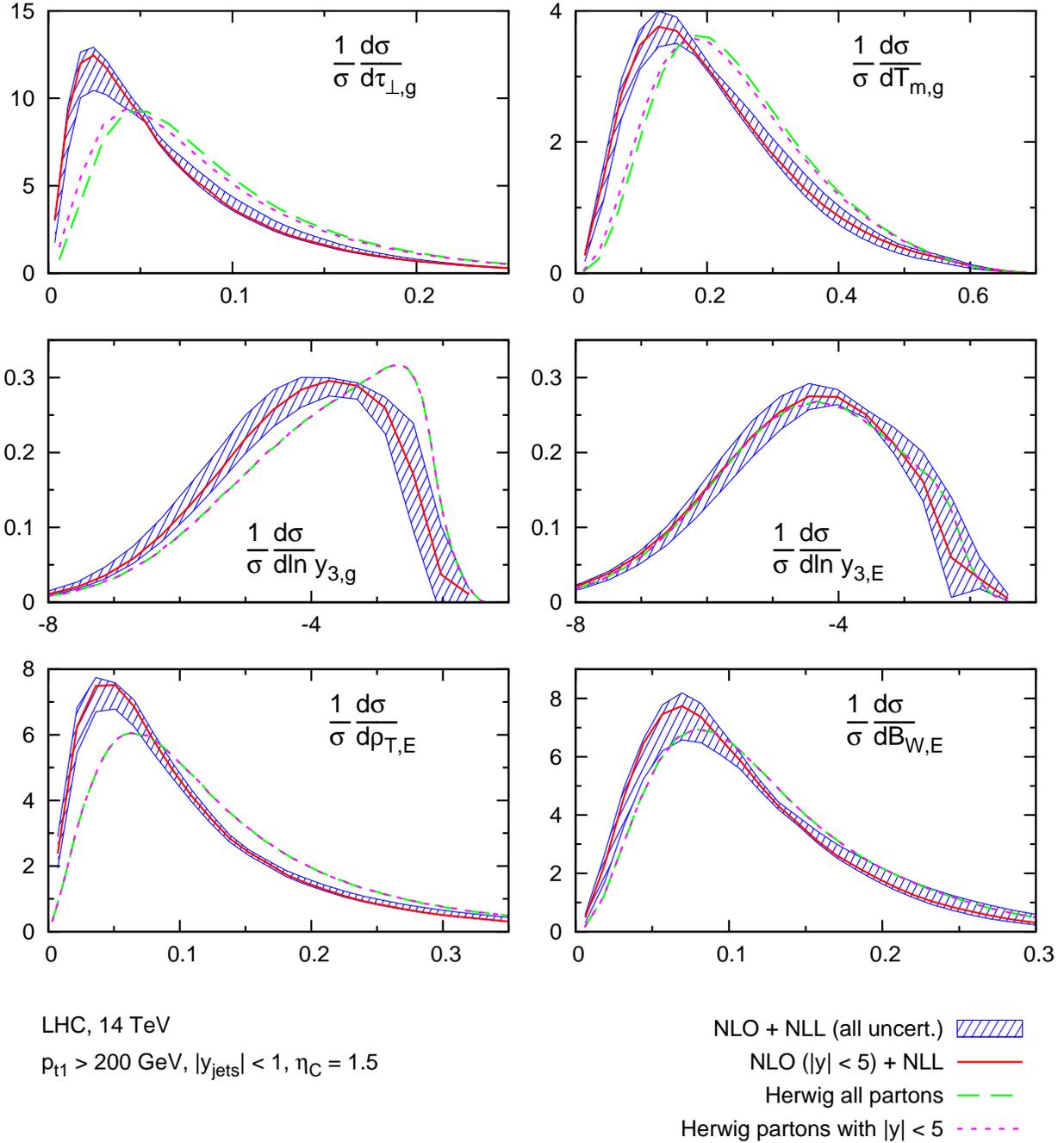}
  \caption{Same as fig.~\ref{fig:ycut-tev200} for the low-$p_t$ sample
    at the LHC.}
  \label{fig:ycut-lhc200}
\end{figure}

Fig.~\ref{fig:ycut-tev200}  shows  comparisons
between NLO+NLL with (solid line) and without the cut (full
uncertainty band), as well as the corresponding Monte
Carlo predictions obtained with  Herwig (without UE) at parton level at the
Tevatron ($p_{t1} > 200$ GeV). 
Fig.~\ref{fig:ycut-lhc200} shows the corresponding results at LHC
($p_{t1}>200$ GeV).
The results with the rapidity 
cut are always contained in the full uncertainty band of the results without.
Furthermore, there is in general very little difference between the
two Monte Carlo predictions, with the exception of the directly global
transverse thrust, which is the observable most sensitive to
forward emissions, as the weight of emissions in the forward region is
large compared to that of emissions inside the jets. We note that this is also
the one observable where the difference between Monte Carlo
predictions and NLO+NLL is largest.

% remember to eliminate this newpage once the spacings are settled
\newpage

%======================================================================


\begin{thebibliography}{99}



\bibitem{SU3}
S.~Kluth et al.,
%``A measurement of the QCD colour factors using event shape distributions  at s**(1/2) = 14-GeV to 189-GeV,''
Eur.\ Phys.\ J.\ C {\bf 21}, 199 (2001) and references therein.
%\epjc{21}{2001}{199} [hep-ex/0012044] and references therein.
%%CITATION = HEP-EX 0012044;%%

\bibitem{Bethke:2002rv}
S.~Bethke,
%``alpha(s) 2002,''
Nucl.\ Phys.\ Proc.\ Suppl.\  {\bf 121} (2003) 74
%\npps{121}{2003}{74}
[hep-ex/0211012].
%%CITATION = HEP-EX 0211012;%%


\bibitem{Herwig}
G.~Marchesini, B.~R.~Webber, G.~Abbiendi, I.~G.~Knowles, M.~H.~Seymour
and L.~Stanco, 
%``HERWIG: A Monte Carlo event generator for simulating hadron emission reactions with interfering gluons. Version 5.1 - April 1991,''
Comput.\ Phys.\ Commun.\  {\bf 67} (1992) 465.
%\cpc{67}{1992}{465};
%%CITATION = CPHCB,67,465;%%
%\bibitem{Corcella:2000bw}
G.~Corcella {\it et al.},
%``HERWIG 6: An event generator for hadron emission reactions with  interfering gluons (including supersymmetric processes),''
JHEP {\bf 0101} (2001) 010
%\jhep{01}{2001}{010}
[hep-ph/0011363].
%%CITATION = HEP-PH 0011363;%%


%\cite{Sjostrand:2006za}
\bibitem{Sjostrand:2006za}
  T.~Sjostrand, S.~Mrenna and P.~Skands,
  %``PYTHIA 6.4 Physics and Manual,''
  JHEP {\bf 0605} (2006) 026
  [arXiv:hep-ph/0603175].
  %%CITATION = JHEPA,0605,026;%%



\bibitem{Ariadne}
L.~L\"onnblad,
%``ARIADNE version 4: A Program for simulation of QCD cascades implementing the color dipole model,''
Comput.\ Phys.\ Commun.\ {\bf 71} (1992) 15.
%\cpc{71}{1992}{15}.
%%CITATION = CPHCB,71,15;

\bibitem{TuningDelphi}
P.~Abreu {\it et al.}  (DELPHI Collaboration),
%``Tuning and test of fragmentation models based on identified particles  and precision event shape data,''
Z.\ Phys.\ {\bf C73} (1996) 11.
%\zpc{73}{1996}{11}.
%%CITATION = ZEPYA,C73,11;%%

\bibitem{DMW} 
  G.~P.~Korchemsky and G.~Sterman,
  % ``Nonperturbative corrections in resummed cross-sections,''
  Nucl.\ Phys.\ B {\bf 437} (1995) 415
  [hep-ph/9411211];
  %% CITATION = HEP-PH 9411211;%%
  %
  Y.~L.~Dokshitzer and B.~R.~Webber,
  % ``Calculation of power corrections to hadronic event shapes,''
  Phys.\ Lett.\ B {\bf 352} (1995) 451
  [hep-ph/9504219];
  %% CITATION = HEP-PH 9504219;%%
  %
  M.~Beneke and V.~M.~Braun,
  % ``Power corrections and renormalons in Drell-Yan production,''
  Nucl.\ Phys.\ B {\bf 454} (1995) 253
  [hep-ph/9506452];
  %% CITATION = HEP-PH 9506452;%%
  %
  R.~Akhoury and V.~I.~Zakharov,
  % ``Leading power corrections in QCD: From renormalons to phenomenology,''
  Nucl.\ Phys.\ B {\bf 465} (1996) 295
  [hep-ph/9507253].
  %% CITATION = HEP-PH 9507253;%%
  %
  Yu.~L.~Dokshitzer, G.~Marchesini and B.~R.~Webber,
  % ``Dispersive Approach to Power-Behaved Contributions in QCD Hard Processes,''
  Nucl.\ Phys.\ B {\bf 469}, 93 (1996) [hep-ph/9512336]; 
  % \npb{469}{1996}{93}  [hep-ph/9512336]; 
  %% CITATION = HEP-PH 9512336;%%
  %
  M.~Dasgupta and B.~R.~Webber,
  % ``Power corrections to event shapes in deep inelastic scattering,''
  Eur.\ Phys.\ J.\ C {\bf 1} (1998) 539
  [hep-ph/9704297];
  %% CITATION = HEP-PH 9704297;%%
  %
  Y.~L.~Dokshitzer and B.~R.~Webber,
  % ``Power corrections to event shape distributions,''
  Phys.\ Lett.\ B {\bf 404} (1997) 321
  [hep-ph/9704298].
  %% CITATION = HEP-PH 9704298;%%
  % 
  Y.~L.~Dokshitzer, A.~Lucenti, G.~Marchesini and G.~P.~Salam,
  % ``On the universality of the Milan factor for 1/Q power corrections to  jet
  % shapes,''
  JHEP {\bf 9805}, 003 (1998)
  % \jhep{05}{1998}{003} 
  [hep-ph/9802381];
  %% CITATION = HEP-PH 9802381;%%
  % 
  G.~P.~Korchemsky and G.~Sterman,
  % ``Power corrections to event shapes and factorization,''
  Nucl.\ Phys.\ B {\bf 555} (1999) 335
  [hep-ph/9902341];
  %% CITATION = HEP-PH 9902341;%%
  % 
  M.~Dasgupta, L.~Magnea and G.~Smye,
  % ``Universality of 1/Q corrections revisited,''
  JHEP {\bf 9911} (1999) 025
  [hep-ph/9911316];
  %% CITATION = HEP-PH 9911316;%
  % 
  E.~Gardi and J.~Rathsman,
  %``The thrust and heavy-jet mass distributions in the two-jet region,''
  Nucl.\ Phys.\  B {\bf 638} (2002) 243
  [arXiv:hep-ph/0201019];
  %%CITATION = NUPHA,B638,243;%%
  %
  C.~Lee and G.~Sterman,
  %``Universality of nonperturbative effects in event shapes,''
  {\it In the Proceedings of FRIF workshop on first principles non-perturbative QCD of hadron jets, LPTHE, Paris, France, 12-14 Jan 2006, pp A001}
  [arXiv:hep-ph/0603066].
  %%CITATION = ECONF,C0601121,A001;%%

\bibitem{Beneke}
M.~Beneke,
%``Renormalons,''
Phys.\ Rept.\  {\bf 317}, 1 (1999).
%\prep{317}{1999}{1}
[hep-ph/9807443].
%%CITATION = HEP-PH 9807443;%%

\bibitem{DasSalReview}
M.~Dasgupta and G.~P.~Salam,
%``Event shapes in e+ e- annihilation and deep inelastic scattering,''
J.\ Phys.\ G {\bf 30} (2004) R143
%\jphg{30}{2004}{R143}
[hep-ph/0312283].
%%CITATION = HEP-PH 0312283;%%

\bibitem{Broad}
  Y.~L.~Dokshitzer, A.~Lucenti, G.~Marchesini and G.~P.~Salam,
  %``On the {QCD} analysis of jet broadening,''
  JHEP {\bf 9801} (1998) 011
  [arXiv:hep-ph/9801324].
  %%CITATION = JHEPA,9801,011;%%


\bibitem{NG} 
M.~Dasgupta and G.~P.~Salam,
%``Resummation of non-global QCD observables,''
Phys.\ Lett.\ B {\bf 512}, 323 (2001)
%\plb{512}{2001}{323} 
[hep-ph/0104277];
%%CITATION = HEP-PH 0104277;%%
%M.~Dasgupta and G.~P.~Salam,
%``Accounting for coherence in interjet E(t) flow: A case study,''
JHEP {\bf 0203} (2002) 017
%\jhep{03}{2002}{017}
[hep-ph/0203009].

%\cite{Banfi:2001bz}
%\bibitem{Banfi:2001bz}
\bibitem{y3}
A.~Banfi, G.~P.~Salam and G.~Zanderighi,
  %``Semi-numerical resummation of event shapes,''
  JHEP {\bf 0201} (2002) 018
  [arXiv:hep-ph/0112156].
  %%CITATION = JHEPA,0201,018;%%

\bibitem{DISENTError}
  V.~Antonelli, M.~Dasgupta and G.~P.~Salam,
  %``Resummation of thrust distributions in DIS,''
  JHEP {\bf 0002}, 001 (2000)
  [arXiv:hep-ph/9912488].
  %%CITATION = JHEPA,0002,001;%%


%\cite{GehrmannDeRidder:2007hr}
%\bibitem{GehrmannDeRidder:2007hr}
\bibitem{NNLOZurich}
A.~Gehrmann-De Ridder, T.~Gehrmann, E.~W.~N.~Glover and G.~Heinrich,
  %``NNLO corrections to event shapes in $e^+e^-$ annihilation,''
  JHEP {\bf 0712} (2007) 094
  [arXiv:0711.4711 [hep-ph]].
  %%CITATION = JHEPA,0712,094;%%


%\cite{Becher:2008cf}
%\bibitem{Becher:2008cf}
\bibitem{NNNLLthrust}
  T.~Becher and M.~D.~Schwartz,
  %``A Precise determination of $\alpha_s$ from LEP thrust data using effective
  %field theory,''
  JHEP {\bf 0807} (2008) 034
  [arXiv:0803.0342 [hep-ph]].
  %%CITATION = JHEPA,0807,034;%%


%\cite{Weinzierl:2008iv}
%\bibitem{Weinzierl:2008iv}
\bibitem{NNLOWeinzierl}
  S.~Weinzierl,
  %``NNLO corrections to 3-jet observables in electron-positron annihilation,''
  Phys.\ Rev.\ Lett.\  {\bf 101} (2008) 162001
  [arXiv:0807.3241 [hep-ph]].
  %%CITATION = PRLTA,101,162001;%%

\bibitem{DSreview}
  M.~Dasgupta and G.~P.~Salam,
  %``Event shapes in e+ e- annihilation and deep inelastic scattering,''
  J.\ Phys.\ G {\bf 30} (2004) R143
  [arXiv:hep-ph/0312283].
  %%CITATION = JPHGB,G30,R143;%%

\bibitem{CDF-Broadening}
F.~Abe {\it et al.}  [CDF Collaboration],
 %``Measurement of QCD jet broadening in p anti-p collisions at S**(1/2) =
%1.8-TeV,''
Phys.\ Rev.\ D {\bf 44} (1991) 601.
%\prd{44}{1991}{601}.
%%CITATION = PHRVA,D44,601;%%

\bibitem{D0Thrust}
I.~A.~Bertram  [D0 Collaboration],
%``Jet Results At The D0 Experiment,''
Acta Phys.\ Polon.\ B {\bf 33}, 3141 (2002).
%\appol{B 33}{2002}{3141}.
%%CITATION = APPOA,B33,3141;%%

\bibitem{NewCDFMesropian}
Christina Mesropian, Talk given at the ``XVII International Workshop on Deep-Inelastic Scattering and Related Subjects'',
DIS 2009, 26-30 April 2009, Madrid, see
  \url{http://indico.cern.ch/materialDisplay.py?contribId=286\&sessionId=3\&materialId=slides\&confId=53294}


\bibitem{NLOJET}
Z.~Nagy,
%``Three-jet cross sections in hadron hadron collisions at next-to-leading  order,''
Phys.\ Rev.\ Lett.\  {\bf 88}, 122003 (2002);
%\prl{88}{2002}{122003}
[hep-ph/0110315];\\
%%CITATION = HEP-PH 0110315;%%
  Z.~Nagy,
  %``Next-to-leading order calculation of three jet observables in hadron hadron
  %collision,''
  Phys.\ Rev.\  D {\bf 68} (2003) 094002
  [arXiv:hep-ph/0307268].
  %%CITATION = PHRVA,D68,094002;%%


\bibitem{KoutZ0}
A.~Banfi, G.~Marchesini, G.~Smye and G.~Zanderighi,
%``Out-of-plane QCD radiation in hadronic Z0 production,''
JHEP {\bf 0108} (2001) 047
%\jhep{08}{2001}{047}
[hep-ph/0106278].
%%CITATION = HEP-PH 0106278;%%

%\cite{Banfi:2004nk}
\bibitem{Banfi:2004nk}
  A.~Banfi, G.~P.~Salam and G.~Zanderighi,
  %``Resummed event shapes at hadron hadron colliders,''
  JHEP {\bf 0408} (2004) 062
  [arXiv:hep-ph/0407287].
  %%CITATION = JHEPA,0408,062;%%

%\cite{Stewart:2009yx}
\bibitem{Stewart:2009yx}
  I.~W.~Stewart, F.~J.~Tackmann and W.~J.~Waalewijn,
  %``Factorization at the LHC: From PDFs to Initial State Jets,''
  arXiv:0910.0467 [hep-ph].
  %%CITATION = ARXIV:0910.0467;%%


%\cite{Dissertori:2008es}
\bibitem{Dissertori:2008es}
  G.~Dissertori, F.~Moortgat and M.~A.~Weber,
  %``Hadronic Event-Shape Variables at CMS,''
  arXiv:0810.3208 [hep-ph].
  %%CITATION = ARXIV:0810.3208;%%

\bibitem{Thaler:2008ju}
  J.~Thaler and L.~T.~Wang,
  %``Strategies to Identify Boosted Tops,''
  JHEP {\bf 0807} (2008) 092
  [arXiv:0806.0023 [hep-ph]].
  %%CITATION = JHEPA,0807,092;%%

\bibitem{Almeida:2008tp}
  L.~G.~Almeida, S.~J.~Lee, G.~Perez, I.~Sung and J.~Virzi,
  %``Top Jets at the LHC,''
  Phys.\ Rev.\  D {\bf 79}, 074012 (2009)
  [arXiv:0810.0934 [hep-ph]].
  %%CITATION = PHRVA,D79,074012;%%

%\cite{Almeida:2008yp}
\bibitem{Almeida:2008yp}
  L.~G.~Almeida, S.~J.~Lee, G.~Perez, G.~Sterman, I.~Sung and J.~Virzi,
  %``Substructure of high-p_T Jets at the LHC,''
  Phys.\ Rev.\  D {\bf 79}, 074017 (2009)
  [arXiv:0807.0234 [hep-ph]].
  %%CITATION = PHRVA,D79,074017;%%

\bibitem{Salam:2009jx}
  G.~P.~Salam,
  %``Towards Jetography,''
  arXiv:0906.1833 [hep-ph].
  %%CITATION = ARXIV:0906.1833;%%

%\cite{Brown:1990nm}
\bibitem{Brown:1990nm}
  N.~Brown and W.~J.~Stirling,
  %``Jet cross-sections at leading double logarithm in e+ e- annihilation,''
  Phys.\ Lett.\  B {\bf 252} (1990) 657.
  %%CITATION = PHLTA,B252,657;%%


%\cite{Banfi:2004yd}
%\bibitem{Banfi:2004yd}
\bibitem{caesar}
  A.~Banfi, G.~P.~Salam and G.~Zanderighi,
  %``Principles of general final-state resummation and automated
  %implementation,''
  JHEP {\bf 0503} (2005) 073
  [arXiv:hep-ph/0407286].
  %%CITATION = JHEPA,0503,073;%%


\bibitem{CTTW}
  S.~Catani, L.~Trentadue, G.~Turnock and B.~R.~Webber,
  %``Resummation of large logarithms in e+ e- event shape distributions,''
  Nucl.\ Phys.\  B {\bf 407} (1993) 3.
  %%CITATION = NUPHA,B407,3;%%

%\cite{deFlorian:2004mp}
\bibitem{deFlorian:2004mp}
  D.~de Florian and M.~Grazzini,
  %``The back-to-back region in e+ e- energy energy correlation,''
  Nucl.\ Phys.\  B {\bf 704} (2005) 387
  [arXiv:hep-ph/0407241].
  %%CITATION = NUPHA,B704,387;%%



%\cite{deFlorian:2000pr}
\bibitem{deFlorian:2000pr}
  D.~de Florian and M.~Grazzini,
  %``Next-to-next-to-leading logarithmic corrections at small transverse
  %momentum in hadronic collisions,''
  Phys.\ Rev.\ Lett.\  {\bf 85} (2000) 4678
  [arXiv:hep-ph/0008152].
  %%CITATION = PRLTA,85,4678;%%

%\cite{Bozzi:2003jy}
\bibitem{Bozzi:2003jy}
  G.~Bozzi, S.~Catani, D.~de Florian and M.~Grazzini,
  %``The q(T) spectrum of the Higgs boson at the LHC in QCD perturbation
  %theory,''
  Phys.\ Lett.\  B {\bf 564} (2003) 65
  [arXiv:hep-ph/0302104].
  %%CITATION = PHLTA,B564,65;%%

\bibitem{Gehrmann:2008kh}
  T.~Gehrmann, G.~Luisoni and H.~Stenzel,
  %``Matching NLLA+NNLO for event shape distributions,''
  Phys.\ Lett.\  B {\bf 664} (2008) 265
  [arXiv:0803.0695 [hep-ph]].
  %%CITATION = PHLTA,B664,265;%%

%%CITATION = HEP-PH 0203009;%%
%\cite{Banfi:2002hw}
\bibitem{Banfi:2002hw}
  A.~Banfi, G.~Marchesini and G.~Smye,
  %``Away-from-jet energy flow,''
  JHEP {\bf 0208} (2002) 006
  [arXiv:hep-ph/0206076].
  %%CITATION = JHEPA,0208,006;%%

\bibitem{Weigert:2003mm}
  H.~Weigert,
  %``Non-global jet evolution at finite N(c),''
  Nucl.\ Phys.\  B {\bf 685} (2004) 321
  [arXiv:hep-ph/0312050].
  %%CITATION = NUPHA,B685,321;%%


%\cite{Forshaw:2006fk}
\bibitem{Forshaw:2006fk}
  J.~R.~Forshaw, A.~Kyrieleis and M.~H.~Seymour,
  %``Super-leading logarithms in non-global observables in QCD,''
  JHEP {\bf 0608} (2006) 059
  [arXiv:hep-ph/0604094].
  %%CITATION = JHEPA,0608,059;%%

\bibitem{Forshaw:2008cq}
  J.~R.~Forshaw, A.~Kyrieleis and M.~H.~Seymour,
  %``Super-leading logarithms in non-global observables in QCD: Colour basis
  %independent calculation,''
  JHEP {\bf 0809} (2008) 128
  [arXiv:0808.1269 [hep-ph]].
  %%CITATION = JHEPA,0809,128;%%

\bibitem{website}
  A.~Banfi, G.~P.~Salam and G.~Zanderighi,\\
  \url{http://www.lpthe.jussieu.fr/~salam/pp-event-shapes/}\,.


\bibitem{DISresum}
%\cite{Dasgupta:2002dc}
%\bibitem{Dasgupta:2002dc}
  M.~Dasgupta and G.~P.~Salam,
  %``Resummed event-shape variables in DIS,''
  JHEP {\bf 0208} (2002) 032
  [arXiv:hep-ph/0208073].
  %%CITATION = JHEPA,0208,032;%%

\bibitem{mweber}
M.~Weber, private communication. 





\bibitem{Farhi}
E.~Farhi,
%``A QCD Test For Jets,''
Phys.\ Rev.\ Lett.\  {\bf 39} (1977) 1587.
%%CITATION = PRLTA,39,1587;%%

\bibitem{FoxWolf}
G.~C.~Fox and S.~Wolfram,
%``Observables For The Analysis Of Event Shapes In E+ E- Annihilation And Other Processes,''
Phys.\ Rev.\ Lett.\  {\bf 41} (1978) 1581;
%%CITATION = PRLTA,41,1581;%%
ibid.,
%``Event Shapes In E+ E- Annihilation,''
Nucl.\ Phys.\ B {\bf 149} (1979) 413
[Erratum-ibid.\ B {\bf 157} (1979) 543].
%%CITATION = NUPHA,B149,413;%%

\bibitem{BSZFpar} A.~Banfi, G.~P.~Salam, G.~Zanderighi, unpublished
  2004, at \url{http://www.qcd-caesar.org/} and
  \url{http://home.fnal.gov/~zanderi/Caesar/Observables/Obs_ee/node7.html}.

\bibitem{KtHH}
  S.~Catani, Y.~L.~Dokshitzer, M.~H.~Seymour and B.~R.~Webber,
  ``Longitudinally invariant K(t) clustering algorithms for hadron hadron
  collisions,''
  Nucl.\ Phys.\ B {\bf 406}, 187 (1993).
  %\npb{406}{1993}{187}.
  %%CITATION = NUPHA,B406,187;%%
  %

\bibitem{Kt-EllisSoper}
  S.~D.~Ellis and D.~E.~Soper,
  ``Successive Combination Jet Algorithm For Hadron Collisions,''
  Phys.\ Rev.\ D {\bf 48}, 3160 (1993)
  %\prd{48}{1993}{3160} 
  [hep-ph/9305266]. 
  %%CITATION = HEP-PH 9305266;%%



%\cite{Blazey:2000qt}
\bibitem{RunII-jet-physics}
G.~C.~Blazey {\it et al.},
%``Run II jet physics,''
hep-ex/0005012.
%%CITATION = HEP-EX 0005012;%%

\bibitem{Clavelli}
L.~Clavelli,
%``Jet Invariant Mass In Quantum Chromodynamics,''
Phys.\ Lett.\ B {\bf 85} (1979) 111;
%\plb{85}{1979}{111};
%%CITATION = PHLTA,B85,111;%%
%\bibitem{Chandramohan:1980ry}
T.~Chandramohan and L.~Clavelli,
%``Consequences Of Second Order QCD For Jet Structure In E+ E- Annihilation,''
Nucl.\ Phys.\ B {\bf 184} (1981) 365;
%\npb{184}{1981}{365};
%%CITATION = NUPHA,B184,365;%%
L.~Clavelli and D.~Wyler,
%``Kinematical Bounds On Jet Variables And The Heavy Jet Mass Distribution,''
Phys.\ Lett.\ B {\bf 103} (1981) 383.
%\plb{103}{1981}{383}.
%%CITATION = PHLTA,B103,383;%%

\bibitem{Aad:2009wy}
  G.~Aad {\it et al.}  [The ATLAS Collaboration],
  %``Expected Performance of the ATLAS Experiment - Detector, Trigger and
  %Physics,''
  arXiv:0901.0512 [hep-ex].
  %%CITATION = ARXIV:0901.0512;%%

\bibitem{Ball:2007zza}
  G.~L.~Bayatian {\it et al.}  [CMS Collaboration],
  %``CMS technical design report, volume II: Physics performance,''
  J.\ Phys.\ G {\bf 34} (2007) 995.
  %%CITATION = JPHGB,G34,995;%%

\bibitem{DiscontGlobal}
M.~Dasgupta and G.~P.~Salam,
%``Resummed event-shape variables in DIS,''
JHEP {\bf 0208}, 032 (2002)
%\jhep{08}{2002}{032}
[hep-ph/0208073].
%%CITATION = HEP-PH 0208073;%%

\bibitem{jetflav}
  A.~Banfi, G.~P.~Salam and G.~Zanderighi,
  %``Infrared safe definition of jet flavour,''
  Eur.\ Phys.\ J.\  C {\bf 47} (2006) 113
  [arXiv:hep-ph/0601139].
  %%CITATION = EPHJA,C47,113;%%

\bibitem{CSS}
J.~C.~Collins, D.~E.~Soper and G.~Sterman,
 %``Transverse Momentum Distribution In Drell-Yan Pair And W And Z Boson
%Production,''
Nucl.\ Phys.\ B {\bf 250} (1985) 199.
%\npb{250}{1985}{199}.
%%CITATION = NUPHA,B250,199;%%



%----------------------------------------------
% not cited... 

% \bibitem{KtJet}
% J.~M.~Butterworth, J.~P.~Couchman, B.~E.~Cox and B.~M.~Waugh,
% %``KtJet: A C++ implementation of the K(T) clustering algorithm,''
% Comput.\ Phys.\ Commun.\  {\bf 153}, 85 (2003)
% %\cpc{153}{2003}{85} 
% [hep-ph/0210022]. 
% %%CITATION = HEP-PH 0210022;%%


\bibitem{BottsSterman}
J.~Botts and G.~Sterman,
%``Hard Elastic Scattering In QCD: Leading Behavior,''
Nucl.\ Phys.\ B {\bf 325}, 62 (1989).
%%CITATION = NUPHA,B325,62;%%
%;\\

\bibitem{KS}
N.~Kidonakis and G.~Sterman,
%``Subleading logarithms in QCD hard scattering,''
Phys.\ Lett.\ B {\bf 387} (1996) 867;
%%CITATION = PHLTA,B387,867;%%
%N.~Kidonakis and G.~Sterman,
%``Resummation for QCD hard scattering,''
Nucl.\ Phys.\ B {\bf 505} (1997) 321
[hep-ph/9705234].
%%CITATION = HEP-PH 9705234;%%

\bibitem{KOS}
N.~Kidonakis, G.~Oderda and G.~Sterman,
%``Evolution of color exchange in {QCD} hard scattering,''
Nucl.\ Phys.\ B {\bf 531}, 365 (1998)
[hep-ph/9803241]. 
%%CITATION = HEP-PH 9803241;%%

\bibitem{Oderda}
G.~Oderda,
%``Dijet rapidity gaps in photoproduction from perturbative {QCD},''
Phys.\ Rev.\ D {\bf 61} (2000) 014004
[hep-ph/9903240].
%%CITATION = HEP-PH 9903240;%%

\bibitem{KidonakisOwens}
N.~Kidonakis and J.~F.~Owens,
%``Effects of higher-order threshold corrections in high-E(T) jet  production,''
Phys.\ Rev.\ D {\bf 63}, 054019 (2001)
[hep-ph/0007268].
%%CITATION = HEP-PH 0007268;%%

\bibitem{KodairaTrentadue}
  J.~Kodaira and L.~Trentadue,
  % ``Summing Soft Emission In QCD,''
  Phys.\ Lett.\ B {\bf 112} (1982) 66;
  % \plb{112}{1982}{66};
  %% CITATION = PHLTA,B112,66;%%
  J.~Kodaira and L.~Trentadue,
  % ``Single Logarithm Effects In Electron - Positron Annihilation,''
  Phys.\ Lett.\ B {\bf 123} (1983) 335.
  % \plb{123}{1983}{335}.
  %% CITATION = PHLTA,B123,335;%%

\bibitem{CollinsSoper}
  J.~C.~Collins and D.~E.~Soper,
  % ``Back-To-Back Jets In QCD,''
  Nucl.\ Phys.\ B {\bf 193} (1981) 381
  %\npb{193}{1981}{381}
  [Erratum-ibid.\ B {\bf 213} (1983) 545];
  %[Erratum \ibid{B 213}{1983}{545}];
  %% CITATION = NUPHA,B193,381;%%
  J.~C.~Collins and D.~E.~Soper,
  % ``Back-To-Back Jets: Fourier Transform From B To K-Transverse,''
  Nucl.\ Phys.\ B {\bf 197} (1982) 446.
  %\npb{197}{1982}{446}.
  %% CITATION = NUPHA,B197,446;%%

\bibitem{Cam}
  Y.~L.~Dokshitzer, G.~D.~Leder, S.~Moretti and B.~R.~Webber,
  ``Better jet clustering algorithms,''
  JHEP {\bf 9708}, 001 (1997)
  [hep-ph/9707323];
  %%CITATION = HEP-PH 9707323;%%

\bibitem{Salam:2008qg}
  G.~P.~Salam and J.~Rojo,
  %``A Higher Order Perturbative Parton Evolution Toolkit (HOPPET),''
  Comput.\ Phys.\ Commun.\  {\bf 180}, 120 (2009)
  [arXiv:0804.3755 [hep-ph]].
  %%CITATION = CPHCB,180,120;%%

\bibitem{BSZ03}
A.~Banfi, G.~P.~Salam and G.~Zanderighi,
%``Generalized resummation of QCD final-state observables,''
Phys.\ Lett.\ B {\bf 584}, 298 (2004)
%\plb{584}{2004}{298} 
[hep-ph/0304148].
%%CITATION = HEP-PH 0304148;%%

\bibitem{Coherence}
V.~S.~Fadin,
%``Double Logarithmic Asymptotics Of The Cross-Sections Of E+ E- Annihilation
%Into Quarks And Gluons. (In Russian),''
Yad.\ Fiz.\  {\bf 37} (1983) 408;
%\yf{37}{1983}{408} [\sjnp{37}{1983}{245}];\\
%%CITATION = YAFIA,37,408;%%
%
B.~I.~Ermolaev and V.~S.~Fadin,
%``Log - Log Asymptotic Form Of Exclusive Cross-Sections In Quantum
%Chromodynamics,''
%\jetpl{33}{1981}{269}
JETP Lett.\  {\bf 33} (1981) 269
[Pisma Zh.\ Eksp.\ Teor.\ Fiz.\  {\bf 33} (1981) 285];\\
%%CITATION = JTPLA,33,269;%%
%
%\bibitem{Mueller:1981ex}
A.~H.~Mueller,
%``On The Multiplicity Of Hadrons In QCD Jets,''
%\plb{104}{1981}{161};\\
Phys.\ Lett.\ B {\bf 104}, 161 (1981);
%%CITATION = PHLTA,B104,161;%%
%
% following is an application of coherence, not proof (says Yuri!)
Y.~L.~Dokshitzer, V.~S.~Fadin and V.~A.~Khoze,
%``Double Logs Of Perturbative QCD For Parton Jets And Soft Hadron Spectra,''
Z.\ Phys.\ C {\bf 15} (1982) 325;
%\zpc{15}{1982}{325};\\
%%CITATION = ZEPYA,C15,325;%%
A.~Bassetto, M.~Ciafaloni and G.~Marchesini,
%``Jet Structure And Infrared Sensitive Quantities In Perturbative QCD,''
Phys.\ Rept.\  {\bf 100} (1983) 201.
%\prep{100}{1983}{201}.
%%CITATION = PRPLC,100,201;%%

\bibitem{Salam:2007xv}
  G.~P.~Salam and G.~Soyez,
  %``A Practical Seedless Infrared-Safe Cone jet algorithm,''
  JHEP {\bf 0705} (2007) 086
  [arXiv:0704.0292 [hep-ph]].
  %%CITATION = JHEPA,0705,086;%%

\bibitem{Pumplin:2002vw}
  J.~Pumplin, D.~R.~Stump, J.~Huston, H.~L.~Lai, P.~M.~Nadolsky and W.~K.~Tung,
  %``New generation of parton distributions with uncertainties from global QCD
  %analysis,''
  JHEP {\bf 0207} (2002) 012
  [arXiv:hep-ph/0201195].
  %%CITATION = JHEPA,0207,012;%%

\bibitem{FastJet} 
  M.~Cacciari and G.~P.~Salam,
  ``Dispelling the $N^3$ myth for the k(t) jet-finder,''
  Phys.\ Lett.\  B {\bf 641} (2006) 57
  [arXiv:hep-ph/0512210];\\
  %%CITATION = PHLTA,B641,57;%%
  M.~Cacciari, G.~P.~Salam and G.~Soyez,
  \url{http://fastjet.fr/}\,.

%\cite{Chekanov:2001fw}
\bibitem{Chekanov:2001fw}
  S.~Chekanov {\it et al.}  [ZEUS Collaboration],
  %``Dijet production in neutral current deep inelastic scattering at HERA,''
  Eur.\ Phys.\ J.\  C {\bf 23} (2002) 13
  [arXiv:hep-ex/0109029].
  %%CITATION = EPHJA,C23,13;%%

% \cite{Aktas:2003ja}
\bibitem{Aktas:2003ja}
  A.~Aktas {\it et al.}  [H1 Collaboration],
  % ``Inclusive dijet production at low Bjorken-x in deep inelastic
  % scattering,''
  Eur.\ Phys.\ J.\  C {\bf 33} (2004) 477
  [arXiv:hep-ex/0310019].
  %% CITATION = EPHJA,C33,477;%%


\bibitem{KK}
  M.~Klasen and G.~Kramer,
  % ``Dijet cross-sections at o (alpha alpha-s**2) in photon - proton
  % collisions,''
  Phys.\ Lett.\ B {\bf 366}, 385 (1996)
  % \plb{366}{1996}{385} 
  [hep-ph/9508337].
  %% CITATION = HEP-PH 9508337;%%


\bibitem{symm-cuts}
  S.~Frixione and G.~Ridolfi,
  % ``Jet photoproduction at HERA,''
  Nucl.\ Phys.\ B {\bf 507} (1997) 315
  % \npb{507}{1997}{315}
  [hep-ph/9707345].
%%CITATION = HEP-PH 9707345;%%

%\cite{Banfi:2003jj}
\bibitem{Banfi:2003jj}
  A.~Banfi and M.~Dasgupta,
  %``Dijet rates with symmetric E(t) cuts,''
  JHEP {\bf 0401} (2004) 027
  [arXiv:hep-ph/0312108].
  %%CITATION = JHEPA,0401,027;%%

%\cite{Parisi:1979se}
\bibitem{Parisi:1979se}
  G.~Parisi and R.~Petronzio,
  %``Small Transverse Momentum Distributions In Hard Processes,''
  Nucl.\ Phys.\  B {\bf 154} (1979) 427.
  %%CITATION = NUPHA,B154,427;%%

\bibitem{Dasgupta:2001eq}
  M.~Dasgupta and G.~P.~Salam,
  %``Resummation of the jet broadening in DIS,''
  Eur.\ Phys.\ J.\  C {\bf 24} (2002) 213
  [arXiv:hep-ph/0110213].
  %%CITATION = EPHJA,C24,213;%%

\bibitem{Catani:1997xc}
  S.~Catani and B.~R.~Webber,
  %``Infrared safe but infinite: Soft gluon divergences inside the physical
  %region,''
  JHEP {\bf 9710} (1997) 005
  [arXiv:hep-ph/9710333].
  %%CITATION = JHEPA,9710,005;%%


\bibitem{alpgen}
%\cite{Mangano:2002ea}
%\bibitem{Mangano:2002ea}
  M.~L.~Mangano, M.~Moretti, F.~Piccinini, R.~Pittau and A.~D.~Polosa,
  %``ALPGEN, a generator for hard multiparton processes in hadronic
  %collisions,''
  JHEP {\bf 0307} (2003) 001
  [arXiv:hep-ph/0206293].
  %%CITATION = JHEPA,0307,001;%%


\bibitem{Alwall:2007fs}
  J.~Alwall {\it et al.},
  %``Comparative study of various algorithms for the merging of parton showers
  %and matrix elements in hadronic collisions,''
  Eur.\ Phys.\ J.\  C {\bf 53} (2008) 473
  [arXiv:0706.2569 [hep-ph]].
  %%CITATION = EPHJA,C53,473;%%

\bibitem{Catani:2001cc}
  S.~Catani, F.~Krauss, R.~Kuhn and B.~R.~Webber,
  %``QCD Matrix Elements + Parton Showers,''
  JHEP {\bf 0111} (2001) 063
  [arXiv:hep-ph/0109231].
  %%CITATION = JHEPA,0111,063;%%

\bibitem{Lai:1999wy}
  H.~L.~Lai {\it et al.}  [CTEQ Collaboration],
  %``Global QCD analysis of parton structure of the nucleon: CTEQ5 parton
  %distributions,''
  Eur.\ Phys.\ J.\  C {\bf 12} (2000) 375
  [arXiv:hep-ph/9903282].
  %%CITATION = EPHJA,C12,375;%%

\bibitem{Skands:2007zg}
  P.~Skands and D.~Wicke,
  %``Non-perturbative QCD effects and the top mass at the Tevatron,''
  Eur.\ Phys.\ J.\  C {\bf 52} (2007) 133
  [arXiv:hep-ph/0703081].

\bibitem{Skands:2009zm}
  P.~Z.~Skands,
  %``The Perugia Tunes,''
  arXiv:0905.3418 [hep-ph].
  %%CITATION = ARXIV:0905.3418;%%

\bibitem{Buckley:2009vk}
  A.~Buckley, H.~Hoeth, H.~Lacker, H.~Schulz and E.~von Seggern,
  %``Monte Carlo tuning and generator validation,''
  arXiv:0906.0075 [hep-ph].
  %%CITATION = ARXIV:0906.0075;%%


%\cite{Banfi:2001aq}
\bibitem{Banfi:2001aq}
  A.~Banfi, G.~Marchesini, G.~Smye and G.~Zanderighi,
  %``Out-of-plane QCD radiation in hadronic Z0 production,''
  JHEP {\bf 0108} (2001) 047
  [arXiv:hep-ph/0106278].
  %%CITATION = JHEPA,0108,047;%%

%\cite{Dasgupta:2007hr}
\bibitem{Dasgupta:2007hr}
  M.~Dasgupta and Y.~Delenda,
  %``Aspects of power corrections in hadron-hadron collisions,''
  JHEP {\bf 0711} (2007) 013
  [arXiv:0709.3309 [hep-ph]].
  %%CITATION = JHEPA,0711,013;%%


%\cite{Dasgupta:2007wa}
\bibitem{Dasgupta:2007wa}
  M.~Dasgupta, L.~Magnea and G.~P.~Salam,
  %``Non-perturbative QCD effects in jets at hadron colliders,''
  JHEP {\bf 0802} (2008) 055
  [arXiv:0712.3014 [hep-ph]].
  %%CITATION = JHEPA,0802,055;%%

%\cite{Dasgupta:2009tm}
\bibitem{Dasgupta:2009tm}
  M.~Dasgupta and Y.~Delenda,
  %``On the universality of hadronisation corrections to QCD jets,''
  JHEP {\bf 0907} (2009) 004
  [arXiv:0903.2187 [hep-ph]].
  %%CITATION = JHEPA,0907,004;%%

%\cite{Butterworth:1996zw}
\bibitem{Butterworth:1996zw}
  J.~M.~Butterworth, J.~R.~Forshaw and M.~H.~Seymour,
  %``Multiparton interactions in photoproduction at HERA,''
  Z.\ Phys.\  C {\bf 72} (1996) 637
  [arXiv:hep-ph/9601371].
  %%CITATION = ZEPYA,C72,637;%%

\bibitem{Albrow:2006rt}
  M.~G.~Albrow {\it et al.}  [TeV4LHC QCD Working Group],
  %``Tevatron-for-LHC Report of the QCD Working Group,''
  arXiv:hep-ph/0610012.
  %%CITATION = HEP-PH/0610012;%%

% uses thrust, sphericity, jet mass on all event particles
% seems not to have dealt too much with proper background studies??
\bibitem{Han:2007ae}
  T.~Han, Z.~Si, K.~M.~Zurek and M.~J.~Strassler,
  %``Phenomenology of Hidden Valleys at Hadron Colliders,''
  JHEP {\bf 0807} (2008) 008
  [arXiv:0712.2041 [hep-ph]].
  %%CITATION = JHEPA,0807,008;%%

%  some discussion of thrust and cluster mass (== hemisphere mass?);
%  comments on issues related to UE; dicusses use of hardish jets (>25
%  GeV) as input to the ev.shp to reduce UE. [== dedicated Sec II.F]
\bibitem{Strassler:2008fv}
  M.~J.~Strassler,
  %``On the Phenomenology of Hidden Valleys with Heavy Flavor,''
  arXiv:0806.2385 [hep-ph].
  %%CITATION = ARXIV:0806.2385;%%

\bibitem{Randall:2008rw}
  L.~Randall and D.~Tucker-Smith,
  %``Dijet Searches for Supersymmetry at the LHC,''
  Phys.\ Rev.\ Lett.\  {\bf 101} (2008) 221803
  [arXiv:0806.1049 [hep-ph]].
  %%CITATION = PRLTA,101,221803;%%

\bibitem{Banfi:2007gu}
  A.~Banfi, G.~P.~Salam and G.~Zanderighi,
  %``Accurate QCD predictions for heavy-quark jets at the Tevatron and LHC,''
  JHEP {\bf 0707} (2007) 026
  [arXiv:0704.2999 [hep-ph]].
  %%CITATION = JHEPA,0707,026;%%




\bibitem{MCFM}
John Campbell and Keith Ellis, http://mcfm.fnal.gov/. 




% used in testing our implementation of the transverse spherocity
\bibitem{Kleiss:1985gy}
  R.~Kleiss, W.~J.~Stirling and S.~D.~Ellis,
  %``A New Monte Carlo Treatment Of Multiparticle Phase Space At
  %High-Energies,''
  % aka RAMBO
  Comput.\ Phys.\ Commun.\  {\bf 40} (1986) 359.
  %%CITATION = CPHCB,40,359;%%

\bibitem{Catani:1996vz}
  S.~Catani and M.~H.~Seymour,
  %``A general algorithm for calculating jet cross sections in NLO QCD,''
  Nucl.\ Phys.\  B {\bf 485} (1997) 291
  [Erratum-ibid.\  B {\bf 510} (1998) 503]
  [arXiv:hep-ph/9605323].
  %%CITATION = NUPHA,B485,291;%%

\end{thebibliography}
\end{document}